\documentclass[prb,letterpaper,aps,superscriptaddress,floatfix,twocolumn,nofootinbib,amsmath,showkeys]{revtex4-2}
\pdfoutput=1
\usepackage{multirow}
\usepackage{graphicx}
\usepackage{array}
\usepackage{amsmath}
\usepackage{amsfonts}
\usepackage{amssymb}
\usepackage[normalem]{ulem}
\usepackage[utf8]{inputenc}
\usepackage{dsfont}
\usepackage{xcolor}
\usepackage{slashed}
\usepackage{soul}
\usepackage{comment}
\usepackage{tikz-cd}
\usepackage[hidelinks]{hyperref}
\hypersetup{
    linkcolor=blue,
    citecolor=blue,
    filecolor=blue,
    urlcolor=blue,
    colorlinks=true
}
\usepackage{bm}
\usepackage{makecell}
\usepackage[caption=false]{subfig}

\usepackage{amsthm}
\usepackage{color}

\begin{document}

\title{Crystalline axion electrodynamics in charge-ordered Dirac semimetals}

\author{Julian May-Mann}
\affiliation{Department of Physics, Stanford University, Stanford, CA 94305, USA}
\author{Mark R. Hirsbrunner}
\affiliation{Department of Physics, University of Illinois Urbana-Champaign, Urbana, IL 61801, USA}
\author{Lei Gioia}
\affiliation{Walter Burke Institute for Theoretical Physics, Caltech, Pasadena, CA, USA}
\affiliation{Department of Physics, Caltech, Pasadena, CA, USA}
\affiliation{Perimeter Institute for Theoretical Physics, Waterloo, Ontario N2L 2Y5, Canada}
\author{Taylor L. Hughes}
\affiliation{Department of Physics, University of Illinois Urbana-Champaign, Urbana, IL 61801, USA}

\begin{abstract}
Three-dimensional Dirac semimetals can be driven into an insulating state by coupling to a charge density wave (CDW) order. Here, we consider the quantized crystalline responses of such charge-ordered Dirac semimetals, which we dub Dirac-CDW insulators, in which charge is bound to disclination defects of the lattice. Using analytic and numeric methods we show the following. First, when the CDW is lattice-commensurate, disclination-line defects of the lattice have a quantized charge per length. Second, when the CDW is inversion-symmetric, disclinations of the lattice have a quantized electric polarization. Third, when the CDW is lattice-commensurate and inversion-symmetric, disclinations are characterized by a ``disclination filling anomaly'' --- a quantized difference in the total charge bound to disclination-lines of Dirac-CDW with open and periodic boundaries. We construct an effective response theory that captures the topological responses of the Dirac-CDW insulators in terms of a total derivative term, denoted the $R\wedge F$ term. The $R\wedge F$ term describes the crystalline analog of the axion electrodynamics that are found in Weyl semimetal-CDW insulators. We also use the crystalline responses and corresponding response theories to classify the strongly correlated topological phases of three-dimensions Dirac-semimetals.
\end{abstract}
\maketitle

\section{Introduction}
\label{sec:intro}

Three-dimensional (3D) Dirac semimetals (DSMs) and Weyl semimetals (WSMs) are quintessential examples of gapless phases of matter that harbor topological features~\cite{armitage2018weyl, wan2011topological, burkov2011weyl, yang2011quantum, xu2011chern, young2012dirac,  wang2012dirac, zyuzin2012topological, zyuzin2012weyl, wang2013three, weng2015weyl, jia2016weyl}. These systems are characterized by linear bulk band crossings (nodal points) that are well-described at low energies by the 3D Dirac and Weyl equations, respectively. The gaplessness of Weyl nodes can be viewed as a consequence of band topology. Each node is a monopole of Berry curvature in momentum space, and, since the total number of Berry curvature monopoles must vanish, single Weyl nodes cannot be added or removed. This reasoning also indicates that Weyl nodes come in pairs having opposite monopole charge~\cite{nielsen1983adler,wan2011topological, burkov2011weyl, burkov2018weyl}. The gaplessness of Dirac nodes is also protected, but requires additional symmetry, e.g., time-reversal, inversion, and spatial rotations~\cite{klinkhamer2005emergent, murakami2007phase}. While the individual nodes of the DSM and WSM are locally stable in momentum space, pairs of nodes can acquire a gap. Normally, this cannot occur since the nodes are located at different points in the Brillouin zone. However, if, for example, translation symmetry is broken (such as by charge density wave order) then nodes at different momenta can couple to each other and form a correlated insulator.

It has been shown that when an inversion-symmetric WSM is gapped out via a translation symmetry breaking charge density wave (CDW), the resulting insulator exhibits axion electrodynamics~\cite{wilczek1987two, wang2013chiral, maciejko2014weyl, you2016response, wieder2020axionic}. The axion electrodynamics is described by a 3D $\Theta$ term~\cite{qi2008topological, essin2009magnetoelectric, witten2016fermion}, and when the WSM has inversion symmetry, $\Theta$ is quantized to values of $0$ or $\pi$~\cite{hughes2011inversion, varnava2018surfaces, wieder2018axion}. In addition to axion electrodynamics, the inversion-symmetric Weyl-CDW insulator also exhibits a quantized 3D anomalous Hall response~\cite{halperin1987possible, kohmoto1992diophantine, haldane2004berry}. When inversion-symmetric open boundaries are present, the $\Theta = 0$ and $\Theta = \pi$ Weyl-CDW insulators can be distinguished by their total Hall conductances, which differ by $e^2/h$~\cite{wieder2020axionic}. 

DSMs can similarly be gapped out via a CDW to produce a Dirac-CDW insulator~\cite{zhang2016topological}. The resulting insulator can be understood by treating the DSM as two copies of a WSM related by time-reversal symmetry~\cite{young2012dirac,  wang2012dirac}. This intuition reveals that the Dirac-CDW insulator cannot display a quantized 3D Hall response, as the 3D Hall response is odd under time-reversal symmetry. The Dirac-CDW insulator also has trivial axion electrodynamics (i.e., a vanishing 3D $\Theta$ term) since $\Theta$ is defined $\mod(2\pi)$. Despite the absence of these effects, there is more to the story of the Dirac-CDW insulator topological responses. In this work, we consider the crystalline-electromagnetic (CEM) responses of Dirac-CDW insulators. Similar to how electromagnetic responses describe charge fluctuations induced by electromagnetic fields, crystalline-electromagnetic responses describe how charge fluctuations are induced by lattice deformations (e.g., shears, strains, and defects)~\cite{chaikin1995principles, ran2009one, teo2010topological, jurivcic2012universal, barkeshli2012topological, asahi2012topological, chung2016dislocation, teo2017topological,ramamurthy2017electromagnetic, roy2021dislocation, gioia2021unquantized, teo2013existence,gopalakrishnan2013disclination, benalcazar2014classification,li2020fractional, manjunath2021crystalline, schindler2022topological, hirsbrunner2023crystalline, hirsbrunner2023anomalous}. 

Since Dirac-CDW insulators have rotation symmetry, they inherit a natural sensitivity to disclination defects, which are line-like fluxes of rotation symmetry. However, the exact nature of the CEM response to disclinations depends on: (i) if the CDW is in/commensurate with the lattice and (ii) if the Dirac-CDW insulator has inversion symmetry. When the CDW is commensurate with the lattice, the Dirac-CDW insulator has residual discrete translation symmetry determined by the CDW period. In this case we will show that disclination loops in the 3D Dirac-CDW insulator bind a quantized charge per length. The charge per length is determined by a combination of the Frank angle of the disclination and the length of the CDW period. This response can be considered as a CEM analog of the 3D Hall effect, as the former describes how charge per length is bound to disclination lines, while the latter describes how charge per length is bound to magnetic flux lines. In fact, the disclination response is essentially a layered version of the 2D discrete Wen-Zee response that describes how charge is bound to point-like disclinations of 2D lattices~\cite{wen1992shift, ruegg2013bound, ruegg2013corner, liu2019shift, han2019generalized, li2020fractional, manjunath2020classification, manjunath2021crystalline, may2022crystalline,peterson2021trapped, zhang2022fractional}. Hence, we refer to this response of the 3D Dirac-CDW insulator as the 3D discrete Wen-Zee (dWZ) response. Furthermore, we will show that in the effective response theory, the 3D dWZ response is captured by a layered version of the 2D Wen-Zee term.

Just like the case for Weyl-CDW insulators, we find that there are two distinct classes of Dirac-CDW insulators when inversion symmetry is preserved. The difference between the two classes of insulators manifest as a difference in the parity of charge bound to disclination lines that terminate on the open boundaries of a system. In analogy to axion electrodynamics, we find that the effective response theories for the two classes of Dirac-CDW insulators differ by a quantized total derivative term. This term, called the $R\wedge F$ term, was previously discussed by some of us in Ref.~\onlinecite{may2022topological}. Similar to the 3D $\Theta$ term that describes axion electrodynamics, the $R\wedge F$ term is a total derivative term that leads to anomalous boundary physics that we discuss in detail below. 

Generically we show that when the CDW is commensurate and preserves the inversion symmetry of the DSM, it is possible to define a ``disclination filling anomaly''~\cite{benalcazar2019quantization, khalaf2021boundary}. A disclination filling anomaly occurs when it is impossible to change from periodic to (gapped) open boundary conditions while preserving both inversion symmetry and conserving charge on the disclination line. Alternatively put, we show that when the disclination filling anomaly is present, disclination lines of an inversion-symmetric insulator must bind different amounts of charge for open and periodic boundary conditions. This difference in charge is quantized and cannot be removed by boundary effects, provided they do not break inversion symmetry. This serves to further classify crystalline insulators by considering filling anomalies in the presence of topological defects such as a disclination. 

The rest of this paper is organized as follows: in Section~\ref{sec:DSMintro} we provide an overview of Dirac semimetals, including lattice models, protecting crystalline symmetries, and their unquantized quasi-topological responses. In Section~\ref{sec:DSMCDW} we describe how adding a CDW to the DSM model can give rise to correlated topological crystalline insulators. In Section~\ref{sec:insulatorresponse} we discuss relevant topological response terms, such as the discrete Wen-Zee and $R\wedge F$ terms. Equipped with this knowledge, in Section~\ref{sec:DSMCrystalResp} we analytically and numerically demonstrate the topological crystalline-electromagnetic responses of Dirac-CDW insulators. Finally, in Section~\ref{sec:conclusion} we conclude with a discussion and outlook, and include technical details in the appendices.

\section{Dirac semimetals: lattice model, topology, and responses}
\label{sec:DSMintro}
In this section we will review the relevant properties of DSMs. We will pay particular attention to the symmetries that protect the DSMs. We will also discuss the anomalous responses of the gapless DSM in the presence of probe gauge fields. Physically, we will demonstrate that these anomalous responses manifest as charge bound to disclination defects of the lattice.
\label{sec:DSManomaly}

\subsection{Lattice model}

In this subsection we briefly review some of the band theory of DSMs. Generically, DSMs are classified into two groups: symmorphic DSMs whose low-energy gapless features are protected by symmorphic symmetries and translation symmetry; and non-symmorphic DSMs where Dirac nodes arise from non-symmorphic symmetry-enforcement of band crossings with four-dimensional irreducible representations at high-symmetry points at the Brillouin zone edge~\cite{armitage2018weyl,wang2012dirac,Nagaosa2014,young2012dirac}. We will focus on symmorphic DSMs, and study only a minimal modle hosting a pair of Dirac nodes that are protected by time-reversal, inversion, rotation, and translation symmetry. Such DSMs are known to possess unquantized anomalous responses~\cite{ramamurthy2015patterns,gioia2021unquantized}. 

Our starting point for the symmorphic DSM will be the following Bloch Hamiltonian for spin-1/2 fermions on an orthorhombic lattice~\cite{wang2012dirac},
\begin{equation}
    \begin{split}
        \mathcal{H}_{\text{DSM}}(\bm{k}) &= \sin (k_x a_x) \Gamma_1 + \sin (k_y a_y)  \Gamma_2\\
        & + [\cos (k_za_z)-\cos(\mathcal{K}a_z)]\Gamma_3 \\
        & - b_{xy}[2 - \cos (k_xa_x) - \cos (k_ya_y) ]\Gamma_3,
    \end{split}
    \label{eq:DSMham}
\end{equation} 
where $a_{x,y,z}$ are the lattice constants in the $x$, $y$, and $z$ directions, respectively. The $\Gamma_a$ matrices are defined as 
\begin{equation}
    \begin{split}
        &\Gamma_1 = \sigma^x s^z, \phantom{=} \Gamma_2 = - \sigma^y, \phantom{=} \Gamma_3 = \sigma^z, \\ & \Gamma_4 = \sigma^x s^x ,\phantom{=} \Gamma_5 = \sigma^x s^y.
    \end{split}
    \label{eq:gamma}
\end{equation}
where $\bm{\sigma}$ and $\bm{s}$ are Pauli matrices acting on the sublattice and spin degrees of freedom, respectively. We note that this identification will not be important for our further discussion. Here and throughout we leave any $2\times 2$ identity matrices implicit, (i.e.,  $\Gamma_{2} = - \sigma^{y} s^0$, and $\Gamma_{3} =  \sigma^{z} s^0$, where $s^0$ is the $2\times 2$ identity in spin space). 

The (doubly-degenerate) energy bands of this model are given by
\begin{equation}
    \begin{split}
    E_{\pm}(\bm{k})= &\pm \Big [ \sin^2 (k_x a_x) +  \sin^2 (k_y a_y)\\
    &+ \big(\cos (k_za_z)-\cos(\mathcal{K}a_z)  \\
    &- b_{xy}[2 - \cos (k_xa_x) - \cos (k_ya_y)]\big)^2 \Big]^{1/2}.
    \end{split}
\end{equation}
From this equation we see that the parameter $\mathcal{K}$ determines the location of the Dirac nodes at which the bands become four-fold degenerate: $\bm{k}_{\text{DN}} \equiv  (0,0,\mathcal{K}),$ $E_{\pm}(\bm{k}_{\text{DN}}) = E_{\pm}(-\bm{k}_{\text{DN}}) = 0$. Intuitively, $\mathcal{K}$ parameterizes a process wherein two pairs of doubly-degenerate bands are inverted. At $\mathcal{K}= 0$ the bands are not inverted and have a quadratic band touching at $\bm{k} = 0$. At $\mathcal{K}= \pi$ the bands are fully inverted and have a quadratic band touching at $\bm{k} = (0,0,\pi)$. For other values of $\mathcal{K}$, the two pairs of bands are partially inverted and have linear crossings at $\bm{k} =  (0,0,\pm \mathcal{K})$. Additionally, the parameter $b_{xy}$ determines the gap away from the high symmetry line $k_x=k_y = 0$. As expected, the Bloch Hamiltonian near the four-fold degenerate nodal points takes on the Dirac-form
\begin{equation}
\begin{split}
    \mathcal{H}_{\text{DSM}}(\pm \bm{k}_{\text{DN}}+\bm{q}) &\approx q_x a_x \Gamma_1 + q_y a_y \Gamma_2 \\
    & \pm q_z a_z \sin(\mathcal{K}a_x)\Gamma_3. 
\label{eq:DiracHamEff}\end{split}    
\end{equation}
Here, and throughout, we will take the system to be half-filled, i.e., two electrons per unit cell, such that the chemical potential intersects the Dirac points. 

\subsection{Symmetries}
\label{sec:DSMsymm}

For generic values of $\mathcal{K}$, the Hamiltonian in Eq.~\ref{eq:DSMham} consists of two doubly degenerate bands that meet at a pair of four-fold degenerate Dirac nodes. The degeneracies are stabilized by the combination of: C$_{4z}$ symmetry (four-fold rotation symmetry around the z-axis), $T^z$ translation symmetry (translation symmetry in the z-direction), U$(1)$ charge conservation symmetry, time-reversal symmetry, and inversion symmetry.

C$_{4z}$ rotation symmetry, $T^z$ translation symmetry, and U$(1)$ charge conservation symmetry are all necessary for the Dirac nodes in Eq.~\ref{eq:DiracHamEff} to remain gapless. The C$_{4z}$ operator is given by 
\begin{equation}
    U_4 = \exp\left(i\frac{\pi}{2} [\tfrac{1}{2}\sigma^zs^z - s^z ]\right).
    \label{eq:rotDef}
\end{equation}
where $(U_4)^4 = -1$, since we have spin-$1/2$ fermions. This symmetry prohibits mass terms $\propto \Gamma_{4}$, and $\propto \Gamma_5$. The $T^z$ translation symmetry prevents the Dirac node at $+\bm{k}_{\text{DN}}$ from hybridizing with the Dirac node at $-\bm{k}_{\text{DN}}$. The U$(1)$ charge symmetry prevents the inclusion of superconducting pairing terms. 

The two-fold degeneracy of the bands of Eq.~\ref{eq:DSMham} away from the Dirac nodes is ensured by the combination of time-reversal and inversion symmetry: time-reversal symmetry guarantees that a state with momentum $\bm{k}$ is accompanied by a degenerate state at $-\bm{k}$ with opposite spin ($\bm{\sigma}$) projection. Similarly, inversion symmetry guarantees that for a state with momentum $\bm{k}$, there is a degenerate state at $-\bm{k}$ with the \textit{same} spin ($\bm{s}$) projection. The combination of time-reversal and inversion symmetries therefore guarantees that all states are at least two-fold degenerate. Furthermore, time-reversal and inversion symmetry also protect the four-fold degeneracy of the Dirac nodes at $\bm{k} = \pm \bm{k}_{\text{DN}}$, and prevent the DSM phase from being deformed into a Weyl semimetal phase. For example, adding a time-reversal breaking term, such as a magnetic Zeeman term $\propto s^z$, will split a Dirac nodes into one positive-chirality and one negative-chirality Weyl node. If inversion symmetry is preserved, then this splitting of Dirac nodes must occur in pairs, and if there is a positive chirality Weyl node at $\bm{k}$, there will be a negative chirality Weyl node at  $-\bm{k}$. This may be understood via the requirement that the Berry curvature satisfies $\mathbf{\Omega}(\bm{k})=-\mathbf{\Omega}(-\bm{k})$ because of inversion symmetry. This Dirac node splitting procedure results in a magnetic Weyl semimetal that possesses an anomalous Hall conductivity. Alternatively, if inversion symmetry is broken and time-reversal symmetry is preserved, we may arrive at a time-reversal invariant Weyl semimetal, where a positive chirality Weyl node at $\bm{k}$, is accompanied by a second positive chirality Weyl node at  $-\bm{k}$. This naturally arises as a consequence of time-reversal symmetry which implies $\mathbf{\Omega}(\bm{k})=\mathbf{\Omega}(-\bm{k})$. Analogous to the magnetic WSM case, such a material is known to possess an unquantized momentum anomaly response that describes non-trivial charge responses to dislocations~\cite{PhysRevX.12.031007,gioia2021unquantized,dubinkin2021higher,hirsbrunner2023anomalous} .

The time-reversal operator is given by $\mathcal{T} = -is^y\kappa$, where $\kappa$ is the complex conjugation operator ($\mathcal{T}^2 = -1$ for spin-$1/2$ fermions). For the lattice model in Eq.~\ref{eq:DSMham} there are multiple inequivalent choices of inversion symmetry, which correspond to different choices of inversion center. The different inversion centers are related by a half-translation in a given direction. Since we are going to consider breaking translation symmetry in the $z$-direction, we will pay special attention to the two inversion symmetry definitions that differ by a half-translation in the $z$-direction: 
\begin{equation}
    \begin{split}
        &\mathcal{I}_s = \sigma^z,\\
        &\mathcal{I}_b = e^{ik_z a_z }\sigma^z.
        \label{eq:inversionsDef}
    \end{split}
\end{equation}
 We will call $\mathcal{I}_s$ the ``site-centered" inversion symmetry since it sends $\bm{r} = (r_x,r_y,r_z)\rightarrow (-r_x,-r_y,-r_z)$ and leaves the site $(0,0,0)$ invariant. Similarly, we will call $\mathcal{I}_b$ the ``bond-centered" inversion symmetry since it sends $(r_x,r_y,r_z)\rightarrow (-r_x,-r_y,-r_z+a_z)$ and leaves the bond $(0,0,a_z/2)$ invariant. There are other choices of inversion symmetry that differ by half a translation in either the $x$ or $y$-directions from those above. However, these other inversion symmetry definitions are unimportant for our forthcoming analysis.
 
 The distinction between the site- and bond-centered inversion symmetries has important implications when considering systems with open boundary conditions. Namely, for a system of length $L_z = a_z N_z $ in the $z$-direction (i.e., $N_z$ lattice sites along the $z$-direction) and open boundary conditions, site-centered inversion symmetry necessitates that $N_z$ is odd, while bond-centered inversion symmetry necessitates that $N_z$ is even. We will discuss the interplay between inversion symmetry and CDWs in Sec. \ref{sec:DSMCDW} below.

\subsection{Anomalous topological response}\label{ssec:anomTopResp}
As shown in Ref. \cite{gioia2021unquantized}, the DSM in Eq.~\ref{eq:DSMham} possess an anomalous topological response to fluxes of C$_{4z}$ rotation symmetry. This response can be expressed in terms of probe (non-dynamic) gauge fields for the U$(1)$ charge symmetry ($A_\mu$) and C$_{4z}$ rotation symmetry ($\omega_\mu$) as
\begin{equation}
    \mathcal{L}_{\text{anom}} =\frac{\nu}{2\pi^2} G_{z} \epsilon^{ijk}  \omega_{i}\partial_j A_k,
    \label{eq:DSManomaly}
\end{equation}
where $i,j,k$ run over $t,x,y$, $\nu=2\mathcal{K}/2\pi$, and $G_z = 2\pi/a_z$ is the reciprocal lattice vector along the $z$ direction. The origin of such a term may be understood via the lowest Landau level (LLL) picture wherein a uniform magnetic flux is turned on in the $z$-direction: such a system possesses two LLLs with different C$_{4z}$ charges, giving rise to a filling-type anomaly that protects the bands from gapping out~\cite{gioia2021unquantized}. We also note that Eq.~\ref{eq:DSManomaly} is not gauge invariant for arbitrary values of $\nu$, and gauge invariance is restored by the gapless degrees of freedom at the nodes of the DSM. 

\begin{figure*}
    \centering
    \includegraphics[width=0.9\textwidth]{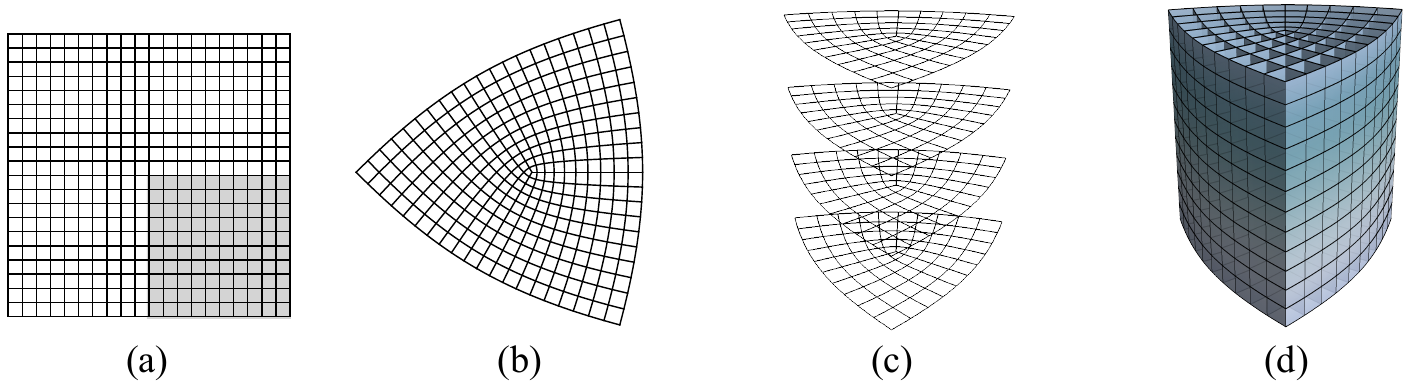}
    \caption[]{(a) A lattice with C$_{4z}$ symmetry and no disclinations. Upon cutting out the greyed-out quadrant and re-gluing, we arrive at (b) a disclinated lattice. (c) A stack of disclinations of 2D layers forms a disclination line of a 3D crystal, as shown in (d).}
    \label{fig:disc}
\end{figure*}

Fluxes of the rotation gauge field, $\omega$, encode disclinations with a Frank-vector $\propto {\hat{z}}$. In 3D lattices, the disclinations are line-defects that extend parallel to their Frank-vector. Because of this, we will implicitly take all disclinations to stretch along the $z$-direction in this work. One can think of these disclination lines as stacks of 2D disclinations of the $xy$-planes, as shown in Fig. \ref{fig:disc}. At the level of gauge field configurations, considering disclinations that stretch along only the $z$-direction amounts to the condition that $\oint_{C_{xz}} \omega = \oint_{C_{yz}} \omega  = 0$, where the loop integrals are over curves $C_{ab}$ in the $ab$-plane that do not intersect the cores of any disclinations. 

While the (discrete) gauge field $\omega$ is formally characterized only by its holonomies, previous works~\cite{may2022topological, may2022crystalline} have shown that it is possible to treat $\omega$ as a continuous gauge field, but having only quantized fluxes, $\oint \omega \in [0,\frac{\pi}{2},\pi, \frac{3\pi}{2})$. This flux quantization can be thought of as arising from a Higgs mechanism, similar to the electromagnetic flux quantization in superconductors, where the analogy of the electric charge is replaced by angular momentum.  Here and throughout, we shall treat $\omega$ as smoothly varying, but with quantized fluxes.

\begin{figure}
\centering
    \includegraphics[width=0.41\textwidth]{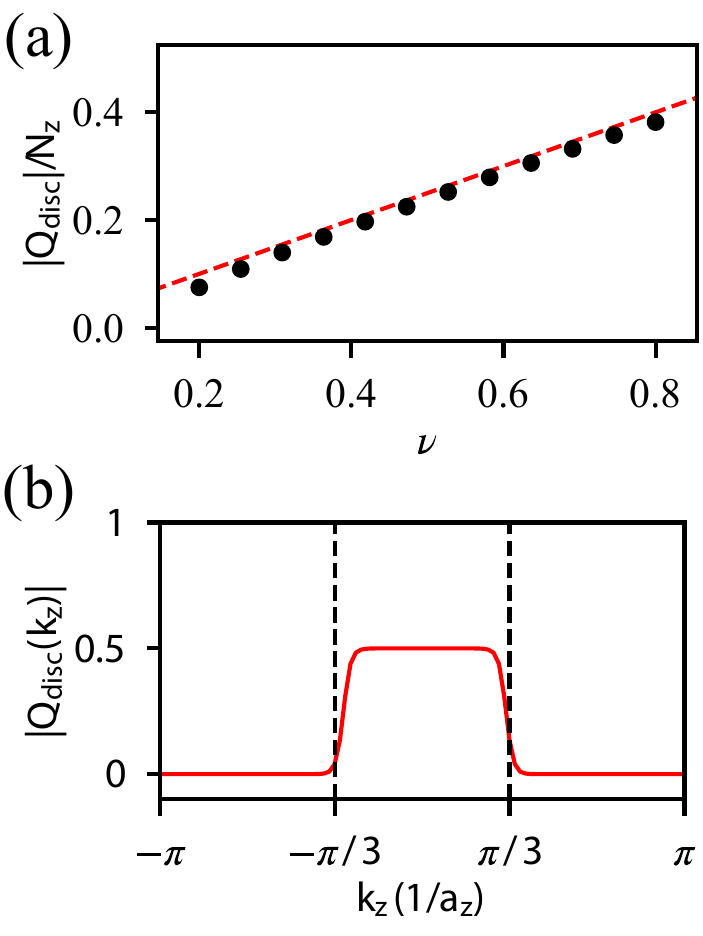}
    \caption[]{(a) The numerically computed charge per layer bound to a $\Theta_F = \pi/2$ disclination as a function of $\nu$ for the Hamiltonian Eq.~\ref{eq:DSMham} with  $b_{xy}=1$  (black circles). The dashed line for comparison has slope $\frac{1}{\pi} \Theta_F\nu$ and vanishing intercept, indicating the theoretical prediction. The disclinated lattice is constructed as in Ref.~\cite{may2022topological} for $N_x = N_y=39$ lattice sites in the $xy$-plane, and 250 momentum points along the $k_z$ axis. (b) The bound charge resolved in $k_z$ (momentum parallel to the disclination line) for $\mathcal{K}= \pi/3$. The deviations from 0 and 1/2 near $k_z = \pm \frac{\pi}{3}\frac{1}{a_z}$ are the result of finite size effects and the small size of the gap near the Dirac nodes.}
    \label{fig:q_disc_vs_nu}
\end{figure}

Now that we have discussed the probe fields we can interpret the anomalous response. We find that Eq. \ref{eq:DSManomaly} implies that disclinations in a DSM bind charge per length determined by $\nu$. To demonstrate, let us take a functional derivative of the action generated by Eq. \ref{eq:DSManomaly} with respect to $A_0$. Hence, the contribution to the charge density from the DSM anomalous response is 
\begin{equation}
    \rho_{\text{anom}} = \frac{\nu}{2\pi^2} G_z [\partial_x \omega_y - \partial_y \omega_x].
\end{equation}
For a thin-core disclination parallel to the $z$-axis, $\partial_x \omega_y- \partial_y \omega_x = \Theta_F \delta(x)\delta(y)$, where $\Theta_F$ is the Frank angle and is quantized to $\Theta_F=2\pi s/4$ with $s\in\mathbb{Z}$. For a system of length $L_z = a_z N_z$ with $N_z \in \mathbb{Z}$, the disclination has total charge 
\begin{equation}\begin{split}
Q_{\text{disc}} = \frac{\Theta_F}{\pi} \nu N_z + \delta Q.
\label{eq:chargedisclination2}\end{split}\end{equation}
Here we have assumed that the system is charge neutral in the absence of any disclinations. Physically, this amounts to adding negatively-charge background ions to each site of the lattice. In addition to the topological contribution $\propto \Theta_F$, we have included an additional contribution to the disclination charge, $\delta Q$. This contribution is $\mathcal{O}(1)$ and arises from the gapless particles that are necessary to restore the gauge invariance of the anomalous DSM response~\cite{gioia2021unquantized}. In Fig. \ref{fig:q_disc_vs_nu}(a) we plot the disclination charge as a function of $\nu$ using the DSM tight binding model in Eq. \ref{eq:DSMham}, and find that the disclination charge follows the predicted trend, $Q_{\text{disc}}/N_z \approx  \frac{\Theta_F}{\pi} \nu$.

A useful way of understanding the disclination response of the DSM is to treat the momentum $k_z$ of Eq.~\ref{eq:DSMham} as a tunable parameter, such that gapped planes with fixed $k_z$ are trivial 2D insulators for $|k_z| > \mathcal{K}$ and quantum spin Hall (QSH) insulators for $|k_z|<\mathcal{K}$~\cite{kane2005quantum, bernevig2006quantum}. The Dirac nodes at $|k_z| = \mathcal{K}$ mark the band crossings that connect the topologically distinct insulators. In this way, the DSM with $\mathcal{K}\neq 0,\pi$ can be viewed as an intermediate phase between a trivial insulator ($\mathcal{K}= 0$), and a weak 3D topological insulator ($\mathcal{K}= \pi$)~\cite{fu2007topologicalIn,ramamurthy2015patterns}. As we shall discuss in Sec.~\ref{sssec:2DdWZ}, disclinations of the 2D QSH insulator bind charge $\Theta_F/\pi$. For a system of length $L_z = a_z N_z$, $k_z$ is discretized in steps of size $2\pi/N_z$. Each $k_z$ slice will bind charge $\Theta_F/\pi$ to a disclination if $|k_z| < \mathcal{K}$, and charge $0$ if $|k_z| > \mathcal{K}$, as shown in Fig.~\ref{fig:q_disc_vs_nu}(b). The disclination will therefore bind a total charge 
\begin{equation}
    Q_{\text{disc}} = \sum^{\kappa}_{k_z = -\kappa} \frac{\Theta_F}{\pi} \approx \frac{\Theta_F}{\pi} 2 \mathcal{K}\frac{N_z}{2\pi} = \frac{\Theta_F}{\pi} \nu N_z.
\end{equation}
If we add in the extra $\mathcal{O}(1)$ contribution, $\delta Q$, that arises from the gapless modes at $k_z = \pm \mathcal{K}$,  we arrive at Eq.~\ref{eq:chargedisclination2}.

\section{Topological crystalline-electromagnetic responses of crystalline insulators}
\label{sec:insulatorresponse}

In preparation for our discussion of the topological crystalline-electromagnetic responses of the Dirac-CDW insulator in Sec.~\ref{sec:DSMCrystalResp}, we will first discuss the crystalline-electromagnetic responses that are relevant for insulators having C$_{4z}$ rotation symmetry, as well as $T_z$ discrete translation and/or inversion symmetry. These responses are similar to the anomalous response of the DSM discussed in Sec.~\ref{ssec:anomTopResp}. However, unlike the DSM responses, the topological responses of insulators are quantized because of gauge invariance and symmetry (DSMs can have unquantzed responses since the there are low energy gapless degrees of freedom that restore gauge invariance). We will mainly consider insulators composed of spin-1/2 fermions with TRS, which is the relevant case for DSMs. Similar responses also exist for spinless fermions, and systems without TRS, and the quantization of the responses will be different~\cite{li2020fractional, may2022topological}. 

For spin-1/2 systems with TRS and C$_{4z}$ rotation symmetry, we will show that crystalline-electromagnetic responses lead to: a $\mathbb{Z}_4$ classification when the system has additional $T^z$ discrete translation symmetry; a $\mathbb{Z}_2$ classification when the insulator has additional inversion symmetry; and a $\mathbb{Z}_4 \times \mathbb{Z}_2$ classification when the system has both $T^z$ discrete translation and inversion symmetry. We will also show that for a system with discrete translation $r_z \rightarrow r_z + a_z$, the $\mathbb{Z}_4$ index indicates that disclinations with Frank angle $\Theta_F$ bind a charge per length of $2n \frac{\Theta_F}{2\pi a_z}$ ($n \in \mathbb{Z}_4$) when the system has periodic boundaries in the $z$-direction. The effective response theory for the insulator with a non-trivial $\mathbb{Z}_4$ index also contains a 3D discrete Wen-Zee (dWZ) term. The $\mathbb{Z}_2$ index corresponds to the charge parity of a disclination when inversion-symmetric open boundary conditions are present. Insulators with different $\mathbb{Z}_2$ indices have effective response theories that differ by a quantized total derivative term that was referred to as the $R\wedge F$ term in Ref.~\onlinecite{may2022topological}. When the insulator has both $T^z$ discrete translation and inversion symmetry, the $\mathbb{Z}_2$ index can be equivalently understood as the presence or absence of a filling anomaly~\cite{benalcazar2019quantization, khalaf2021boundary} of the 1D disclination lines, which we will define more formally in Sec. \ref{ssec:RFResponse}.

To prevent confusion we note that here we are discussing only the topological crystalline-electromagnetic responses of insulators in various symmetry classes. This is not an exhaustive list of all topological responses. Notably, we are ignoring any purely electromagnetic responses in the section. 

\subsection{The discrete Wen-Zee response}\label{ssec:dWZ}
To begin, we will consider the dWZ response in 2D and 3D. The dWZ response of a 2D system describes how disclinations bind a quantized charge~\cite{wen1992shift, ruegg2013bound, ruegg2013corner, liu2019shift, han2019generalized, li2020fractional, manjunath2020classification, manjunath2021crystalline, may2022crystalline,peterson2021trapped, zhang2022fractional}. The 3D dWZ response is essentially a layered version of the 2D dWZ response, and describes how disclination lines bind a quantized charge per length.

\subsubsection{The discrete Wen-Zee response in 2D}\label{sssec:2DdWZ}
The 2D dWZ response corresponds to the following topological term in the effective field theory,
\begin{equation}
    \mathcal{L}_{\text{2D-WZ}} = \frac{\mathcal{S}_{\text{2D}}}{2\pi} \epsilon^{ijk}\omega_i \partial_j A_k,
\label{eq:2DWZ}\end{equation}
where $i,j,k$ run over $t,x,y$. Here, fluxes of $\omega$ (which are point-like in 2D) are disclinations of the 2D lattice. For spin-$1/2$ fermions with TRS (which will be the focus of this work) $\mathcal{S}_{\text{2D}}$ is quantized to be an even integer~\footnote{For insulators composed of spinless fermions with TRS or spin-1/2 fermions without TRS, $\mathcal{S}_{\text{2D}}$ is quantized as an integer. For insulators composed of spinless fermions without TRS, $\mathcal{S}_{\text{2D}} = C/2\mod(1)$ where $C$ is the Chern number.} and defined $\mod(8)$ in the case of insulators with C$_{4}$ symmetry (four-fold rotations of the plane)~\footnote{For spin-1/2 fermions with TRS, $\mathcal{S}_{\text{2D}}$ is defined $\mod(2n)$ in $C_n$ symmetric insulators. For spinless fermions either with or without TRS, or spin-1/2 fermions without TRS, $\mathcal{S}_{\text{2D}}$ is defined $\mod(n)$ in $C_n$ symmetric insulators}. Therefore the allowed inequivalent values are $\mathcal{S}_{\text{2D}} = 0$, $2$, $4$, or $6$ for time-reversal and C$_{4}$ symmetric insulators. 

A representative 2D insulator with $\mathcal{S}_{\text{2D}} = 2$, is realized by the following 4-band Hamiltonian,
\begin{equation}\begin{split}
\mathcal{H}_{\text{QSH}}(k_x,k_y) &= \sin (k_x) \Gamma_1 + \sin (k_y)  \Gamma_2\\ & - [m - \cos (k_x) - \cos (k_y) ]\Gamma_3,
\end{split}\label{eq:QSHham}\end{equation}
with C$_{4}$ rotation given by Eq.~\ref{eq:rotDef}, and TRS given by $\mathcal{T} = -is^y\kappa$. This Hamiltonian is a QSH insulator for $0<|m|<2$ and a trivial insulator for $2<|m|$~\cite{kane2005quantum, bernevig2006quantum}. The QSH insulator has a 2D dWZ response with $\mathcal{S}_{\text{2D}} = 2$, while the 2D dWZ response vanishes ($\mathcal{S}_{\text{2D}} = 0$) in the trivial phase\footnote{In Ref.~\cite{zhang2022fractional}, a trivial insulator has a non-zero discrete shift. This difference arises because Ref.~\cite{zhang2022fractional} considers the charge contribution from only the electrons. Here we are considering the contribution from the electrons, and the contribution from the background ions, which are effectively a trivial insulator composed of negative charges.}~\cite{may2022crystalline}. We note that Eq. \ref{eq:QSHham} can be extended to Eq. \ref{eq:DSMham} by fixing $b_{xy} = 1$ and setting $M = 2+\cos(k_z) - \cos(Q).$

The 2D dWZ response indicates that  a disclination with Frank angle $\Theta_F$ binds charge
\begin{equation}
Q_{\text{disc}} = \mathcal{S}_{\text{2D}}\frac{\Theta_{F}}{2\pi} \mod(2),
\end{equation}
where we have added additional negative background charges such that the system is charge neutral in the absence of disclinations. The $\mod(2)$ ambiguity in defining $Q_{\text{disc}}$ arises from the fact that for spin-1/2 insulators with TRS, it is possible to add a Kramers pair of particles to the disclination core without changing any topological properties. Since $\Theta_F$ is a multiple of $\pi/2$, the disclination charge is always trivial if $\mathcal{S}_{\text{2D}} = 8$. This is why $\mathcal{S}_{\text{2D}}$ is defined $\mod(8)$ as noted before.

\subsubsection{The discrete Wen-Zee response in 3D}\label{sssec:3DdWZ}
For insulators having $T^z$ discrete translation, $z \rightarrow z + a_z$, the 2D dWZ response can be extended to 3D as a layered response. The 3D layered response can be expressed as the following topological term in the effective response theory,
\begin{equation}
    \mathcal{L}_{\text{3D-WZ}} = \mathcal{S}_{\text{3D}}\frac{G_{z}}{4\pi^2} \epsilon^{ijk}  \omega_i \partial_j A_k,
\label{eq:3DWZ2}
\end{equation}
where $G_{z} = \frac{2\pi}{a_z}$ is the reciprocal lattice vector along the $z$-direction. Here, as in Sec.~\ref{sec:DSManomaly}, flux-lines of $\omega$ are disclination lines with Frank-vector $\propto {\hat{z}}$. Again, we take all disclination lines to extend along the $z$-direction. This response is a layered version of the 2D dWZ response having one 2D layer per lattice period, $a_z$. The coefficient $\mathcal{S}_{\text{3D}}$, which we will call the 3D discrete shift, is defined $\mod(8)$ and quantized as an even integer for spin-$1/2$ fermions with TRS and C$_{4z}$ rotation symmetry, i.e. $\mathcal{S}_{\text{3D}}= 0$, $2$, $4$, or $6$. Because of the this, the value of $\mathcal{S}_{\text{3D}}$ defines a $\mathbb{Z}_4$ index for insulators in this symmetry class.

This quantization of $\mathcal{S}_{\text{3D}}$ is directly inherited from the quantization of $\mathcal{S}_{\text{2D}}$. The 3D dWZ term, Eq.~\ref{eq:3DWZ2}, has the same form as the unquantized anomaly of the DSM, Eq.~\ref{eq:DSManomaly}. However, the coefficient of the 3D dWZ term, $\mathcal{S}_{\text{3D}}$, is quantized, while the coefficient of the DSM anomaly equation, $\nu$, is not (recall that $\nu$ is not quantized due to the gaplessness of the DSM).

\begin{figure}
    \centering
    \includegraphics[width=0.45\textwidth]{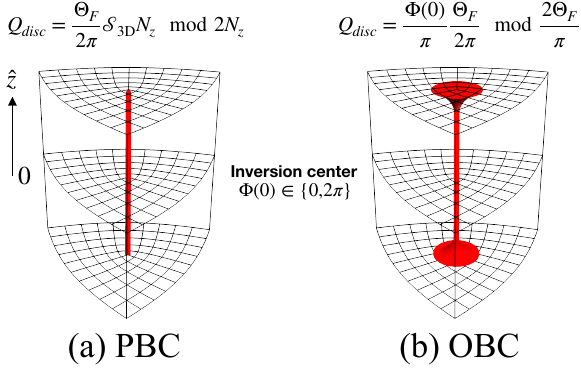}
    \caption[]{(a) The total charge per length, $Q_{\text{disc}}/N_z$, of an insulator with a disclination with Frank angle $\Theta_F$ is proportional to $\mathcal{S}_{3D}$, the coefficient of the dWZ term.  
    (b) The disclination charge parity, $Q_{\text{disc}}\mod(2\Theta_F/\pi)$, of a $\Theta_F$-disclination of an inversion-symmetric insulator with open boundary conditions is determined by $\Phi(0) = \{0 ,2\pi\}$, the value of the coefficient of the $R\wedge F$ term at the inversion center.}\label{fig:responses}
\end{figure}

The 3D dWZ response physically manifests as a quantized charge per length bound to disclination lines as illustrated in Fig. \ref{fig:responses}a. If we take a system of length $L_z = a_z N_z$ that includes a disclination line with Frank angle $\Theta_F$, Eq. \ref{eq:3DWZ2} indicates that the total charge on the disclination line is
\begin{equation}\begin{split}
Q_{\text{disc}}=\frac{\Theta_F}{2\pi} \mathcal{S}_{\text{3D}} N_z \mod(2N_z).
\label{eq:WZDiscCharge}\end{split}\end{equation}
Here we have again implicitly added negatively charged background ions, such that the system is charge neutral without any disclinations. The $\mod(2N_z)$ ambiguity arises from the fact that it is possible to change the disclination charge by embedding a 1D insulator in the disclination core without changing any topological properties of the system. Due to the Kramers degeneracy and the Lieb-Schultz-Mattis theorem~\cite{lieb1961two}, the charge per unit length must be an even integer for translation invariant spin-1/2 1D insulators with TRS. From the $\mod(2N_z)$ ambiguity, and the fact that the Frank angles of disclinations of a C$_{4z}$ symmetric insulators are quantized in units of $\pi/2$, we also find that $\mathcal{S}_{\text{3D}} = 8$ leads to a trivial charge of the disclination, confirming our earlier assertion that $\mathcal{S}_{\text{3D}}$ is defined $\mod(8)$.

Without $T^z$ discrete translation symmetry, the charge per length of a disclination line is not a well-defined quantity, as it is possible to add local charges to arbitrary points along the disclination. An important consequence of this is that the charge per length of a disclination is not a topological, quantized quantity for insulators with incommensurate CDWs, as it is possible to shift the charge per length of a disclination by an arbitrary amount. 
However, for systems that are finite in the $z$-direction, the \textit{total} charge on a disclination line $Q_{\text{disc}}$ is still a quantized multiple of $\Theta_F/\pi$. This can be understood by treating the $z$-coordinate as an internal degree of freedom of an effectively 2D system with TRS and C$_{4z}$ rotation symmetry. A disclination-line of the finite 3D system is also a disclination of the effective 2D system, and the total charge bound to the disclination is therefore quantized according to the 2D dWZ response (Eq.~\ref{eq:2DWZ}).

\subsection{The \texorpdfstring{$R\wedge F$}{R\^F} Term}\label{ssec:RFResponse}
In Ref.~\onlinecite{may2022topological} some of us showed that the effective response theory of 3D crystalline insulators with rotation symmetry around a fixed axis can contain the following topological term,
\begin{equation}\begin{split}
    \mathcal{L}_{R\wedge F} &= \epsilon^{\mu\nu\lambda\eta}  \frac{\Phi}{4\pi^2}  \partial_\mu \omega_\nu \partial_\lambda A_\eta.
\label{eq:RFTerm}\end{split}\end{equation}
where $\mu,\nu,\lambda,\eta$ run over $t,x,y,z,$ and $\Phi$ can generally be a function of position and time. We denote this response as the $R\wedge F$ term, since it couples the lattice curvature $R_{\mu\nu} = \partial_\mu \omega_\nu - \partial_\nu \omega_\mu$ to the electromagnetic field strength $F_{\mu\nu} = \partial_\mu A_\nu - \partial_\nu A_\mu$~\cite{may2022topological}. The $R\wedge F$ term has a similar form to the $\Theta$ term, $F\wedge F$, that describes axion electrodynamics in 3D topological insulators~\cite{wilczek1987two, qi2008topological}. Like the $\Theta$ term, the $R\wedge F$ term has a periodic coefficient and is a total derivative when the coefficient is constant. For spin-$1/2$ insulators with TRS, $\Phi$ is $4\pi$ periodic. In general, the periodicity of $\Phi$ depends on the spin of the fermions and the presence of TRS, as discussed in Ref. \onlinecite{may2022topological}. 

The charge responses associated with the $R\wedge F$ term are
\begin{equation}
    j^\mu = \epsilon^{\mu\nu\lambda\eta}  \frac{1}{4\pi^2} (\partial_\nu \Phi) \partial_\lambda \omega_\eta.
\label{eq:RFCurrents1}\end{equation}
If we fixed $\omega$ to the  configuration of a disclination line, $\partial_x \omega_y - \partial_y \omega_x = \Theta_F \delta(x)\delta(y)$, the charge responses simplify to 
\begin{equation}\begin{split}
    &\rho_{\text{disc}} = \frac{\Theta_F}{4\pi^2} (\partial_z \Phi) \delta(x)\delta(y),\\
    &j^z_{\text{disc}} =  -\frac{\Theta_F}{4\pi^2} (\partial_t \Phi) \delta(x)\delta(y).
\label{eq:RFCurrents2}\end{split}\end{equation} 
The first lines indicates that spatial fluctuations of $\Phi$ bind charge along disclination lines. The second line indicates that temporal fluctuations drive a current along the disclination line. Together, these two responses indicate a charge polarization in the $z$-direction proportional to $\Theta_F$ localized on the disclination line. 

Based on Eq. \ref{eq:RFCurrents2}, if we adiabatically increase $\Phi$ homogeneously by an amount $\delta \Phi$ over some time $T$ (i.e., $\Phi(t) = \Phi_0 + \delta \Phi  \frac{t}{T} $), then the polarization on the disclination line, $P^z_{\text{disc}}$, will change by 
\begin{equation}\begin{split}
    \Delta P^z_{\text{disc}}  &= \int^T_0 \! dt \partial_t P^z_{\text{disc}}= \int^T_0 \!dt j^z_{\text{disc}} = \frac{\delta \Phi}{2\pi}\frac{\Theta_F}{2\pi}.
\label{eq:RFPolChange}\end{split}\end{equation}
We therefore find that if the effective response theories of two insulators \textit{differ} by an $R\wedge F$ term with a constant coefficient $\delta \Phi$, the polarization of disclination lines will \textit{differ} by $\frac{\delta \Phi}{2\pi}\frac{\Theta_F}{2\pi}$ between the two insulators. \footnote{Note that two systems can have a well-defined difference in polarization even if the systems do not have a well-defined polarization individually. For example, a system with a net charge does not have a well-defined polarization, as the polarization will depend on the choice of origin. However, the difference in polarization between two systems with the same net charge is independent of the choice of origin.} 

\subsubsection{\texorpdfstring{$R\wedge F$}{R\^F} term with discrete translation symmetry}

\begin{figure}
    \centering
    \includegraphics[width=0.4\textwidth]{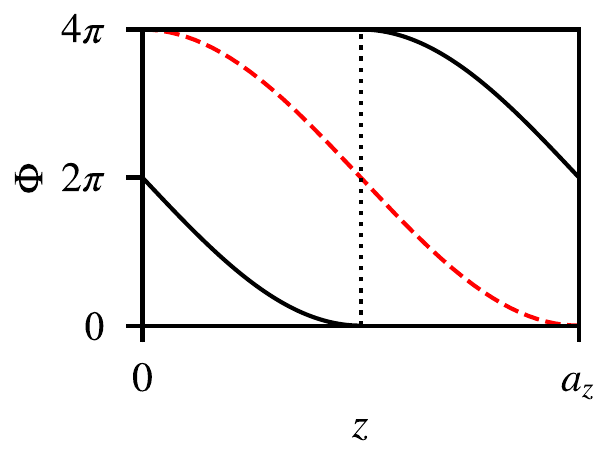}
    \caption[]{Two examples of how $\Phi$ can wind across the unit cell. Both lines wind by $-4\pi$ across the unit cell, but one has $\Phi(0)=0$ (red dashed) and the other has $\Phi(0)=2\pi$ (black solid).}\label{fig:windingsUC}
\end{figure}

We now turn our attention to the connection between the $R\wedge F$ term and the 3D dWZ response. We will consider a system having $T^z$ discrete translation, $z \rightarrow z + a_z$, and, since the disclination responses in Eq. \ref{eq:RFCurrents2} primarily involve the dependence of $\Phi$ on $z$ and $t$,  we will suppress any dependence on $x$ and $y$ here.  Interestingly, the coefficient of the $R\wedge F$ term can be non-constant for a system with $T^z$ discrete translation symmetry, as $\Phi(z) = \Phi(z+a_z) \mod(4\pi)$ has non-constant solutions. 
We can characterize the solutions by their winding across a single unit-cell period $\int^{a_z}_{0} dz \partial_{z} \Phi(z)$, which will be an integer multiple of $4\pi$. Examples of different winding configurations are shown in Fig.~\ref{fig:windingsUC}. 

With this in mind we can compute the total charge bound to a disclination for a periodic system of length $L_z = a_z N_z$ (using Eq. \ref{eq:RFCurrents2}):
\begin{equation}\begin{split}
Q_{\text{disc}} &= \frac{\Theta_F}{2\pi} \int^{L_z}_0 \partial_z \frac{\Phi(z)}{2\pi} dz \mod(2N_z)\\
& = \frac{\Theta_F}{2\pi}  N_z \int^{a_z}_{0} \partial_{z} \frac{\Phi(z)}{2\pi} dz \mod(2N_z),
\end{split}\label{eq:QdiscWentoRF}\end{equation}
where the $\mod(2N_z)$ ambiguity again arises from the fact that it is possible to change the disclination charge per length by embedding a 1D insulator in the disclination core. Comparing to Eq. \ref{eq:WZDiscCharge}, we find a relation for the dWZ response coefficient: 
\begin{equation}
\mathcal{S}_{\text{3D}} = \frac{1}{2\pi}\int^{a_z}_{0} \! dz \partial_{z} \Phi(z),
\label{eq:ShiftFromPhi}\end{equation}
from which we see that the winding of $\Phi$ over one period, $a_z$, generates a 3D dWZ response.  

We can make the connection between the $R \wedge F$ and 3D dWZ terms more concrete by setting
\begin{equation}
    \Phi(z) = \Phi_0 + \mathcal{S}_{\text{3D}} G_z z,
\label{eq:PhidWZ}\end{equation}
where $\Phi_0$ is an arbitrary constant. This solution preserves the $T^z$ discrete translation symmetry when $G_z = 2\pi/a_z$. Plugging this value of $\Phi(z)$ into Eq. \ref{eq:RFTerm} leads to the 3D dWZ term in Eq. \ref{eq:3DWZ2}, after an integration by parts. Because of the $\mod(2N_z)$ ambiguity, the $R\wedge F$ term identifies a $\mathbb{Z}_4$ classification of insulators with $T^z$ discrete translation symmetry, exactly as discussed in Sec. \ref{sssec:3DdWZ}. 

Although it is not manifest in Eq.~\ref{eq:PhidWZ}, it is important to note that the coefficient of the $R\wedge F$ term, $\Phi$, is a $4\pi$-periodic variable. Importantly, when $\Phi$ has a winding there is no discontinuity between $z = 0$ and $z= L_z$, since $\mathcal{S}_{\text{3D}}$ is an even integer and $\mathcal{S}_{\text{3D}} G_z L_z = \mathcal{S}_{\text{3D}} 2\pi N_z = 0 \mod(4\pi)$. The expression for $\Phi(z)$ given in Eq. \ref{eq:PhidWZ} is therefore smooth everywhere with $\partial_z \Phi(z) = \mathcal{S}_{\text{3D}} G_z$. The same would be true if we replaced the linear interpolation in Eq. \ref{eq:PhidWZ} with a more complicated function like those shown in Fig. \ref{fig:windingsUC}.

\subsubsection{\texorpdfstring{$R\wedge F$}{R\^F} term with inversion symmetry}\label{ssec:RFandInversion}
Let us now consider the $R\wedge F$ term for a system with inversion symmetry. We again take $\Phi$ to be independent of the $x$ and $y$ coordinates but allow for $\Phi$ to depend on $z$. We will now show that we can use the $R\wedge F$ term to resolve a $\mathbb{Z}_2$ distinction between inversion-symmetric insulators. Inversion symmetry requires that $\Phi(z) = -\Phi(-z)\mod(4\pi)$, i.e., $\Phi$ must be an odd function of $z$. Here we take the inversion center to be in the $z=0$ plane. It is also possible to have an inversion center at $z = a_z/2$, but the difference between the two inversion centers can be accounted for by a redefinition of the coordinate system. In continuum effective field theory, such a coordinate change is innocuous. However, such a change is not innocuous for lattice systems, as the discrete lattice sites lead to a unique choice of origin (modulo lattice translations). At $z=0$, inversion symmetry requires that $\Phi(0) = -\Phi(0)$. Since $\Phi$ is $4\pi$ periodic, $\Phi(0) = 0$ or $\Phi(0) = 2\pi$ are both allowed values satisfying this condition. The choice of these values defines two classes (i.e., a $\mathbb{Z}_2$ classification) of inversion-symmetric insulators. Alternatively expressed, the response theories of insulators with different $\mathbb{Z}_2$ indices differ by a $R\wedge F$ term with $\Phi(0) = 2\pi$.

 Importantly, the value of $\Phi(0)$ has a direct physical interpretation when switching from periodic boundaries to open boundaries in the $z$-direction. We will show below that with open boundaries, a disclination will have charge $Q_{\text{disc}} = 0 \mod(2\frac{\Theta_F}{\pi})$ when $\Phi(0) = 0,$ and charge $\frac{\Theta_F}{\pi} \mod(2\frac{\Theta_F}{\pi})$ when $\Phi(0) = 2\pi$. We will refer to the quantity $Q_{\text{disc}} \mod(2\frac{\Theta_F}{\pi})$ as the \textit{disclination charge parity}. The disclination charge parity is a natural quantity to consider for inversion-symmetric systems with open boundary conditions, as purely 2D boundary effects can at most change the disclination charge by an integer multiple of $2\frac{\Theta_F}{\pi}$, provided they respect inversion symmetry. Hence the charge parity is a physical observable that determines the $\mathbb{Z}_2$ classification.

To show why the value of $\Phi(0)$ determines the disclination charge parity, let us consider a system of length $L_z$ with periodic boundaries in the $z$-direction, and use local perturbations to ``cut'' the system and generate open boundaries at $z = \pm L_z/2$. We assume that the resulting open boundary system satisfies the following four conditions: 
\begin{enumerate}
\item Changing from periodic to open boundary conditions preserves inversion symmetry around $z=0$.
\item The system with open boundaries is gapped, such that $\Phi(z)$ can be defined over the entire system.
\item The value of $\Phi$ at the inversion center, $\Phi(0)$ remains constant when changing from periodic to open boundaries. This condition is satisfied if the gap at the inversion center remains open when switching the boundary conditions. 
\item At the boundaries of the system, $\Phi$ vanishes  (i.e., $\Phi(z) =  0$ for $|z| \ge L_z/2$). This choice of boundary conditions is equivalent to requiring that the $R\wedge F$ term fully vanishes outside the boundaries of the system.
\end{enumerate}

\begin{figure}
    \centering
    \includegraphics[width=0.45\textwidth]{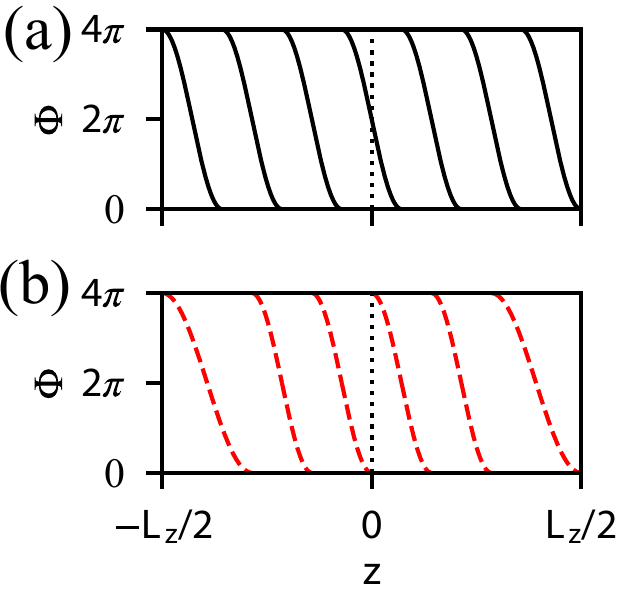}
    \caption[]{The disclination charge parity of inversion-symmetric insulators is determined by the value of $\Phi(z)$ at the inversion center (the $z=0$ plane here). Figures (a) and (b) show two windings of $\Phi(z)$ with $n=3$ but different values of $\Phi(0)$. Figure (a) has $\Phi(0)=2\pi$ and thus winds a total of seven times, producing an odd disclination charge parity. Figure (b) has $\Phi(0)=0$ and therefore winds only six times, yielding an even disclination charge parity.}\label{fig:windingsOBC}
\end{figure}

Let us now add a disclination to the system with open boundary conditions. Using Eq.~\ref{eq:RFCurrents2}, the total charge bound to the disclination is 
\begin{equation}\begin{split}
 Q_{\text{disc}} &= \frac{\Theta_F}{2\pi}\frac{1}{2\pi} \int^{L_z/2}_{-L_z/2} \partial_z \Phi(z) dz \\
  &= \frac{\Phi(0)}{\pi}\frac{\Theta_F}{2\pi} +  2 n \frac{\Theta_F}{\pi},\\
\label{eq:FillingAnomaly}\end{split}\end{equation}
where $n \in \mathbb{Z}$ is the number of times $\Phi$ fully winds by $4\pi$, between $0$ and $L_z/2$, and we have used that $\Phi(z) = -\Phi(-z)$, and $\Phi(L_z/2) = 0$. This equation reflects how if $\Phi(0) =\Phi(L_z/2) = 0$, $\Phi$ must wind by an integer multiple of $4\pi$ between $0$ and $L_z/2$, while if $\Phi(0) = 2\pi$ and $\Phi(L_z/2) = 0$, $\Phi$ must instead wind by a half-integer integer multiple of $4\pi$. This is illustrated in Fig.~\ref{fig:responses}b, and two representative configurations of $\Phi(z)$ are shown in Fig.~\ref{fig:windingsOBC}, where $n = 3$. We therefore find that 
\begin{equation}
    Q_{\text{disc}} = \frac{\Phi(0)}{\pi} \frac{\Theta_F}{2\pi}  \mod(2\frac{\Theta_F}{\pi}).
\end{equation} 
The difference between the $\Phi(0) = 0$ and $\Phi(0) = 2\pi$ insulators is therefore manifest in the value of $Q_{\text{disc}} \mod(2\frac{\Theta_F}{\pi})$ when open boundaries are present. Specifically $Q_{\text{disc}} =0 \mod(2\frac{\Theta_F}{\pi})$ and $\frac{\Theta_F}{\pi} \mod(2\frac{\Theta_F}{\pi})$ for $\Phi(0) = 0$ and $2\pi$ respectively. We also find that $Q_{\text{disc}} =0 \mod(2\frac{\Theta_F}{\pi})$ for $\Phi(0) = 4\pi$, in agreement with $\Phi$ being $4\pi$ periodic.

\subsubsection{The \texorpdfstring{$R\wedge F$}{R\^F} term with discrete translation and inversion symmetry}\label{ssec:discFillingAnom}

If an insulator has both inversion symmetry and $T^z$ discrete translation symmetry ($z \rightarrow z + a_z$), then the values of $\Phi(0)$ and $\mathcal{S}_{\text{3D}}$  from Eq.~\ref{eq:ShiftFromPhi} lead to a $\mathbb{Z}_4 \times \mathbb{Z}_2$ classification based on the phenomena associated to the $R\wedge F$ term. We stress that this classification is not exhaustive, but it is sufficient to characterize the Dirac-CDW insulators we will discuss in the following sections.

If two insulators have the same 3D dWZ response (i.e., the same value of $\mathcal{S}_{\text{3D}}$ in Eq.~\ref{eq:ShiftFromPhi}), but different values of $\Phi(0)$, the polarization of disclination lines will \textit{differ} by $ \frac{\delta \Phi(0)}{2\pi} \frac{\Theta_F}{2\pi}$, where the difference between the value of $\Phi(0)$ between the two insulators, $\delta \Phi(0)$, is equal to either $0$ or $2\pi$ based on our previous discussions.  To show why different values of $\Phi(0)$ lead to a difference in polarization, we take a given $\Phi(z)$ that is invariant under $T^z$ discrete translation and inversion symmetry, and define an adiabatic evolution where $\Phi(z)$ increases by $2\pi$ over a time $T$, $\Phi(z) \rightarrow \Phi(z) + 2\pi t/T$. Inversion symmetry is broken for $0<t<T$ but is restored at $t = T$. During this process, $\Phi(0)$ changes by $2\pi$, and $\int^{a_z}_{0} \! dz \partial_{z} \Phi(z)$ remains constant (i.e., $\mathcal{S}_{\text{3D}}$ remains constant). Based on Eq. \ref{eq:RFCurrents2}, and \ref{eq:RFPolChange} the polarization of a disclination line changes by $\frac{\Theta_F}{2\pi}$ during this process. Recall that disclinations of insulators with a non-zero $\mathcal{S}_{\text{3D}}$ have a finite charge per length. Therefore, the polarization of a disclination is not strictly well-defined for insulators with finite $\mathcal{S}_{\text{3D}}$. However, the difference in the polarization of disclinations between insulators with the same $\mathcal{S}_{\text{3D}}$ is well-defined. This is analogous to discussions of polarization in Chern insulators\cite{coh2009,vaidya2023}.

For insulators having $\mathcal{S}_{\text{3D}} = 2$, the two classes of inversion-symmetric insulators are related to each other by a half-translation $z \rightarrow z + a_z/2$. For $\mathcal{S}_{\text{3D}} = 2$, this is clear if we set $\Phi(z) = 2 G_z z$, for which $\Phi(0) = 0$. Translating the system by $a_z/2$ shifts $\Phi(z) \rightarrow 2\pi + 2 G_z z$, for which $\Phi(0) = 2\pi$. This relationship can be naturally understood if we treat the 3D dWZ response as arising from a stacking of 2D systems having $\mathcal{S}_{\text{2D}} = 2$. In order to preserve inversion symmetry around $z=0$, the 2D systems must be centered at either $z = n a_z$ or $z = n a_z + a_z/2$ with $n \in \mathbb{Z}$. A disclination line of such a system, will have charge $\Theta_F/\pi$ centered at $z = n a_z$ or $z = n a_z + a_z/2$ respectively.  This is the same charge configurations that are realized by the two topologically distinct phases of the 1D inversion symmetric Su–Schrieffer–Heeger (SSH) model~\cite{su1979solitons, su1983erratum}, if we set the electron charge to be $\Theta_F/\pi$ instead of $1$ in the SSH model. 

We similarly have that the two insulators with $\mathcal{S}_{\text{3D}} = 6$ are related by a half-translation. This can be understood using that $\mathcal{S}_{\text{3D}} = 6 = -2 \mod(8)$, and hence the $\mathcal{S}_{\text{3D}} = 6$ dWZ response is captured by $\Phi(z) = -2 G_z z$. A half translation then shifts $\Phi(0)\rightarrow \Phi(0)-2\pi = \Phi(0)+ 2\pi \mod(4\pi)$. We can also invoke a similar stacking argument to the one used above, but instead stacking 2D systems with $\mathcal{S}_{\text{2D}} = -2$.

The two inversion symmetric insulators having $\mathcal{S}_{\text{3D}} = 4$ are related to each other by a quarter translation, $z \rightarrow z + a_z /4$. Here, we can imagine the 3D dWZ response as arising from having \textit{two} 2D systems between $z = n a_z$ and $z = (n+1)a_z$, each with $\mathcal{S}_{\text{2D}} = 2$. To preserve inversion symmetry, these 2D systems must be stacked in one of two classes of configurations. In the first class of stacking configurations, one insulator is at $z = n a_z + \delta_z $ and one is at $z = (n +1 )a_z - \delta_z $ for each $n \in \mathbb{Z}$, where $\delta_z \in [0, a_z)$ is a constant offset. Since inversion symmetry is preserved for all $\delta_z$ here, stacking configurations having different values of $\delta_z$ are adiabatically connected. In the second class of stacking configurations, one insulator is at $z = n a_z $ and one is at $z = n a_z + a_z/2$. It can be directly confirmed that the second stacking configuration is distinct from the first stacking configuration for any value of $\delta_z$. Furthermore, if we take the second stacking configuration and perform a quarter translation, we arrive at the first stacking configuration with $\delta_z = a_z/4$.

As we have discussed, the value of $\Phi(0)$ determines the disclination charge parity for a system having open boundaries. However, as we discuss in Appendix~\ref{app:DiscChargeParity}, the absolute disclination charge parity is not a very practical quantity to consider since it can depend on the choice of inversion center. For insulators having $T^z$ discrete translation symmetry, a more useful quantity to consider is the \textit{difference} in the value of the disclination bound charge $Q_{\text{disc}} \mod(2\frac{\Theta_F}{\pi})$ between a system with open boundaries and a system of the same size with periodic boundaries. For a system of size $L_z = a_z N_z$, the difference in charge is
\begin{equation}
    Q_{\text{FA}} = \frac{\Theta_F}{2\pi} \left[\frac{\Phi(0)}{\pi} - \mathcal{S}_{\text{3D}} N_z \right] \mod(2\frac{\Theta_F}{\pi}).
\label{eq:DiscFACharge}\end{equation}
Since $\mathcal{S}_{\text{3D}}$ is an even integer for spin-1/2 insulators with TRS, $Q_{\text{FA}} = 0$ or $\frac{\Theta_F}{\pi}$ are the distinct values. When $ Q_{\text{FA}}$ is non-zero, we will refer to the system as having a \textit{disclination filling anomaly}. The usual filling anomaly\cite{benalcazar2019quantization, khalaf2021boundary, fang2021filling, rao2023effective} reflects an inability to symmetrically deform from periodic to open boundary conditions while keeping the charge in the system constant. Here, the disclination filling anomaly reflects an inability to change from periodic to open boundary conditions while keeping both inversion symmetry and the charge on disclination lines constant. 

In Appendix~\ref{app:DiscChargeParity} we argue that the disclination-line filling anomaly necessarily vanishes for any layered insulator. Here we are specifically defining a layered insulator as an insulator where electrons from different unit cells in the $z$-direction are fully decoupled.  Indeed, cutting a layered system to change from periodic to open boundary conditions is completely a trivial procedure under this definition. Hence, insulators that exhibit a disclination filling anomaly are not layered in this way.

In summary, we find that we can use the $R\wedge F$ term to characterize the 3D dWZ response, and the disclination filling anomaly of insulators. The response phenomena lead to a $\mathbb{Z}_4 \times \mathbb{Z}_2$ classification for insulators having $T^z$ discrete translation symmetry and inversion symmetry. On the one hand, the $\mathbb{Z}_4$ index is determined by calculating the disclination charge per length for a system with periodic boundary conditions. On the other, the $\mathbb{Z}_2$ index is determined by calculating and comparing the total disinclination charge for a system having periodic boundary conditions and one having open boundary conditions (with the same length). 

For systems that do not have translation symmetry, the value of $\Phi(0)$ remains quantized, since its quantization requires only inversion symmetry. This means that for systems with only inversion symmetry, there should be two distinct insulators characterized by $\Phi(0) = 0$ and $\Phi(0) = 2\pi$ respectively. For open boundary conditions, the two classes of insulators will have different disclination charge parities. However, without translation symmetry, it is not possible to define the disclination filling anomaly. The disclination charge parity will therefore be significant when considering DSMs with an incommensurate CDW, which lack any form of translation symmetry.

\section{Gapping a Dirac Semimetal with a Charge Density Wave}
\label{sec:DSMCDW}

The gapless nodes of the DSM are protected by translation, rotation, and U$(1)$ charge conservation symmetries. By spontaneously breaking one of these symmetries it is possible to drive the system into a gapped phase with charge density wave (CDW), nematic, or superconducting orders, respectively. Here, we consider the gapped insulating phase that is generated by breaking translation symmetry using CDW order (see  Ref.~\cite{zhang2016topological} for a discussion of the instabilities of a DSM to CDW formation). We will pay special attention to the inversion symmetry of the DSM, and show that there are two distinct classes of inversion-symmetric, Dirac-CDW insulators. The differences between these two classes of insulators will be further discussed in Sec.~\ref{sec:DSMCrystalResp}. 

To consider the effects of a CDW distortion on the low-energy degrees of freedom, we start with a DSM with two Dirac nodes at $\pm {\bf{k}}_{DN},$ and expand the fermion creation operators around the two Dirac points, as 
\begin{equation}
c_\mathbf{r}\approx \sum_{\bm{q}} e^{i\mathbf{r}\cdot \bm{q} + i r_z\mathcal{K} } c_{\bm{q},R} +e^{i\mathbf{r}\cdot \bm{q} - i r_z \mathcal{K} } c_{\bm{q},L},
\label{eq:LEFermDef}\end{equation}
and introduce a new set of Pauli matrices $\bm{\tau}$, where $\tau^z = +1$ ($-1$) for the fermions in the $R$ ($L$) valley. In terms of the $R$ and $L$ fermions, the $8\times 8$ low-energy Hamiltonian for the DSM is
\begin{align}
\mathcal{H} = &q_x a_x \Gamma_1 + q_y a_y \Gamma_2  + q_z a_z \sin(\mathcal{K}a_x)\Gamma_3 \tau^z. 
\label{eq:LowEnergyHam}\end{align}

The crystalline symmetries act on the low-energy Hamiltonian as follows. The C$_{4z}$ rotation symmetry acts as 
\begin{equation}
R_4 = \exp\left(-i \frac{\pi}{4} [\sigma^z s^z - 2s^z ]\right).\end{equation}
As discussed previously, there are two possible inversion symmetries, site-centered and bond-centered. The site-centered inversion symmetry acts on the low-energy Hamiltonian as
\begin{equation}\begin{split}
&\mathcal{I}_s = \sigma^z \tau^x=\Gamma_3\tau^x.
\label{eq:InversionOnSiteDef}\end{split}\end{equation}
The bond-centered inversion symmetry can be similarly written as 
\begin{equation}\begin{split}
& \mathcal{I}_b = \mathcal{I}_s e^{i q_z a_z  + i \mathcal{K}a_z \tau^z }.
\end{split}\end{equation}
 This expression can be simplified if we perform a unitary transformation, $U = \exp(-i \mathcal{K}a_z \tau^z )$, which leaves the Hamiltonian unaffected, but reduces the bond-centered inversion symmetry to
\begin{equation}\begin{split}
& \mathcal{I}_b =\mathcal{I}_s  e^{i q_z a_z} . 
\label{eq:InversionBondDef}\end{split}\end{equation}
This unitary transformation will simplify our later discussions, and so we will implicitly assume such a unitary transformation has been performed when discussing continuum models with bond-centered inversion symmetry.

A CDW  can be dynamically generated through interactions, as discussed in Ref. \cite{zhang2016topological}. In the mean field limit, a CDW order corresponds to a translation symmetry breaking term.  We consider two possible types of translation symmetry breaking terms here. First there is an onsite term that is given in terms of the original lattice model by 
\begin{equation}
\hat{H}_{\text{site}} = \sum_{\mathbf{r}} |\Delta_s| \cos( 2\mathcal{K}r_z + \theta_s) [c^\dagger_\mathbf{r} \Gamma_3 c_\mathbf{r}].
\label{eq:MFonsite}\end{equation}
where $|\Delta_s|$ is the amplitude of the onsite CDW, and $\theta_s$ is its phase. Second, there is a bond-order term in the $z$-direction that modulates the hopping terms by 
\begin{equation}
\hat{H}_{\text{bond}} = \sum_{\mathbf{r}} |\Delta_b| \cos(2\mathcal{K}r_z + \theta_b) [c^\dagger_\mathbf{r} \Gamma_3 c_\mathbf{r+{z}}] + H.c.,
\label{eq:MFhop}\end{equation}
where $|\Delta_b|$ is the amplitude of the bond-order CDW, and $\theta_b$ is its phase.
Note that we are explicitly choosing terms that carry momentum equal to the separation of the Dirac nodes ($2\mathcal{K}$). This nesting condition allows the mean field term to couple the low-energy degrees of freedom at the two Dirac nodes. We also could have chosen different $\Gamma$-matrix structure in Eqs. \ref{eq:MFonsite},\ref{eq:MFhop}, but the choice of $\Gamma_3$ ensures these terms can open a mass gap in the continuum Hamiltonian. 

Returning to the low-energy theory, Eq.~\ref{eq:LowEnergyHam}, the CDW mean field terms in Eq.~\ref{eq:MFonsite} and~\ref{eq:MFhop} generically induce the following terms in the continuum Hamiltonian 
\begin{equation}\begin{split}
\mathcal{H}_{\text{mass}} = M(\bm{q}) \Gamma_3 \tau^x  + M'(\bm{q})\Gamma_3 \tau^y.
\label{eq:contMassTerms}\end{split}\end{equation}
The two mass terms are off diagonal in $\tau$ and, hence,  couple the two Dirac-nodes. The spectrum of the continuum Hamiltonian with the $M$ and $M'$ terms is 4-fold degenerate with energy eigenvalues 
\begin{equation}\begin{split}
E_{\pm}(\bm{q}) = \pm \Big[ &(q_x a_x)^2 + (q_y a_y)^2 + \left( q_z a_z \sin(\mathcal{K} a_z)\right)^2 \\ 
&+ M(\bm{q})^2 + M'(\bm{q})^2 \Big ]^{1/2}.
\label{eq:LESpectrum}\end{split}\end{equation} For our analysis, we are interested in the situation where the CDW is weak, and tangibly affects the low-energy modes along only the $k_x = k_y =0$ high symmetry line. In terms of the original lattice model, this leads to the requirement that $|\Delta_{s/b}| \ll b_{xy}$ (recall that $b_{xy}$ determines the gap away from $k_x = k_y =0$).

Now let us consider the effects of inversion symmetry. By comparing Eq.~\ref{eq:InversionBondDef} and~\ref{eq:InversionOnSiteDef} to Eq.~\ref{eq:contMassTerms}, we find that in presence of inversion $M$ and $M'$ must be an even and odd functions of $\bm{q}$, respectively. Therefore $M'(\bm{q})$ must vanish at the Dirac point, $\bm{q} = 0$ in inversion-symmetric systems. This leads to two distinct inversion-symmetric insulators, one with and $M'(0)= 0$ and $M(0)>0,$ and one with $M'(0)= 0$ and $M(0) < 0$. 
We now wish to relate the values of the $M$ and $M'$ to the microscopic mean-field CDW parameters, $|\Delta_s|$, $|\Delta_b|$, $\theta_s$ and $\theta_b$.
Recall that we are implicitly using a unitary transformation when considering bond-centered inversion symmetry, such that the inversion operation takes on the simple form in Eq.~\ref{eq:InversionBondDef}. Because of this, we will have to consider the site-centered and bond-centered cases (i.e. those with and without the unitary transformation) separately when relating $M$ and $M'$ to the mean-field parameters.
Plugging Eq.~\ref{eq:LEFermDef} into Eq.~\ref{eq:MFonsite} and~\ref{eq:MFhop}, we find that for site-centered inversion symmetry $M$ and $M'$ are related to the mean-field terms as 
\begin{equation}\begin{split}
M(0) & = |\Delta_s| \cos(\theta_s) + |\Delta_b| \cos(\theta_b - \mathcal{K}a_z),\\
M'(0) &= |\Delta_s| \sin(\theta_s) + |\Delta_b| \sin(\theta_b - \mathcal{K}a_z).\\
\end{split}\end{equation}
To be compatible with inversion symmetry we need $M'(0)$ to vanish, hence, the onsite term is compatible with site-centered inversion symmetry when $\theta_s = 0$, or $\pi,$ and the hopping term is compatible with site-centered inversion symmetry when  $\theta_b = \mathcal{K}a_z$, or $\mathcal{K}a_z +\pi$.
Similarly, for bond-centered inversion symmetry (using the unitary transformation discussed above) $M$ and $M'$ are related to the mean-field terms as
\begin{equation}\begin{split}
M(0) &= |\Delta_s| \cos(\theta_s + \mathcal{K}a_z) + |\Delta_b| \cos(\theta_b  ),\\
M'(0) &= |\Delta_s| \sin(\theta_s + \mathcal{K}a_z ) + |\Delta_b| \sin(\theta_b ).\\
\end{split}\end{equation}
Hence, the onsite term is compatible with bond-centered inversion symmetry when $\theta_s  = - \mathcal{K}a_z$, or $-\mathcal{K}a_z + \pi,$ and the hopping term is compatible with bond-centered inversion when $\theta_b = 0, \pi$. 
 
We note that the commensurate CDW with period 2 ($\mathcal{K}= \pi/2a_z$) is a special case, since the corresponding onsite term is compatible with site centered inversion symmetry for all values of $\theta_s$, while the hopping term is not compatible with site-centered inversion symmetry for any value of $\theta_b$. Similarly, the hopping term is compatible with bond-centered inversion symmetry for all $\theta_b$, while the onsite term is never compatible with bond-centered inversion symmetry. This can be confirmed by direct inspection of Eq.~\ref{eq:MFonsite} and~\ref{eq:MFhop}.

The two inversion-symmetric Dirac-CDW insulators ($M(0)\lessgtr 0$, $M'(0) = 0$) are anisotropic topological crystalline insulators (TCI). As we shall show in the next section, these have different $\mathbb{Z}_2$ indices using the classification scheme discussed in Sec.~\ref{ssec:RFandInversion}. Although they have different crystalline-electromagnetic responses, both classes of insulators share many important physical properties. Namely, both classes of insulators have symmetry protected gapless surface states on boundaries normal to the $x$ and $y$ directions. This can be understood by noting that the Dirac-CDW insulator is adiabatically connected to two TRS-related copies of the Weyl-CDW insulator. Since the Weyl-CDW insulators have gapless 1D chiral surface modes on the $x$ and $y$ surfaces that circulate around the $z$-axis~\cite{wieder2020axionic}, the Dirac-CDW insulator has helical surface modes that counter-propagate around the $z$-axis~\cite{zhang2016topological}. These edge modes match those that are found in a stack of 2D quantum spin Hall (QSH) layers. The helical surface modes also indicate that edge and screw dislocations of the CDW bind helical modes~\cite{zhang2016topological}. The helical modes of the Dirac-CDW insulator are interesting in their own right, but they do not have an immediate impact on our forthcoming discussion of the disclination responses of the Dirac-CDW insulator, provided that disclinations are far away from any boundaries normal to the $x$ or $y$ directions. We note that while the surfaces normal to the $x$ and $y$ directions are gapless due to time-reversal symmetry, surfaces  normal to the $z$ direction can be gapped while preserving all relevant symmetries.

\section{crystalline-electromagnetic responses of the Dirac-CDW insulator}\label{sec:DSMCrystalResp}
We now turn our attention to the topological responses of Dirac-CDW insulators. The usual topological electromagnetic responses, namely the 3D Hall response and axion electrodynamics, are trivial for a Dirac-CDW insulator since such an insulator is essentially two time-reversal symmetry related, Weyl-CDW insulators. However, even though Dirac-CDW insulators have trivial electromagnetic responses, they can host non-trivial crystalline-electromagnetic responses. The exact nature of these responses depends on if the CDW preserves a subgroup of the $T^z$ discrete translation symmetry of the lattice (i.e., if the CDW is commensurate with the lattice) and if the CDW preserves inversion symmetry. 

Below we will show the following results. If the CDW is commensurate, the Dirac-CDW insulators have a quantized 3D dWZ response. If the CDW preserves inversion symmetry but is incommensurate, we find there are two distinct inversion-symmetric Dirac-CDW insulators that differ by a $\Phi = 2\pi$ $R\wedge F$ term. Based on our previous arguments,  two insulators differing by an $R\wedge F$ term will have different disclination charge parities. If the CDW is both commensurate and preserves inversion symmetry, we find that there are two distinct inversion-symmetric Dirac-CDW insulators that have the same 3D dWZ response, but differ by a $\Phi = 2\pi$ $R\wedge F$ term. One of these insulators has a disclination filling anomaly, while the other does not. We will establish these results analytically, and confirm them through explicit numerical lattice model calculations. 

\subsection{The 3D discrete Wen-Zee response of the Dirac-CDW insulator}\label{ssec:3DWZDSMCDW}

In this subsection, we will show that the Dirac-CDW insulator realizes the 3D dWZ response when the CDW is commensurate. Such a response cannot occur for an incommensurate CDW, as a well-defined 3D dWZ response requires discrete translation symmetry. 

To this end, let us consider a Dirac-CDW insulator generated by a commensurate CDW with a period $\frac{p}{q} a_z$, where $a_z$ is lattice constant along the $z$-direction ($\mathcal{K} = \frac{q}{p}\frac{\pi}{a_z}$ in Eq. \ref{eq:DSMham}). Such a system has an enlarged unit cell of length $\tilde{a}_z = p a_z$, and a reduced $T^z$ discrete translation symmetry $z \rightarrow z + p a_z$. The 3D dWZ response of such a system can be written analogously to Eq. \ref{eq:3DWZ2} as
\begin{equation}\begin{split}
    \mathcal{L}_{\text{3D-WZ CDW}} &= \mathcal{S}_{\text{3D}}\frac{\tilde{G}_{z}}{4\pi^2} \epsilon^{ijk}  \omega_i \partial_j A_k,\\
    \tilde{G}_{z} &= \frac{2\pi}{p a_z},\\
    \mathcal{S}_{\text{3D}} &= 2q.
\label{eq:CDW3DWZ}
\end{split}\end{equation}
Here, $\tilde{G}_{z}$ is the reciprocal lattice vector along the $z$-direction with respect to the enlarged lattice. This response indicates that a disclination line of a system of size $L_z = a_z N_z$ binds charge
\begin{equation}\begin{split}
    Q_{\text{disc}} &= \mathcal{S}_{\text{3D}} \frac{\Theta_F}{2\pi} N_z/p \mod(2 N_z/p),\\ &= 2 \frac{q}{p}\frac{\Theta_F}{2\pi} N_z \mod(2 N_z/ p), 
\label{eq:CDW3DWZCharge}\end{split}\end{equation} 
where the $\mod(2 N_z/p)$ ambiguity comes from adding a local Kramers degenerate pair of electrons to each enlarged unit cell. 
Note that we assume $N_z$ is a multiple of $p$ in order for the system to be invariant under the reduced translation symmetry. 

We can arrive at Eq.~\ref{eq:CDW3DWZ} by starting with the anomaly equation of the DSM, Eq.~\ref{eq:DSManomaly}, and enlarging the unit cell, 
\begin{equation}\begin{split}
    &a_z \rightarrow \tilde{a}_z= p a_z,\\
    &G_z \rightarrow \tilde{G}_z = G_z / p
\end{split}\end{equation}
Using that $\nu = q/p$, we then arrive at Eq.~\ref{eq:CDW3DWZ}. Note that after this redefinition of the unit cell, Eq. \ref{eq:DSManomaly} has an integer coefficient and is therefore gauge invariant and anomaly free. The system no longer needs to have gapless degrees of freedom to compensate. Indeed, the Dirac nodes can be gapped out after enlarging the unit cell, as discussed previously. The 3D dWZ response of the Dirac-CDW insulator is therefore the descendant of the anomalous DSM response, when the anomaly is removed by breaking translation symmetry. 

We can also argue that the commensurate Dirac-CDW insulator realizes the response in Eq.~\ref{eq:CDW3DWZ} by treating the DSM (without a CDW) as a family of 2D insulators that are parameterized by $k_z$. We recall from above that  in this interpretation, the 2D insulators with $|k_z| < \mathcal{K}$ are QSH insulators with $\mathcal{S}_{\text{2D}} = 2$, the insulators with $|k_z| > \mathcal{K}$ are trivial ($\mathcal{S}_{\text{2D}} = 0$), and the system is gapless at $|k_z| = \mathcal{K}$. The total charge on a disclination in the DSM is then simply the sum of contributions coming from the QSH insulators, and the gapless degrees of freedom at $k_z = \pm \mathcal{K}$.  To get an insulator we can turn on the CDW which will couple the 2D system indexed by $k_z$ to the 2D system indexed by $k_z + 2\mathcal{K}$. If the CDW is small but finite, the coupling will affect only the gapless systems at $k_z = \pm \mathcal{K}$. The effect of the CDW on the gapped states that are initially indexed $|k_z|<\mathcal{K}$ and $|k_z|>\mathcal{K}$ will be negligible and will therefore not affect the topological responses of these states. In particular, at weak coupling the topological responses of the $|k_z| < \mathcal{K}$ (i.e., the QSH insulators) will not be affected. Therefore, if we add a disclination to the system, each of the $|k_z| < \mathcal{K}$  insulators will still bind charge $\frac{\Theta_F}{\pi}$. Additionally, a direct calculation shows that the two hybridized states at $k_z = \pm \mathcal{K}$, bind a net charge of $\frac{\Theta_F}{\pi}$. The states for $|k_z| > \mathcal{K}$ correspond to trivial 2D insulators and therefore do not bind any charge. For a system of size $L_z = a_z N_z,$ where $k_z$ is quantized as a multiple of $2\pi/N_z$ (assuming periodic boundary conditions), there will be a total of $\frac{q}{p} N_z -1$ states with $|k_z| < \mathcal{K}$ (recall that $N_z$ must be a multiple of $q$). A disclination therefore binds a total charge of $ 2\frac{q}{p} \frac{\Theta_{F}}{2\pi} N_z$, in agreement with Eq.~\ref{eq:CDW3DWZCharge}.

\subsection{The disclination filling anomaly of the Dirac semimetal-charge density wave insulator}\label{ssec:DSMCDWRF}

In Sec. \ref{sec:DSMCDW} we showed that for a given $\mathcal{K}$, there are two distinct inversion-symmetric Dirac-CDW insulators. In this section, we will show that the effective response theories for these insulators differ by an $R\wedge F$ term with $\Phi = 2\pi$. If the CDW is commensurate, such that the system has $T^z$ discrete translation symmetry, one of these insulators will have a disclination filling anomaly, and the other will not (see Sec.~\ref{ssec:RFResponse}). More generally, i.e., even in incommensurate CDWs, the difference in the $R\wedge F$ term indicates that the disclination charges of the two insulators will differ by $\Theta_F/\pi \mod(2\Theta_F/\pi)$ when open boundaries are present. 

To show that the two inversion-symmetric insulators differ by a $\Phi = 2\pi$ $R\wedge F$ term, we consider the low-energy Lagrangian for the DSM subject to a mean-field CDW term (see Sec. \ref{sec:DSMCDW} and Eq.~\ref{eq:contMassTerms}), and minimally couple the Dirac fermions to the U$(1)$ charge and C$_{4z}$ rotation gauge fields,
\begin{equation}\begin{split}
\mathcal{L}&=\bar{\Psi}[i\bar{\Gamma}^\mu D_\mu + M + M'\bar{\Gamma}^5] \Psi,\\
D_\mu&=\partial_\mu-iA_\mu-i\omega_{\mu}\left[ \frac{1}{2}\sigma^z s^z -s^z\right],
\label{eq:ContLag}\end{split}\end{equation}
where $\Psi$ is an 8 component spinor, $\bar{\Psi} = \Psi^\dagger \bar{\Gamma}^0$, and 
\begin{equation}\begin{split}
&\bar\Gamma^x=\sigma^y s^z \tau^x, \phantom{=} \bar\Gamma^y=\sigma^x  \tau^x, \phantom{=} \bar\Gamma^z= \tau^y, \\ & \bar\Gamma^0=\sigma^z\tau^x, \phantom{=}  \bar\Gamma^5 =\tau^z.
\end{split}\end{equation}
The covariant derivative $D_\mu$ minimally couples the low-energy fermions to the U$(1)$ gauge field, $A_\mu$, and the C$_{4z}$ gauge field, $\omega_\mu$. The latter couples via the C$_{4z}$ angular momentum operator, $\frac{1}{2}\sigma^z s^z - s^z$~\cite{may2022topological}. In this minimal coupling procedure, we are treating $\omega$ as a continuous gauge field with fluxes that are quantized in multiples of $\pi/2$ as discussed before. Compared to Eq.~\ref{eq:contMassTerms}, we are suppressing any momentum dependence of the CDW mass terms $M$ and $M'$, as it is either irrelevant at this order, or can be absorbed into a redefinition of the Fermi velocity. 

We are interested in the difference between the topological response of the two gapped inversion-symmetric phases, i.e., inversion symmetry sets $M' = 0$, and we want to compare the insulators having  $M>0$ or $M<0$. To this end, we set $M = \bar{M}\cos(\theta)$ and $M' = \bar{M}\sin(\theta)$, and consider a process where $\theta$ is smoothly increased from $0$ to $\pi$. The fermions remain gapped during this process (Eq.~\ref{eq:LESpectrum}), and the effective response theory (treated as a function of $\theta$) is found by integrating out the massive fermions. The resulting effective response theory contains the topological term,
\begin{align}
\mathcal{L}_{\mathrm{top}}[\theta] =\frac{\theta}{2\pi^2} \epsilon^{\mu\nu\lambda\eta}\partial_\mu \omega_{\nu}\partial_\lambda A_\eta\quad,
\label{eq:RFTermEff}\end{align}
which arises from a triangle diagram with external legs $A_\mu$, $\omega_{\mu}$, and $\theta$. The $\theta = 0$ and $\theta= \pi$ inversion-symmetric insulators therefore differ by a $\Phi = 2\pi$ $R\wedge F$ term (Eq. \ref{eq:RFTerm}).

\begin{figure}
    \centering
   \includegraphics[width=0.45\textwidth]{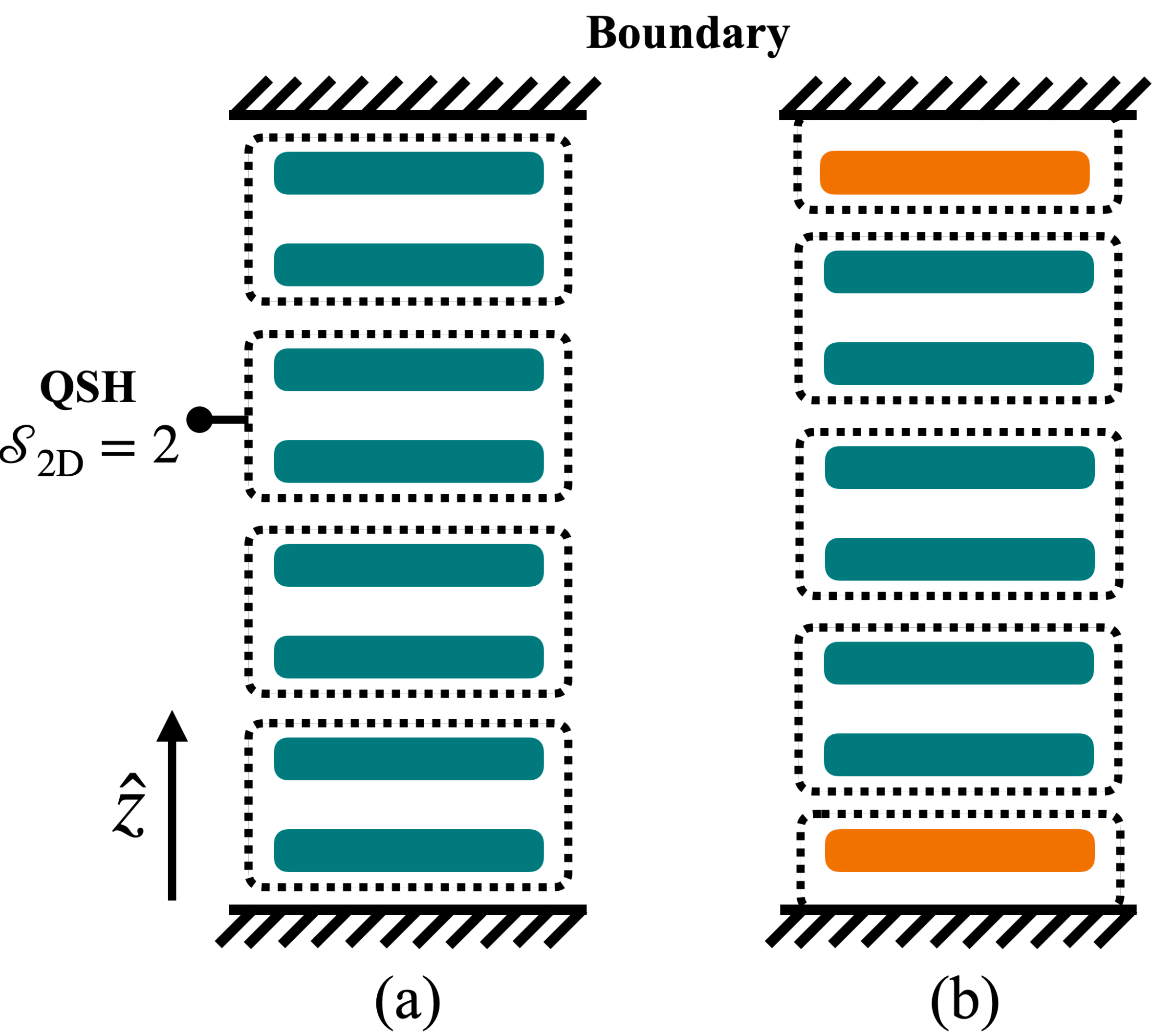}
    \caption[]{
   Layered construction of the period-2 Dirac-CDW insulators, where a two adjacent $z$-layers (teal) pair up to form a QSH insulator (dashed box) with $\mathcal{S}_{\text{3D}} = 2$. For a system of length $L_z = a_z N_z$ with open boundaries, this stacking can be done such that there are $N_z/2$ QSH insulators (a) or $N_z/2-1$ QSH insulator layers (b). This describes Eq.~\ref{eq:realSpaceP2} with $|\Delta_b| = 1$, and $\theta_b = 0$ and $\pi$ respectively.
   When there $N_z/2-1$ QSH insulator layers, there are a pair of decoupled $z$-layers on the top and bottom surfaces (orange).
    }\label{fig:HalfTrans}
\end{figure}

Based on this, and our discussions in Sec.~\ref{ssec:RFResponse} and~\ref{ssec:3DWZDSMCDW}, we therefore conclude that the two distinct inversion-symmetric Dirac-CDW insulators differ by a $\Phi = 2\pi$ $R\wedge F$ term. For commensurate CDWs, this difference is related to the presence or absence of a disclination filling anomaly. For incommensurate CDWs, the difference in $R\wedge F$ term indicates that the two insulators have different disclination charge parities.

\subsection{Analytic solution for the  period 2 charge density wave}\label{ssec:P2CDW}
To demonstrate the Dirac-CDW insulator responses we will first consider the insulators that form when a DSM with $\mathcal{K}= \pi/2a_z$ is driven into a massive phase by a period-2 CDW with inversion symmetry. As noted in Sec. \ref{sec:DSManomaly}, there are two types of inversion symmetries to consider, the site-centered inversion symmetry that sends $z\rightarrow -z,$ and the bond-centered inversion symmetry that sends $z\rightarrow -z+a_z$. For a system with open boundaries, the site-centered inversion symmetry requires an odd number of sites, while the bond-centered inversion symmetry requires an even number of sites. For an insulator with a period-2 CDW, it is natural to require an even number of sites for commensurability, and so we will consider a lattice model with bond-centered inversion symmetry.

The mean field limit of the DSM (Eq.~\ref{eq:DSMham}) having a period 2 CDW hopping term (Eq.~\ref{eq:MFhop}) is 
\begin{equation}
\begin{split}
&H_{\text{DSM-CDW}} =\begin{bmatrix} H^{xy} & H^{z}\\
 H^{z\dagger}& H^{xy} \end{bmatrix}\\
 &H^{xy}  = \sin(k_x a_x) \Gamma_1 + \sin(k_y a_y) \Gamma_2 \\
 &\phantom{====}-b_{xy}(2-\cos(k_x a_x )-\cos(k_y a_y)) \Gamma_3\\
 &H^{z} =  (1+|\Delta_b|\cos(\theta_b))\Gamma_3 \\
 &\phantom{===}+ e^{i \tilde{k}_z \tilde{a}_z}(1-|\Delta_b|\cos(\theta_b))\Gamma_3
\label{eq:MFHamPeriod2}\end{split}
\end{equation}
where $\tilde{k}_z = k_z/2$ is the momentum of the folded Brillouin zone with conjugate position, $\tilde{z} = 2z.$ The unit cells indexed by $\tilde{z}$ are twice as large as in the gapless DSM, and have length $\tilde{a}_z = 2 a_z$. We will consider a weak CDW where $|\Delta_b| \ll b_{xy}$. The bond-centered inversion symmetry acts on $H_{\text{DSM-CDW}}$ as
\begin{align}
\mathcal{I}_b = \sigma^z  t^x
\end{align}
where the $\bm{t}$ Pauli matrices act on the $4\times 4$ blocks that make up $H_{\text{DSM-CDW}}$. The even-$z$ (odd-$z$) sublattices of Eq. \ref{eq:DSMham} correspond to $t^z = +1 (-1)$ in Eq. \ref{eq:MFHamPeriod2}. The $\bm{t}$ Pauli matrices should not to be confused with the $\bm{\tau}$ Pauli matrices that differentiate the two Dirac nodes in the low-energy Hamiltonian, Eq. \ref{eq:LowEnergyHam}. 

There are two inversion-symmetric gapped phases to consider, $\cos(\theta_b) >0$ and $\cos(\theta_b) < 0$. For simplicity, we will restrict our attention to the cases where $\theta_b = 0$ or $\theta_b = \pi$. For the period-2 CDW, other values of $\theta_b$ are equivalent to one of these two values by a redefinition of $|\Delta_b|$.
To demonstrate that the $\theta_b = 0$ and $\theta_b = \pi$ insulators differ by a $\Phi = 2\pi$ $R\wedge F$ term we can add an inversion symmetry breaking onsite term
\begin{equation}
H' = 2\Delta' \Gamma_3 t^z
\label{eq:MFsite2}\end{equation}
which is simply the onsite CDW term, Eq.~\ref{eq:MFonsite} with $\Delta' = |\Delta_s|\cos(\theta_s)$. Indeed, in the continuum limit, $H_{\text{DSM-CDW}} + H'$ is equivalent to Eq.~\ref{eq:ContLag} with $M \propto |\Delta_b| \cos(\theta_b)$ and $M' \propto \Delta'$. Hence, based on Sec. \ref{ssec:DSMCDWRF} the $\theta_b = 0$ and $\theta_b = \pi$ phases differ by a $\Phi = 2\pi$ $R\wedge F$ term.

To analyze the full lattice model, it is useful to consider $|\Delta_b| = 1 $ and $\theta_b = 0$ where the Hamiltonian is independent of $\tilde{k}_{z}$, and can be treated as a stack of 2D insulating layers. The $t^x = \pm 1$ sectors of each layer are equivalent to QSH Hamiltonians, Eq. \ref{eq:QSHham},  with masses $m = 2 \mp \frac{2}{b_{xy}}$ respectively. Recalling that we have set $b_{xy}\gg |\Delta_b| = 1$ , the $t^x = -1$ sector is in the QSH phase, while the $t^x = +1$ sector is in the trivial phase. We can therefore conclude that for $|\Delta_b| = 1 $ and $\theta_b = 0$ each CDW period hosts a single 2D QSH insulator, in agreement with our earlier arguments. 

To determine the 3D dWZ response in this layered limit, let us consider adding a disclination to a lattice of length $L_{z} = a_z N_{z}$ having periodic boundary conditions. In order for the system to preserve the reduced translation symmetry, $z\rightarrow z + 2a_z$, $N_z$ must be an even integer such that there are $N_z/2$ CDW periods. Since each CDW period contains a single QSH insulator which has a $\mathcal{S}_{\text{2D}} = 2$ 2D dWZ response, the charge bound to a disclination is 
\begin{equation}
    Q_{\text{disc}} = \frac{1}{2}\frac{\Theta_F}{\pi} N_z \mod(N_z). 
\label{eq:Period2DiscChargePeriodic}\end{equation}
Comparing this to Eq. \ref{eq:CDW3DWZCharge}, we find that the full 3D system has a 3D dWZ response with $\mathcal{S}_{\text{3D}} = 2$, as expected for a layered system. 

Next let us consider the $|\Delta_b| = 1 $ and $\theta_b = \pi$ insulator. This insulator is related to the $\theta_b = 0$ insulating phase by a translation in the $z$-direction by half of the doubled unit cell (i.e., translation by one unit cell of the original DSM lattice model) 
\begin{equation}
    T_{z-1/2} =  \begin{bmatrix}
        0 & e^{i k_z} \\
        1 & 0
    \end{bmatrix}.
\end{equation}
The $|\Delta_b| = 1 $ and $\theta_b = \pi$ insulator can therefore also be treated as a stack of 2D insulating layers. The 2D insulators equally occupy the $t^z = +1$ sublattice of layer $\tilde{z}$ and the $t^z = -1$ sublattice of layer $\tilde{z}+1$. Based on our previous results, the bulk of the $|\Delta_b| = 1 $ and $\theta_b = \pi$ insulator also hosts a single 2D QSH insulator per CDW period, and therefore has $\mathcal{S}_{\text{3D}} = 2$. Hence, both the $|\Delta_b| = 1 $ and $\theta_b = 0,\pi$ insulators have an identical 3D dWZ response with $\mathcal{S}_{\text{3D}} = 2$. Since the 3D dWZ response is quantized for insulators with rotation symmetry and charge conservation, we conclude that all insulating phases of our model with $b_{xy} \gg |\Delta_b| \neq 0$ have this response.

While these phases are not distinguished by their dWZ response, based our previous discussions in Sec.~\ref{ssec:RFResponse}, we expect two distinct classes of inversion-symmetric insulators can be distinguished by another response. In fact, we have shown that the effective response theories for these insulators differ by a $\Phi = 2\pi$ $R\wedge F$ term. Based on our results from Sec.~\ref{ssec:DSMCDWRF}, the $\theta_b = 0$, and $\theta_b = \pi$ insulators should fall into different classes, and one of these insulators should have a disclination filling anomaly, while the other should not. As we shall show, the $\theta_b = \pi$ insulator has a disclination filling anomaly, while the $\theta_b = 0$ insulator does not. 

To show this, let us consider a system described by Eq. \ref{eq:MFHamPeriod2} of length $L_{z} = a N_{z}$ ($N_{z}$ even)  having open boundaries in the $z$-direction. Using position space in the $z$-direction, and momentum space in the $x$ and $y$-directions, the Hamiltonian is
\begin{equation}\begin{split}
\hat{H}_{\text{DSM-CDW}} &= \sum^{N_z/2}_{\tilde{z}=1}  [ c^\dagger_{k_x,k_y,\tilde{z}} H^{xy} c_{k_x,k_y,\tilde{z}} \\ 
&+ (1 +|\Delta_b|\cos(\theta_b)) c^\dagger_{k_x,k_y,\tilde{z}} \Gamma_3 t^x c_{k_x,k_y,\tilde{z}}\\
&+ (1 - |\Delta_b|\cos(\theta_b)) c^\dagger_{k_x,k_y,\tilde{z}} \Gamma_3 t^x c_{k_x,k_y,\tilde{z}+1}],
\label{eq:realSpaceP2}\end{split}\end{equation}
where $\tilde{z} = 2z$ labels the doubled unit cells along the $z$-direction. As before, we consider the representative insulators with $|\Delta_b| = 1$ and $\theta_b = 0$ or $\pi$.

For $|\Delta_b| = 1$ and $\theta_b = 0$, the spectrum is fully gapped, and different $\tilde{z}$-layers are fully decoupled. Furthermore, each of these layers has $\mathcal{S}_{\text{2D}} = 2,$ and constitutes a QSH insulator, i.e. one non-trivial QSH per $\tilde{z}$. This stacking configuration is shown in Fig.~\ref{fig:HalfTrans}(a). If we add a $\pi/2$ disclination-line to the insulator, each $\tilde{z}$-layer will bind charge $\Theta_F/\pi$. For a system of length $L_z = a_z N_z$, the total charge on the disclination line is the sum of the charges on the 2D layers, 
 \begin{equation}\begin{split}
Q_{\text{disc}} &= \frac{1}{2}\frac{\Theta_F}{\pi} N_{z} \mod(2\frac{\Theta_F}{\pi}),
\label{eq:DiscCharge1P2}\end{split}\end{equation}
where the $2\Theta_F/\pi$ ambiguity comes from the ability to add 2D insulators having $\mathcal{S}_{\text{2D}} \in 2\mathbb{Z}$  to the top and bottom surfaces while preserving inversion symmetry (recall that $\mathcal{S}_{\text{2D}}$ is an even integer for spin-1/2 insulators with TRS). Comparing Eq.~\ref{eq:DiscCharge1P2} and \ref{eq:Period2DiscChargePeriodic}, we find that the disclination charge of the $|\Delta_b| = 1$ and $\theta_b = 0$ insulator changes by $0 \mod(2\Theta_F/\pi)$ when changing boundary conditions, indicating that this system does not have a disclination filling anomaly.

We can similarly find the charge on the disclination line when $|\Delta_b| = 1$ and $\theta_b = \pi$. 
For open boundaries, the $t^z=-1$ sublattice of the $\tilde{z}=1$ layer is fully decoupled.  The Hamiltonian for the fermions in this sublattice/layer sector is just $H^{xy}$ from Eq.~\ref{eq:MFHamPeriod2}, which has a gapless point at $k_x = k_y = 0$. The fermions on the $t^z=+1$ sublattice of the $\tilde{z}=N_{z}/2$ layer at the top are also decoupled from the other layers/sublattices and are gapless. These sectors can be gapped out, with inversion symmetry preserving mass terms $\propto \Gamma^3$. When such a perturbation has been applied we can determine the charge distribution on a $\pi/2$-disclination line. As noted previously, the bulk of the $|\Delta_b| = 1$, $\theta_b = \pi$ insulator is composed of layers of QSH insulators having $\mathcal{S}_{\text{2D}} =2$. These insulators have equal weight on the $t^z = +1$ sublattice of $\tilde{z}$ and the $t^z = -1$ sublattice of $\tilde{z}+1$. Since these QSH insulators live \textit{between} $\tilde{z}$ layers, a system of length $L_{z} = a N_{z}$ with open boundaries in the z-direction will have $N_{z}/2 - 1$ QSH layers. This stacking configuration is shown in Fig.~\ref{fig:HalfTrans}(b).
To compute the disclination charge we also note that the gapping perturbation on the surfaces can generate boundary insulators that also have 2D dWZ responses with $\mathcal{S}_{\text{2D-bnd}} \in 2\mathbb{Z}$. However, the discrete Wen-Zee shift of the boundary insulators at the top and bottom must be the same because of inversion symmetry. Counting up the contributions from the bulk QSH layers and the boundary insulating layers, the total charge on the disclination line is 
\begin{equation}\begin{split}
Q_{\text{disc}} = \frac{1}{2}\frac{\Theta_F}{\pi} (N_{z}- 2 + 2\mathcal{S}_{\text{2D-bnd}})  \mod(2 \frac{\Theta_F}{\pi} ).
\label{eq:DiscCharge2P2}\end{split}\end{equation}
Using Eqs. \ref{eq:DiscCharge2P2} and \ref{eq:Period2DiscChargePeriodic}, we find that the charge on the disclination lines of the $|\Delta_b| = 1$, $\theta_b = \pi$ insulator changes by $\Theta_F/\pi \mod(2 \Theta_F/\pi )$ when changing boundary conditions. The $\theta_b = \pi$ insulator therefore has a disclination filling anomaly. Physically, the difference in charge can be traced back to the fact that if we place the QSH insulators between $\tilde{z}$ layers (as we did for the $\theta_b = \pi$ insulator), there will be one less QSH insulator than if we place the QSH insulators on the $\tilde{z}$ layers (as we did for the $\theta_b = 0$ insulator). Boundary effects can, effectively, add a QSH insulator to the top and bottom layers. But even with the effect, the parity of the number of QSH insulators is fixed in each stacking configuration. This leads to the quantized difference in disclination charge found above.

In summary, we find two distinct inversion-symmetric insulators can arise from the coupling a $\mathcal{K} = \pi/2a_z$ DSM to a CDW. As expected, both insulators have a $\mathcal{S}_{\text{3D}} = 2$ 3D dWZ response, and one has a disclination filling anomaly while the other does not.

\subsection{Numeric analysis}\label{ssec:numerics}

\begin{figure*}
    \centering
    \subfloat[][]{\label{fig:disc_charge_vs_z_a}\includegraphics[width=0.31\textwidth]{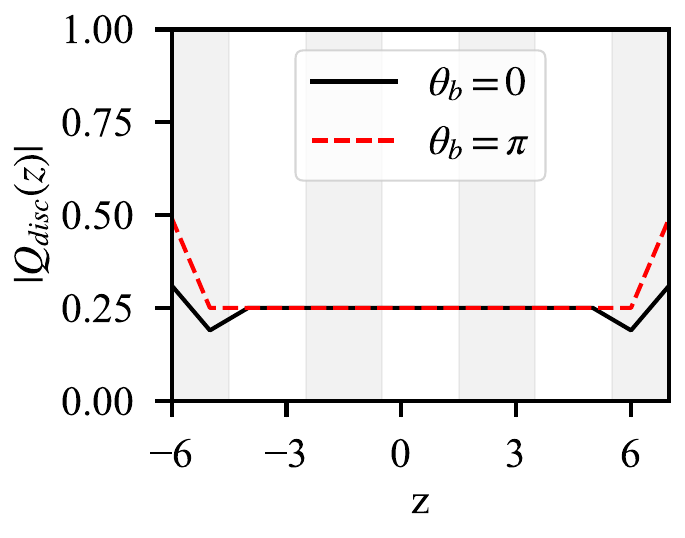}}
    \hfill
    \subfloat[][]{\label{fig:disc_charge_vs_z_b}\includegraphics[width=0.31\textwidth]{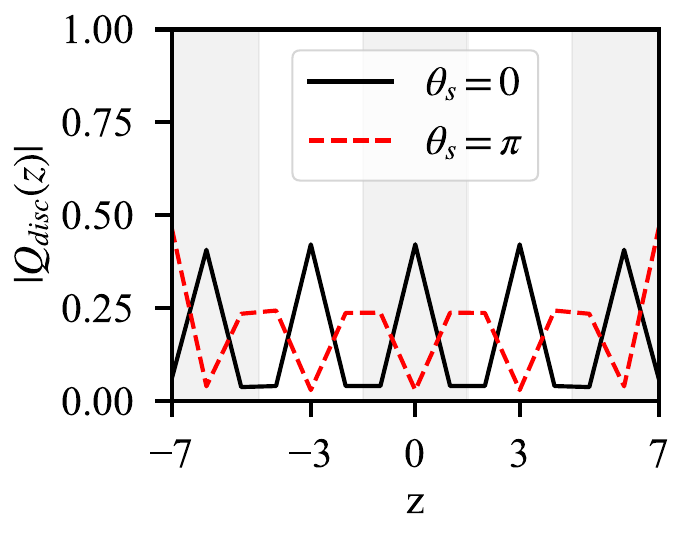}}
    \hfill
    \subfloat[][]{\label{fig:disc_charge_vs_z_c}\includegraphics[width=0.31\textwidth]{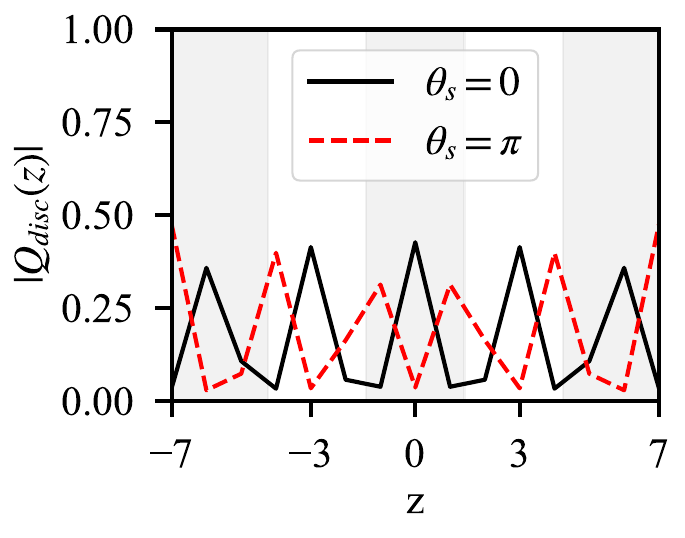}}
    \\
    \subfloat[][]{\label{fig:disc_charge_vs_z_d}\includegraphics[width=0.31\textwidth]{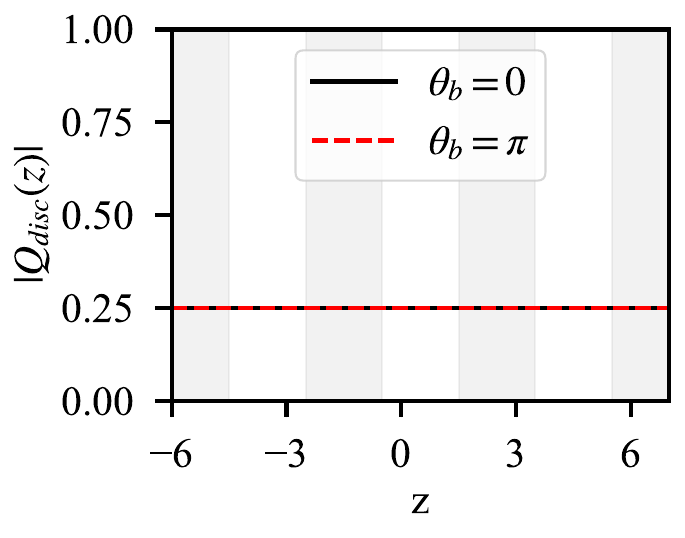}}
    \hfill
    \subfloat[][]{\label{fig:disc_charge_vs_z_e}\includegraphics[width=0.31\textwidth]{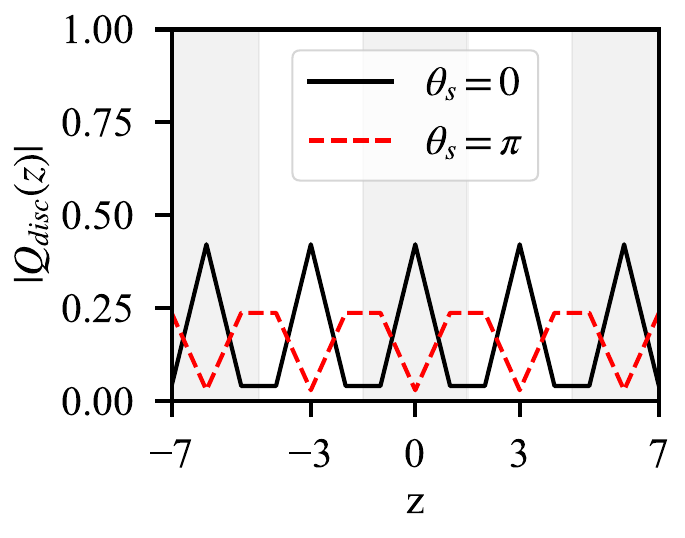}}
    \hfill
    \subfloat[][]{\label{fig:disclination_charge_distribution}\includegraphics[width=0.31\textwidth]{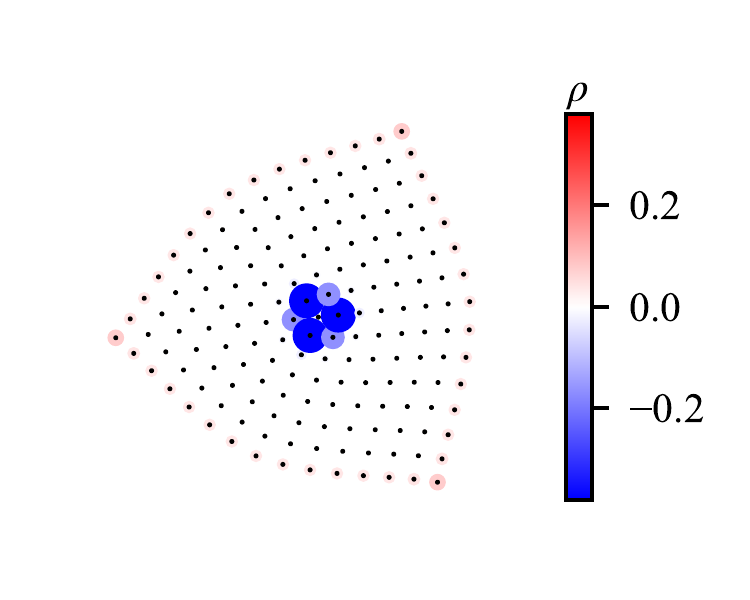}}
    \caption[]{(a, b) The charge bound to a disclination as a function of $z$ for $\mathcal{K}=\pi/2a_z$, $\Delta_s=0$, $\Delta_b=0.5$, $n_x = 15$, and $n_z = 14$ with (a) open and (d) periodic boundary conditions.  (b, e) The charge bound to a disclination as a function of $z$ for $\mathcal{K}=\pi/3a_z$, $\Delta_s=0.5$, $\Delta_b=0$, $n_x = 15$, and $n_z = 15$ with (b) open and (e) periodic boundary conditions. (c) The charge bound to a disclination as a function of $z$ for $\mathcal{K}=\sqrt{2}\pi/4$, $\Delta_s=0.5$, $\Delta_b=0$, $n_x = 15$, and $n_z = 15$ with open boundary conditions. In all cases we set $b_{xy}=1.0$ and $\Delta_{\text{surf}}=0.25$. The solid black and dashed red lines indicate $\theta_{b/s}=0$ and $\theta_{b/s}=\pi$, respectively. The alternating gray and white shading indicates unit cells of the CDWs. (f) The charge density around a disclination summed over the $z$-direction with $\mathcal{K}=\pi/2a_z$, $\theta_b=\pi$, $\Delta_s=0$, $\Delta_b=0.5$, $b_{xy}=1.0$, $\Delta_{\text{surf}}=0.25$, $n_x = 15$, and $n_z = 12$. The charge distribution is qualitatively identical for all other cases considered in this work.} \label{fig:disc_charge_vs_z}
\end{figure*}

\begin{table}
\centering
\setlength{\tabcolsep}{4pt}
\renewcommand{\arraystretch}{1.1}
\begin{tabular}{ c | c  c  c | c | c  c  c } 
    $\mathcal{K}a_z$ & $|\Delta_s|$ & $|\Delta_b|$ & $\theta_{s/b}$ & $N_z$ &\makecell{$Q_{\text{disc}}$ \\ OBC} & \makecell{$Q_{\text{disc}}$ \\ PBC} & \makecell{DL Filling \\ Anomaly} \\[0.25\normalbaselineskip]
    \hline
    \rule{0pt}{1.25\normalbaselineskip}
    $\pi/2$ & 0 & $1/2$ & 0 & 14 & 7/2 & 7/2 & No \\
    $\pi/2$ & 0 & $1/2$ & $\pi$ & 14 & 4 & 7/2 & Yes \\
    \rule{0pt}{1.25\normalbaselineskip}
    $\pi/3$ & $1/2$ & 0 & 0 & 15 & 5/2 & 5/2 & No \\
    $\pi/3$ & $1/2$ & 0 & $\pi$ & 15 & 3 & 5/2 & Yes \\
    \rule{0pt}{1.25\normalbaselineskip}
    $\pi/2\sqrt{2}$ & $1/2$ & 0 & 0 & 15 & 5/2 & N/A& N/A \\
    $\pi/2\sqrt{2}$ & $1/2$ & 0 & $\pi$ & 15 & 3 & N/A & N/A \\
\end{tabular}
\caption{Summary of the numerical analysis of the crystalline-electromagnetic responses of the Dirac-CDW insulator. The Hamiltonian for the DSM is given by Eq.~\ref{eq:DSMham} with Dirac nodes at $\bm{k} = (0,0,\pm \mathcal{K})$ and $b_{xy} = 1$. The CDW is included at the via either an inversion-symmetric onsite potential ($\hat{H}_s$ in Eq.~\ref{eq:MFonsite} with $\theta_s = 0$ or $\pi$) or an inversion-symmetric hopping amplitude ($\hat{H}_b$ in Eq.~\ref{eq:MFhop} with $\theta_b = 0$ or $\pi$) with period $\pi/\mathcal{K}$. For the commensurate values of $\mathcal{K}= \pi/2$ and $\pi/3$) the total charge bound to a disclination is calculated for periodic and open and boundary conditions, the difference of which gives the disclination filling anomaly, defined in Sec.~\ref{ssec:discFillingAnom}. However, periodic boundary conditions are not compatible with the incommensurate value of $\mathcal{K}= \pi/2\sqrt{2}$, precluding the calculation of the disclination filling anomaly.}
\label{table:NumericResults}
\end{table}

In this subsection, we numerically calculate the crystalline-electromagnetic responses for the inversion-symmetric Dirac-CDW insulators that form when a DSM is coupled to: a commensurate $\mathcal{K}=\pi/2a_z$ (period 2) CDW, a commensurate $\mathcal{K}=\pi/3a_z$ (period 3) CDW, and an incommensurate $\mathcal{K}=\pi/2\sqrt{2}a_z$ CDW. These calculations are performed using the DSM Hamiltonian in Eq.~\ref{eq:DSMham} with $b_{xy}=1$. The $\mathcal{K}=\pi/2a_z$ CDW is implemented via the hopping mean field term in Eq.~\ref{eq:MFhop} with $|\Delta_b| = 1/2$ and either $\theta_b = 0$ or $\pi$, both of which preserve bond-centered inversion symmetry. The $\mathcal{K}=\pi/3a_z$ CDW is implemented via the onsite mean field term in Eq.~\ref{eq:MFonsite} with $|\Delta_s| = 1/2$ and either $\theta_s = 0$ or $\pi$, preserving the site-centered inversion symmetry. The $\mathcal{K}=\pi/2\sqrt{2}a_z$ CDW is similarly implemented with the onsite mean field term in Eq.~\ref{eq:MFonsite} with $|\Delta_s| = 1/2$ and either $\theta_s = 0$ or $\pi$, also preserving site-centered inversion symmetry. 

For each system, we evaluate the Hamiltonian on a lattice having open boundary conditions and a $\Theta_F = \pi/2$ disclination line. The disclination line is located at the center of the lattice and stretches straight along the $z$-direction. To reduce the impact of finite size effects, we also add an inversion-preserving perturbation of the form $H'_{\text{surf}} = \Delta_{\text{surf}}\Gamma_3$  ($\Delta_{\text{surf}}=1/4$ on both the top and bottom layers) to increase the gap on the $z$-normal surfaces. Then we calculate the disclination charge by summing the charge density over all sites within some radial distance from the disclination core. To further reduce the impact of finite size effects, we perform these calculations over a range of system sizes and extrapolate to the infinite-size limit. The details of how the integration radius is determined and how the extrapolations are performed, are provided in Appendix~\ref{app:Numerics}.

As expected from our analysis in previous sections, we find that charge is bound to the disclination line for all Dirac-CDW insulators, as shown in Fig. \ref{fig:disclination_charge_distribution}. Specifically, in Fig.~\ref{fig:disc_charge_vs_z} we plot the layer-resolved disclination charge for each insulator at the largest accessible system size. For the commensurate CDWs, $\mathcal{K}=\pi/2a_z$ and $\mathcal{K}=\pi/3a_z$, we find that the charge distribution oscillates with a period matching the CDW period, with deviations at open boundaries.
There is no notable periodicity for the $\mathcal{K}=\pi/2\sqrt{2}a_z$ CDW, as it is incommensurate with the lattice. The total disclination charges for each of these insulators extrapolated to the infinite-size limit are enumerated in Table~\ref{table:NumericResults}.

For $\mathcal{K}=\pi/2a_z$, the total charge bound to the disclination with open boundary conditions differs by $1/2$ between the $\theta_b = 0$ and $\theta_b = \pi$ Dirac-CDW insulators. This difference in disclination charge is a direct manifestation of the difference in the coefficient of the $R\wedge F$ term between the two Dirac-CDW insulators. Similarly, the disclination charge differs by $1/2$ between the $\theta_s = 0$ and $\theta_s = \pi$ Dirac-CDW insulators with $\mathcal{K}=\pi/3a_z,$ and  $\mathcal{K}=\pi/2\sqrt{2}a_z$. We therefore find that the two distinct inversion-symmetric Dirac-CDW insulators for each CDW wavevector have different disclination charge parities, matching our analytic predictions.

For the commensurate $\mathcal{K}=\pi/2a_z$ and $\mathcal{K}=\pi/3a_z$ CDWs, it is also possible to have periodic boundary conditions in the $z$-direction. Hence in these cases we can determine the 3D dWZ response of the Dirac-CDW insulators and whether or not the insulators  host a disclination filling anomaly. We find that for periodic boundaries, both the $\theta_b = 0$ and $\pi$ insulators with $\mathcal{K}=\pi/2a_z$ bind charge $1/2$ per CDW period ($1/4$ per original lattice layer). The $\theta_b = 0$ and $\pi$ insulators with $\mathcal{K}=\pi/3a_z$ also bind charge $1/2$ per CDW period ($1/6$ per original lattice layer). This is exactly the 3D dWZ response that we have previously discussed. Furthermore, we find that the disclination charges differ by $1/2$ with periodic and open boundary conditions for the $\mathcal{K}=\pi/2a_z$, $\theta_b = \pi$ and $\mathcal{K}=\pi/3a_z$, $\theta_s = \pi$ insulators, indicating disclination filling anomalies. The presence of the disclination filling anomaly can also be observed in the charge density profiles plotted in Fig.~\ref{fig:disc_charge_vs_z}, in which it manifests as a deviation at the boundaries of the disclination charge per CDW period. The disclination charges for open and periodic boundary conditions are identical for the $\mathcal{K}=\pi/2a_z$, $\theta_b = 0$ and $\mathcal{K}=\pi/3a_z$, $\theta_s = 0$ insulators, marking the absence of disclination filling anomalies. These results agree with the analytic calculations presented in Sec.~\ref{ssec:P2CDW}. As discussed in Sec.~\ref{ssec:discFillingAnom}, the disclination filling anomaly is ill-defined for insulators with incommensurate $\mathcal{K}$, like the $\mathcal{K}=\pi/2\sqrt{2}a_z$ Dirac-CDW insulators, since they are incompatible with periodic boundaries.
 
\section{Conclusion and outlook}
\label{sec:conclusion}
In this work, we have considered the topological responses of the insulating state that arises from coupling a 3D DSM to a CDW. Unlike the related Weyl-CDW insulators, the Dirac-CDW insulators do not display a Hall effect or axion electrodynamics. Rather, the Dirac-CDW insulators display novel crystalline-electromagnetic responses wherein charge is bound to disclination defects of the lattice. These crystalline-electromagnetic responses are encoded in a 3D discrete Wen-Zee term and an $R\wedge F$ term. Due to the inversion symmetry of the DSM, there are two classes of insulating states where $\Phi = 0$  and $\Phi = 2\pi$ respectively. These two insulating phases can be differentiated by considering the total charge bound to disclination lines. These responses naturally arise from the combination of topology and spatial symmetries that stabilize the DSMs.

In terms of real materials, a potential material candidate for realizing the Dirac-CDW insulator is TaTe$_4$, a DSM that develops commensurate CDW order at finite temperature~\cite{zhang2020eightfold}. The CDW momentum matches the momentum space separation for a pair of Dirac points. Despite this, the CDW phase of TaTe$_4$ is not insulating, but rather has an 8-fold degenerate doubled Dirac point in the folded Brillouin zone. Additionally, TaTe$_4$ has Dirac points at other momenta that do not strongly couple to the CDW, and are expected to remain gapless. The gapless Dirac points prevent TaTe$_4$ from realizing the quantized crystalline-electromagnetic responses discussed, but it does indicate that the combination of Dirac and CDW physics can occur in real materials. 

There is also a much broader family of topological semimetals protected by non-symmorphic crystal symmetries, for which there are many candidate host materials and some experimentally confirmed examples~\cite{bradlyn2016beyond, lv2021experimental}. Understanding the mixed crystalline-electromagnetic response of non-symmorphic topological semimetals is an open question for future work, as the structure of non-symmorphic symmetry fluxes is not currently understood. In addition, it would be interesting to consider the crystalline-electromagnetic responses of higher order Dirac semimetals, both with and without interactions. Previous studies of higher order semimetals~\cite{lin2018topological,ghorashi2020higher, wang2020higher,wieder2020strong,hirsbrunner2023crystalline} have already revealed novel behavior not found in their first order counterparts, but a full understanding of the topological responses of higher order semimetals remains incomplete, even for symmorphic symmetries.

\acknowledgments
The authors thank Anton Burkov, Pranav Rao, Ryan Thorngren, and Chong Wang for helpful discussions. J.M.M. thanks the National Science Foundation Graduate Research Fellowship Program under Grant No. DGE - 1746047 and a startup fund at Stanford University. L.G. thanks the Government of Canada through the Department of Innovation, Science and Economic Development and the Province of Ontario through the Ministry of Economic Development, Job Creation and Trade. M.R.H., J.M.M., and T.L.H. thank ARO MURI W911NF2020166 for support. T.L.H. thanks the U.S. Office of Naval Research (ONR) Multidisciplinary University Research Initiative (MURI) Grant No. N00014-20-1-2325 on Robust Photonic Materials with High-Order Topological Protection for support.


\begin{thebibliography}{83}%
\makeatletter
\providecommand \@ifxundefined [1]{%
 \@ifx{#1\undefined}
}%
\providecommand \@ifnum [1]{%
 \ifnum #1\expandafter \@firstoftwo
 \else \expandafter \@secondoftwo
 \fi
}%
\providecommand \@ifx [1]{%
 \ifx #1\expandafter \@firstoftwo
 \else \expandafter \@secondoftwo
 \fi
}%
\providecommand \natexlab [1]{#1}%
\providecommand \enquote  [1]{``#1''}%
\providecommand \bibnamefont  [1]{#1}%
\providecommand \bibfnamefont [1]{#1}%
\providecommand \citenamefont [1]{#1}%
\providecommand \href@noop [0]{\@secondoftwo}%
\providecommand \href [0]{\begingroup \@sanitize@url \@href}%
\providecommand \@href[1]{\@@startlink{#1}\@@href}%
\providecommand \@@href[1]{\endgroup#1\@@endlink}%
\providecommand \@sanitize@url [0]{\catcode `\\12\catcode `\$12\catcode
  `\&12\catcode `\#12\catcode `\^12\catcode `\_12\catcode `\%12\relax}%
\providecommand \@@startlink[1]{}%
\providecommand \@@endlink[0]{}%
\providecommand \url  [0]{\begingroup\@sanitize@url \@url }%
\providecommand \@url [1]{\endgroup\@href {#1}{\urlprefix }}%
\providecommand \urlprefix  [0]{URL }%
\providecommand \Eprint [0]{\href }%
\providecommand \doibase [0]{http://dx.doi.org/}%
\providecommand \selectlanguage [0]{\@gobble}%
\providecommand \bibinfo  [0]{\@secondoftwo}%
\providecommand \bibfield  [0]{\@secondoftwo}%
\providecommand \translation [1]{[#1]}%
\providecommand \BibitemOpen [0]{}%
\providecommand \bibitemStop [0]{}%
\providecommand \bibitemNoStop [0]{.\EOS\space}%
\providecommand \EOS [0]{\spacefactor3000\relax}%
\providecommand \BibitemShut  [1]{\csname bibitem#1\endcsname}%
\let\auto@bib@innerbib\@empty
\bibitem [{\citenamefont {Armitage}\ \emph {et~al.}(2018)\citenamefont
  {Armitage}, \citenamefont {Mele},\ and\ \citenamefont
  {Vishwanath}}]{armitage2018weyl}%
  \BibitemOpen
  \bibfield  {author} {\bibinfo {author} {\bibfnamefont {N.}~\bibnamefont
  {Armitage}}, \bibinfo {author} {\bibfnamefont {E.}~\bibnamefont {Mele}}, \
  and\ \bibinfo {author} {\bibfnamefont {A.}~\bibnamefont {Vishwanath}},\
  }\href@noop {} {\bibfield  {journal} {\bibinfo  {journal} {Reviews of Modern
  Physics}\ }\textbf {\bibinfo {volume} {90}},\ \bibinfo {pages} {015001}
  (\bibinfo {year} {2018})}\BibitemShut {NoStop}%
\bibitem [{\citenamefont {Wan}\ \emph {et~al.}(2011)\citenamefont {Wan},
  \citenamefont {Turner}, \citenamefont {Vishwanath},\ and\ \citenamefont
  {Savrasov}}]{wan2011topological}%
  \BibitemOpen
  \bibfield  {author} {\bibinfo {author} {\bibfnamefont {X.}~\bibnamefont
  {Wan}}, \bibinfo {author} {\bibfnamefont {A.~M.}\ \bibnamefont {Turner}},
  \bibinfo {author} {\bibfnamefont {A.}~\bibnamefont {Vishwanath}}, \ and\
  \bibinfo {author} {\bibfnamefont {S.~Y.}\ \bibnamefont {Savrasov}},\
  }\href@noop {} {\bibfield  {journal} {\bibinfo  {journal} {Physical Review
  B}\ }\textbf {\bibinfo {volume} {83}},\ \bibinfo {pages} {205101} (\bibinfo
  {year} {2011})}\BibitemShut {NoStop}%
\bibitem [{\citenamefont {Burkov}\ and\ \citenamefont
  {Balents}(2011)}]{burkov2011weyl}%
  \BibitemOpen
  \bibfield  {author} {\bibinfo {author} {\bibfnamefont {A.}~\bibnamefont
  {Burkov}}\ and\ \bibinfo {author} {\bibfnamefont {L.}~\bibnamefont
  {Balents}},\ }\href@noop {} {\bibfield  {journal} {\bibinfo  {journal}
  {Physical review letters}\ }\textbf {\bibinfo {volume} {107}},\ \bibinfo
  {pages} {127205} (\bibinfo {year} {2011})}\BibitemShut {NoStop}%
\bibitem [{\citenamefont {Yang}\ \emph {et~al.}(2011)\citenamefont {Yang},
  \citenamefont {Lu},\ and\ \citenamefont {Ran}}]{yang2011quantum}%
  \BibitemOpen
  \bibfield  {author} {\bibinfo {author} {\bibfnamefont {K.-Y.}\ \bibnamefont
  {Yang}}, \bibinfo {author} {\bibfnamefont {Y.-M.}\ \bibnamefont {Lu}}, \ and\
  \bibinfo {author} {\bibfnamefont {Y.}~\bibnamefont {Ran}},\ }\href@noop {}
  {\bibfield  {journal} {\bibinfo  {journal} {Physical Review B}\ }\textbf
  {\bibinfo {volume} {84}},\ \bibinfo {pages} {075129} (\bibinfo {year}
  {2011})}\BibitemShut {NoStop}%
\bibitem [{\citenamefont {Xu}\ \emph {et~al.}(2011)\citenamefont {Xu},
  \citenamefont {Weng}, \citenamefont {Wang}, \citenamefont {Dai},\ and\
  \citenamefont {Fang}}]{xu2011chern}%
  \BibitemOpen
  \bibfield  {author} {\bibinfo {author} {\bibfnamefont {G.}~\bibnamefont
  {Xu}}, \bibinfo {author} {\bibfnamefont {H.}~\bibnamefont {Weng}}, \bibinfo
  {author} {\bibfnamefont {Z.}~\bibnamefont {Wang}}, \bibinfo {author}
  {\bibfnamefont {X.}~\bibnamefont {Dai}}, \ and\ \bibinfo {author}
  {\bibfnamefont {Z.}~\bibnamefont {Fang}},\ }\href@noop {} {\bibfield
  {journal} {\bibinfo  {journal} {Physical review letters}\ }\textbf {\bibinfo
  {volume} {107}},\ \bibinfo {pages} {186806} (\bibinfo {year}
  {2011})}\BibitemShut {NoStop}%
\bibitem [{\citenamefont {Young}\ \emph {et~al.}(2012)\citenamefont {Young},
  \citenamefont {Zaheer}, \citenamefont {Teo}, \citenamefont {Kane},
  \citenamefont {Mele},\ and\ \citenamefont {Rappe}}]{young2012dirac}%
  \BibitemOpen
  \bibfield  {author} {\bibinfo {author} {\bibfnamefont {S.~M.}\ \bibnamefont
  {Young}}, \bibinfo {author} {\bibfnamefont {S.}~\bibnamefont {Zaheer}},
  \bibinfo {author} {\bibfnamefont {J.~C.}\ \bibnamefont {Teo}}, \bibinfo
  {author} {\bibfnamefont {C.~L.}\ \bibnamefont {Kane}}, \bibinfo {author}
  {\bibfnamefont {E.~J.}\ \bibnamefont {Mele}}, \ and\ \bibinfo {author}
  {\bibfnamefont {A.~M.}\ \bibnamefont {Rappe}},\ }\href@noop {} {\bibfield
  {journal} {\bibinfo  {journal} {Physical review letters}\ }\textbf {\bibinfo
  {volume} {108}},\ \bibinfo {pages} {140405} (\bibinfo {year}
  {2012})}\BibitemShut {NoStop}%
\bibitem [{\citenamefont {Wang}\ \emph {et~al.}(2012)\citenamefont {Wang},
  \citenamefont {Sun}, \citenamefont {Chen}, \citenamefont {Franchini},
  \citenamefont {Xu}, \citenamefont {Weng}, \citenamefont {Dai},\ and\
  \citenamefont {Fang}}]{wang2012dirac}%
  \BibitemOpen
  \bibfield  {author} {\bibinfo {author} {\bibfnamefont {Z.}~\bibnamefont
  {Wang}}, \bibinfo {author} {\bibfnamefont {Y.}~\bibnamefont {Sun}}, \bibinfo
  {author} {\bibfnamefont {X.-Q.}\ \bibnamefont {Chen}}, \bibinfo {author}
  {\bibfnamefont {C.}~\bibnamefont {Franchini}}, \bibinfo {author}
  {\bibfnamefont {G.}~\bibnamefont {Xu}}, \bibinfo {author} {\bibfnamefont
  {H.}~\bibnamefont {Weng}}, \bibinfo {author} {\bibfnamefont {X.}~\bibnamefont
  {Dai}}, \ and\ \bibinfo {author} {\bibfnamefont {Z.}~\bibnamefont {Fang}},\
  }\href@noop {} {\bibfield  {journal} {\bibinfo  {journal} {Physical Review
  B}\ }\textbf {\bibinfo {volume} {85}},\ \bibinfo {pages} {195320} (\bibinfo
  {year} {2012})}\BibitemShut {NoStop}%
\bibitem [{\citenamefont {Zyuzin}\ and\ \citenamefont
  {Burkov}(2012)}]{zyuzin2012topological}%
  \BibitemOpen
  \bibfield  {author} {\bibinfo {author} {\bibfnamefont {A.}~\bibnamefont
  {Zyuzin}}\ and\ \bibinfo {author} {\bibfnamefont {A.}~\bibnamefont
  {Burkov}},\ }\href@noop {} {\bibfield  {journal} {\bibinfo  {journal}
  {Physical Review B}\ }\textbf {\bibinfo {volume} {86}},\ \bibinfo {pages}
  {115133} (\bibinfo {year} {2012})}\BibitemShut {NoStop}%
\bibitem [{\citenamefont {Zyuzin}\ \emph {et~al.}(2012)\citenamefont {Zyuzin},
  \citenamefont {Wu},\ and\ \citenamefont {Burkov}}]{zyuzin2012weyl}%
  \BibitemOpen
  \bibfield  {author} {\bibinfo {author} {\bibfnamefont {A.}~\bibnamefont
  {Zyuzin}}, \bibinfo {author} {\bibfnamefont {S.}~\bibnamefont {Wu}}, \ and\
  \bibinfo {author} {\bibfnamefont {A.}~\bibnamefont {Burkov}},\ }\href@noop {}
  {\bibfield  {journal} {\bibinfo  {journal} {Physical Review B}\ }\textbf
  {\bibinfo {volume} {85}},\ \bibinfo {pages} {165110} (\bibinfo {year}
  {2012})}\BibitemShut {NoStop}%
\bibitem [{\citenamefont {Wang}\ \emph {et~al.}(2013)\citenamefont {Wang},
  \citenamefont {Weng}, \citenamefont {Wu}, \citenamefont {Dai},\ and\
  \citenamefont {Fang}}]{wang2013three}%
  \BibitemOpen
  \bibfield  {author} {\bibinfo {author} {\bibfnamefont {Z.}~\bibnamefont
  {Wang}}, \bibinfo {author} {\bibfnamefont {H.}~\bibnamefont {Weng}}, \bibinfo
  {author} {\bibfnamefont {Q.}~\bibnamefont {Wu}}, \bibinfo {author}
  {\bibfnamefont {X.}~\bibnamefont {Dai}}, \ and\ \bibinfo {author}
  {\bibfnamefont {Z.}~\bibnamefont {Fang}},\ }\href@noop {} {\bibfield
  {journal} {\bibinfo  {journal} {Physical Review B}\ }\textbf {\bibinfo
  {volume} {88}},\ \bibinfo {pages} {125427} (\bibinfo {year}
  {2013})}\BibitemShut {NoStop}%
\bibitem [{\citenamefont {Weng}\ \emph {et~al.}(2015)\citenamefont {Weng},
  \citenamefont {Fang}, \citenamefont {Fang}, \citenamefont {Bernevig},\ and\
  \citenamefont {Dai}}]{weng2015weyl}%
  \BibitemOpen
  \bibfield  {author} {\bibinfo {author} {\bibfnamefont {H.}~\bibnamefont
  {Weng}}, \bibinfo {author} {\bibfnamefont {C.}~\bibnamefont {Fang}}, \bibinfo
  {author} {\bibfnamefont {Z.}~\bibnamefont {Fang}}, \bibinfo {author}
  {\bibfnamefont {B.~A.}\ \bibnamefont {Bernevig}}, \ and\ \bibinfo {author}
  {\bibfnamefont {X.}~\bibnamefont {Dai}},\ }\href@noop {} {\bibfield
  {journal} {\bibinfo  {journal} {Physical Review X}\ }\textbf {\bibinfo
  {volume} {5}},\ \bibinfo {pages} {011029} (\bibinfo {year}
  {2015})}\BibitemShut {NoStop}%
\bibitem [{\citenamefont {Jia}\ \emph {et~al.}(2016)\citenamefont {Jia},
  \citenamefont {Xu},\ and\ \citenamefont {Hasan}}]{jia2016weyl}%
  \BibitemOpen
  \bibfield  {author} {\bibinfo {author} {\bibfnamefont {S.}~\bibnamefont
  {Jia}}, \bibinfo {author} {\bibfnamefont {S.-Y.}\ \bibnamefont {Xu}}, \ and\
  \bibinfo {author} {\bibfnamefont {M.~Z.}\ \bibnamefont {Hasan}},\ }\href@noop
  {} {\bibfield  {journal} {\bibinfo  {journal} {Nature materials}\ }\textbf
  {\bibinfo {volume} {15}},\ \bibinfo {pages} {1140} (\bibinfo {year}
  {2016})}\BibitemShut {NoStop}%
\bibitem [{\citenamefont {Nielsen}\ and\ \citenamefont
  {Ninomiya}(1983)}]{nielsen1983adler}%
  \BibitemOpen
  \bibfield  {author} {\bibinfo {author} {\bibfnamefont {H.~B.}\ \bibnamefont
  {Nielsen}}\ and\ \bibinfo {author} {\bibfnamefont {M.}~\bibnamefont
  {Ninomiya}},\ }\href@noop {} {\bibfield  {journal} {\bibinfo  {journal}
  {Physics Letters B}\ }\textbf {\bibinfo {volume} {130}},\ \bibinfo {pages}
  {389} (\bibinfo {year} {1983})}\BibitemShut {NoStop}%
\bibitem [{\citenamefont {Burkov}(2018)}]{burkov2018weyl}%
  \BibitemOpen
  \bibfield  {author} {\bibinfo {author} {\bibfnamefont {A.}~\bibnamefont
  {Burkov}},\ }\href@noop {} {\bibfield  {journal} {\bibinfo  {journal} {Annual
  Review of Condensed Matter Physics}\ }\textbf {\bibinfo {volume} {9}},\
  \bibinfo {pages} {359} (\bibinfo {year} {2018})}\BibitemShut {NoStop}%
\bibitem [{\citenamefont {Klinkhamer}\ and\ \citenamefont
  {Volovik}(2005)}]{klinkhamer2005emergent}%
  \BibitemOpen
  \bibfield  {author} {\bibinfo {author} {\bibfnamefont {F.}~\bibnamefont
  {Klinkhamer}}\ and\ \bibinfo {author} {\bibfnamefont {G.}~\bibnamefont
  {Volovik}},\ }\href@noop {} {\bibfield  {journal} {\bibinfo  {journal}
  {International Journal of Modern Physics A}\ }\textbf {\bibinfo {volume}
  {20}},\ \bibinfo {pages} {2795} (\bibinfo {year} {2005})}\BibitemShut
  {NoStop}%
\bibitem [{\citenamefont {Murakami}(2007)}]{murakami2007phase}%
  \BibitemOpen
  \bibfield  {author} {\bibinfo {author} {\bibfnamefont {S.}~\bibnamefont
  {Murakami}},\ }\href@noop {} {\bibfield  {journal} {\bibinfo  {journal} {New
  Journal of Physics}\ }\textbf {\bibinfo {volume} {9}},\ \bibinfo {pages}
  {356} (\bibinfo {year} {2007})}\BibitemShut {NoStop}%
\bibitem [{\citenamefont {Wilczek}(1987)}]{wilczek1987two}%
  \BibitemOpen
  \bibfield  {author} {\bibinfo {author} {\bibfnamefont {F.}~\bibnamefont
  {Wilczek}},\ }\href@noop {} {\bibfield  {journal} {\bibinfo  {journal}
  {Physical review letters}\ }\textbf {\bibinfo {volume} {58}},\ \bibinfo
  {pages} {1799} (\bibinfo {year} {1987})}\BibitemShut {NoStop}%
\bibitem [{\citenamefont {Wang}\ and\ \citenamefont
  {Zhang}(2013)}]{wang2013chiral}%
  \BibitemOpen
  \bibfield  {author} {\bibinfo {author} {\bibfnamefont {Z.}~\bibnamefont
  {Wang}}\ and\ \bibinfo {author} {\bibfnamefont {S.-C.}\ \bibnamefont
  {Zhang}},\ }\href@noop {} {\bibfield  {journal} {\bibinfo  {journal}
  {Physical Review B}\ }\textbf {\bibinfo {volume} {87}},\ \bibinfo {pages}
  {161107} (\bibinfo {year} {2013})}\BibitemShut {NoStop}%
\bibitem [{\citenamefont {Maciejko}\ and\ \citenamefont
  {Nandkishore}(2014)}]{maciejko2014weyl}%
  \BibitemOpen
  \bibfield  {author} {\bibinfo {author} {\bibfnamefont {J.}~\bibnamefont
  {Maciejko}}\ and\ \bibinfo {author} {\bibfnamefont {R.}~\bibnamefont
  {Nandkishore}},\ }\href@noop {} {\bibfield  {journal} {\bibinfo  {journal}
  {Physical Review B}\ }\textbf {\bibinfo {volume} {90}},\ \bibinfo {pages}
  {035126} (\bibinfo {year} {2014})}\BibitemShut {NoStop}%
\bibitem [{\citenamefont {You}\ \emph {et~al.}(2016)\citenamefont {You},
  \citenamefont {Cho},\ and\ \citenamefont {Hughes}}]{you2016response}%
  \BibitemOpen
  \bibfield  {author} {\bibinfo {author} {\bibfnamefont {Y.}~\bibnamefont
  {You}}, \bibinfo {author} {\bibfnamefont {G.~Y.}\ \bibnamefont {Cho}}, \ and\
  \bibinfo {author} {\bibfnamefont {T.~L.}\ \bibnamefont {Hughes}},\
  }\href@noop {} {\bibfield  {journal} {\bibinfo  {journal} {Physical Review
  B}\ }\textbf {\bibinfo {volume} {94}},\ \bibinfo {pages} {085102} (\bibinfo
  {year} {2016})}\BibitemShut {NoStop}%
\bibitem [{\citenamefont {Wieder}\ \emph
  {et~al.}(2020{\natexlab{a}})\citenamefont {Wieder}, \citenamefont {Lin},\
  and\ \citenamefont {Bradlyn}}]{wieder2020axionic}%
  \BibitemOpen
  \bibfield  {author} {\bibinfo {author} {\bibfnamefont {B.~J.}\ \bibnamefont
  {Wieder}}, \bibinfo {author} {\bibfnamefont {K.-S.}\ \bibnamefont {Lin}}, \
  and\ \bibinfo {author} {\bibfnamefont {B.}~\bibnamefont {Bradlyn}},\
  }\href@noop {} {\bibfield  {journal} {\bibinfo  {journal} {Phys. Rev. Res.}\
  }\textbf {\bibinfo {volume} {2}},\ \bibinfo {pages} {042010} (\bibinfo {year}
  {2020}{\natexlab{a}})}\BibitemShut {NoStop}%
\bibitem [{\citenamefont {Qi}\ \emph {et~al.}(2008)\citenamefont {Qi},
  \citenamefont {Hughes},\ and\ \citenamefont {Zhang}}]{qi2008topological}%
  \BibitemOpen
  \bibfield  {author} {\bibinfo {author} {\bibfnamefont {X.-L.}\ \bibnamefont
  {Qi}}, \bibinfo {author} {\bibfnamefont {T.~L.}\ \bibnamefont {Hughes}}, \
  and\ \bibinfo {author} {\bibfnamefont {S.-C.}\ \bibnamefont {Zhang}},\
  }\href@noop {} {\bibfield  {journal} {\bibinfo  {journal} {Physical Review
  B}\ }\textbf {\bibinfo {volume} {78}},\ \bibinfo {pages} {195424} (\bibinfo
  {year} {2008})}\BibitemShut {NoStop}%
\bibitem [{\citenamefont {Essin}\ \emph {et~al.}(2009)\citenamefont {Essin},
  \citenamefont {Moore},\ and\ \citenamefont
  {Vanderbilt}}]{essin2009magnetoelectric}%
  \BibitemOpen
  \bibfield  {author} {\bibinfo {author} {\bibfnamefont {A.~M.}\ \bibnamefont
  {Essin}}, \bibinfo {author} {\bibfnamefont {J.~E.}\ \bibnamefont {Moore}}, \
  and\ \bibinfo {author} {\bibfnamefont {D.}~\bibnamefont {Vanderbilt}},\
  }\href@noop {} {\bibfield  {journal} {\bibinfo  {journal} {Physical review
  letters}\ }\textbf {\bibinfo {volume} {102}},\ \bibinfo {pages} {146805}
  (\bibinfo {year} {2009})}\BibitemShut {NoStop}%
\bibitem [{\citenamefont {Witten}(2016)}]{witten2016fermion}%
  \BibitemOpen
  \bibfield  {author} {\bibinfo {author} {\bibfnamefont {E.}~\bibnamefont
  {Witten}},\ }\href@noop {} {\bibfield  {journal} {\bibinfo  {journal}
  {Reviews of Modern Physics}\ }\textbf {\bibinfo {volume} {88}},\ \bibinfo
  {pages} {035001} (\bibinfo {year} {2016})}\BibitemShut {NoStop}%
\bibitem [{\citenamefont {Hughes}\ \emph {et~al.}(2011)\citenamefont {Hughes},
  \citenamefont {Prodan},\ and\ \citenamefont
  {Bernevig}}]{hughes2011inversion}%
  \BibitemOpen
  \bibfield  {author} {\bibinfo {author} {\bibfnamefont {T.~L.}\ \bibnamefont
  {Hughes}}, \bibinfo {author} {\bibfnamefont {E.}~\bibnamefont {Prodan}}, \
  and\ \bibinfo {author} {\bibfnamefont {B.~A.}\ \bibnamefont {Bernevig}},\
  }\href@noop {} {\bibfield  {journal} {\bibinfo  {journal} {Physical Review
  B}\ }\textbf {\bibinfo {volume} {83}},\ \bibinfo {pages} {245132} (\bibinfo
  {year} {2011})}\BibitemShut {NoStop}%
\bibitem [{\citenamefont {Varnava}\ and\ \citenamefont
  {Vanderbilt}(2018)}]{varnava2018surfaces}%
  \BibitemOpen
  \bibfield  {author} {\bibinfo {author} {\bibfnamefont {N.}~\bibnamefont
  {Varnava}}\ and\ \bibinfo {author} {\bibfnamefont {D.}~\bibnamefont
  {Vanderbilt}},\ }\href@noop {} {\bibfield  {journal} {\bibinfo  {journal}
  {Physical Review B}\ }\textbf {\bibinfo {volume} {98}},\ \bibinfo {pages}
  {245117} (\bibinfo {year} {2018})}\BibitemShut {NoStop}%
\bibitem [{\citenamefont {Wieder}\ and\ \citenamefont
  {Bernevig}(2018)}]{wieder2018axion}%
  \BibitemOpen
  \bibfield  {author} {\bibinfo {author} {\bibfnamefont {B.~J.}\ \bibnamefont
  {Wieder}}\ and\ \bibinfo {author} {\bibfnamefont {B.~A.}\ \bibnamefont
  {Bernevig}},\ }\href@noop {} {\bibfield  {journal} {\bibinfo  {journal}
  {arXiv preprint arXiv:1810.02373}\ } (\bibinfo {year} {2018})}\BibitemShut
  {NoStop}%
\bibitem [{\citenamefont {Halperin}(1987)}]{halperin1987possible}%
  \BibitemOpen
  \bibfield  {author} {\bibinfo {author} {\bibfnamefont {B.~I.}\ \bibnamefont
  {Halperin}},\ }\href@noop {} {\bibfield  {journal} {\bibinfo  {journal}
  {Japanese Journal of Applied Physics}\ }\textbf {\bibinfo {volume} {26}},\
  \bibinfo {pages} {1913} (\bibinfo {year} {1987})}\BibitemShut {NoStop}%
\bibitem [{\citenamefont {Kohmoto}\ \emph {et~al.}(1992)\citenamefont
  {Kohmoto}, \citenamefont {Halperin},\ and\ \citenamefont
  {Wu}}]{kohmoto1992diophantine}%
  \BibitemOpen
  \bibfield  {author} {\bibinfo {author} {\bibfnamefont {M.}~\bibnamefont
  {Kohmoto}}, \bibinfo {author} {\bibfnamefont {B.~I.}\ \bibnamefont
  {Halperin}}, \ and\ \bibinfo {author} {\bibfnamefont {Y.-S.}\ \bibnamefont
  {Wu}},\ }\href@noop {} {\bibfield  {journal} {\bibinfo  {journal} {Physical
  Review B}\ }\textbf {\bibinfo {volume} {45}},\ \bibinfo {pages} {13488}
  (\bibinfo {year} {1992})}\BibitemShut {NoStop}%
\bibitem [{\citenamefont {Haldane}(2004)}]{haldane2004berry}%
  \BibitemOpen
  \bibfield  {author} {\bibinfo {author} {\bibfnamefont {F.}~\bibnamefont
  {Haldane}},\ }\href@noop {} {\bibfield  {journal} {\bibinfo  {journal}
  {Physical review letters}\ }\textbf {\bibinfo {volume} {93}},\ \bibinfo
  {pages} {206602} (\bibinfo {year} {2004})}\BibitemShut {NoStop}%
\bibitem [{\citenamefont {Zhang}\ \emph {et~al.}(2016)\citenamefont {Zhang},
  \citenamefont {Hutasoit}, \citenamefont {Sun}, \citenamefont {Yan},
  \citenamefont {Xu},\ and\ \citenamefont {Liu}}]{zhang2016topological}%
  \BibitemOpen
  \bibfield  {author} {\bibinfo {author} {\bibfnamefont {R.-X.}\ \bibnamefont
  {Zhang}}, \bibinfo {author} {\bibfnamefont {J.~A.}\ \bibnamefont {Hutasoit}},
  \bibinfo {author} {\bibfnamefont {Y.}~\bibnamefont {Sun}}, \bibinfo {author}
  {\bibfnamefont {B.}~\bibnamefont {Yan}}, \bibinfo {author} {\bibfnamefont
  {C.}~\bibnamefont {Xu}}, \ and\ \bibinfo {author} {\bibfnamefont {C.-X.}\
  \bibnamefont {Liu}},\ }\href@noop {} {\bibfield  {journal} {\bibinfo
  {journal} {Physical Review B}\ }\textbf {\bibinfo {volume} {93}},\ \bibinfo
  {pages} {041108} (\bibinfo {year} {2016})}\BibitemShut {NoStop}%
\bibitem [{\citenamefont {Chaikin}\ \emph {et~al.}(1995)\citenamefont
  {Chaikin}, \citenamefont {Lubensky},\ and\ \citenamefont
  {Witten}}]{chaikin1995principles}%
  \BibitemOpen
  \bibfield  {author} {\bibinfo {author} {\bibfnamefont {P.~M.}\ \bibnamefont
  {Chaikin}}, \bibinfo {author} {\bibfnamefont {T.~C.}\ \bibnamefont
  {Lubensky}}, \ and\ \bibinfo {author} {\bibfnamefont {T.~A.}\ \bibnamefont
  {Witten}},\ }\href@noop {} {\emph {\bibinfo {title} {Principles of condensed
  matter physics}}},\ Vol.~\bibinfo {volume} {10}\ (\bibinfo  {publisher}
  {Cambridge university press Cambridge},\ \bibinfo {year} {1995})\BibitemShut
  {NoStop}%
\bibitem [{\citenamefont {Ran}\ \emph {et~al.}(2009)\citenamefont {Ran},
  \citenamefont {Zhang},\ and\ \citenamefont {Vishwanath}}]{ran2009one}%
  \BibitemOpen
  \bibfield  {author} {\bibinfo {author} {\bibfnamefont {Y.}~\bibnamefont
  {Ran}}, \bibinfo {author} {\bibfnamefont {Y.}~\bibnamefont {Zhang}}, \ and\
  \bibinfo {author} {\bibfnamefont {A.}~\bibnamefont {Vishwanath}},\
  }\href@noop {} {\bibfield  {journal} {\bibinfo  {journal} {Nature Physics}\
  }\textbf {\bibinfo {volume} {5}},\ \bibinfo {pages} {298} (\bibinfo {year}
  {2009})}\BibitemShut {NoStop}%
\bibitem [{\citenamefont {Teo}\ and\ \citenamefont
  {Kane}(2010)}]{teo2010topological}%
  \BibitemOpen
  \bibfield  {author} {\bibinfo {author} {\bibfnamefont {J.~C.}\ \bibnamefont
  {Teo}}\ and\ \bibinfo {author} {\bibfnamefont {C.~L.}\ \bibnamefont {Kane}},\
  }\href@noop {} {\bibfield  {journal} {\bibinfo  {journal} {Physical Review
  B}\ }\textbf {\bibinfo {volume} {82}},\ \bibinfo {pages} {115120} (\bibinfo
  {year} {2010})}\BibitemShut {NoStop}%
\bibitem [{\citenamefont {Juri{\v{c}}i{\'c}}\ \emph {et~al.}(2012)\citenamefont
  {Juri{\v{c}}i{\'c}}, \citenamefont {Mesaros}, \citenamefont {Slager},\ and\
  \citenamefont {Zaanen}}]{jurivcic2012universal}%
  \BibitemOpen
  \bibfield  {author} {\bibinfo {author} {\bibfnamefont {V.}~\bibnamefont
  {Juri{\v{c}}i{\'c}}}, \bibinfo {author} {\bibfnamefont {A.}~\bibnamefont
  {Mesaros}}, \bibinfo {author} {\bibfnamefont {R.-J.}\ \bibnamefont {Slager}},
  \ and\ \bibinfo {author} {\bibfnamefont {J.}~\bibnamefont {Zaanen}},\
  }\href@noop {} {\bibfield  {journal} {\bibinfo  {journal} {Physical review
  letters}\ }\textbf {\bibinfo {volume} {108}},\ \bibinfo {pages} {106403}
  (\bibinfo {year} {2012})}\BibitemShut {NoStop}%
\bibitem [{\citenamefont {Barkeshli}\ and\ \citenamefont
  {Qi}(2012)}]{barkeshli2012topological}%
  \BibitemOpen
  \bibfield  {author} {\bibinfo {author} {\bibfnamefont {M.}~\bibnamefont
  {Barkeshli}}\ and\ \bibinfo {author} {\bibfnamefont {X.-L.}\ \bibnamefont
  {Qi}},\ }\href@noop {} {\bibfield  {journal} {\bibinfo  {journal} {Physical
  Review X}\ }\textbf {\bibinfo {volume} {2}},\ \bibinfo {pages} {031013}
  (\bibinfo {year} {2012})}\BibitemShut {NoStop}%
\bibitem [{\citenamefont {Asahi}\ and\ \citenamefont
  {Nagaosa}(2012)}]{asahi2012topological}%
  \BibitemOpen
  \bibfield  {author} {\bibinfo {author} {\bibfnamefont {D.}~\bibnamefont
  {Asahi}}\ and\ \bibinfo {author} {\bibfnamefont {N.}~\bibnamefont
  {Nagaosa}},\ }\href@noop {} {\bibfield  {journal} {\bibinfo  {journal}
  {Physical Review B}\ }\textbf {\bibinfo {volume} {86}},\ \bibinfo {pages}
  {100504} (\bibinfo {year} {2012})}\BibitemShut {NoStop}%
\bibitem [{\citenamefont {Chung}\ \emph {et~al.}(2016)\citenamefont {Chung},
  \citenamefont {Chan},\ and\ \citenamefont {Yao}}]{chung2016dislocation}%
  \BibitemOpen
  \bibfield  {author} {\bibinfo {author} {\bibfnamefont {S.~B.}\ \bibnamefont
  {Chung}}, \bibinfo {author} {\bibfnamefont {C.}~\bibnamefont {Chan}}, \ and\
  \bibinfo {author} {\bibfnamefont {H.}~\bibnamefont {Yao}},\ }\href@noop {}
  {\bibfield  {journal} {\bibinfo  {journal} {Scientific reports}\ }\textbf
  {\bibinfo {volume} {6}},\ \bibinfo {pages} {1} (\bibinfo {year}
  {2016})}\BibitemShut {NoStop}%
\bibitem [{\citenamefont {Teo}\ and\ \citenamefont
  {Hughes}(2017)}]{teo2017topological}%
  \BibitemOpen
  \bibfield  {author} {\bibinfo {author} {\bibfnamefont {J.~C.}\ \bibnamefont
  {Teo}}\ and\ \bibinfo {author} {\bibfnamefont {T.~L.}\ \bibnamefont
  {Hughes}},\ }\href@noop {} {\bibfield  {journal} {\bibinfo  {journal} {Annual
  Review of Condensed Matter Physics}\ }\textbf {\bibinfo {volume} {8}},\
  \bibinfo {pages} {211} (\bibinfo {year} {2017})}\BibitemShut {NoStop}%
\bibitem [{\citenamefont {Ramamurthy}\ \emph {et~al.}(2017)\citenamefont
  {Ramamurthy}, \citenamefont {Wang},\ and\ \citenamefont
  {Hughes}}]{ramamurthy2017electromagnetic}%
  \BibitemOpen
  \bibfield  {author} {\bibinfo {author} {\bibfnamefont {S.~T.}\ \bibnamefont
  {Ramamurthy}}, \bibinfo {author} {\bibfnamefont {Y.}~\bibnamefont {Wang}}, \
  and\ \bibinfo {author} {\bibfnamefont {T.~L.}\ \bibnamefont {Hughes}},\
  }\href@noop {} {\bibfield  {journal} {\bibinfo  {journal} {Physical review
  letters}\ }\textbf {\bibinfo {volume} {118}},\ \bibinfo {pages} {146602}
  (\bibinfo {year} {2017})}\BibitemShut {NoStop}%
\bibitem [{\citenamefont {Roy}\ and\ \citenamefont
  {Juri{\v{c}}i{\'c}}(2021)}]{roy2021dislocation}%
  \BibitemOpen
  \bibfield  {author} {\bibinfo {author} {\bibfnamefont {B.}~\bibnamefont
  {Roy}}\ and\ \bibinfo {author} {\bibfnamefont {V.}~\bibnamefont
  {Juri{\v{c}}i{\'c}}},\ }\href@noop {} {\bibfield  {journal} {\bibinfo
  {journal} {Physical Review Research}\ }\textbf {\bibinfo {volume} {3}},\
  \bibinfo {pages} {033107} (\bibinfo {year} {2021})}\BibitemShut {NoStop}%
\bibitem [{\citenamefont {Gioia}\ \emph {et~al.}(2021)\citenamefont {Gioia},
  \citenamefont {Wang},\ and\ \citenamefont {Burkov}}]{gioia2021unquantized}%
  \BibitemOpen
  \bibfield  {author} {\bibinfo {author} {\bibfnamefont {L.}~\bibnamefont
  {Gioia}}, \bibinfo {author} {\bibfnamefont {C.}~\bibnamefont {Wang}}, \ and\
  \bibinfo {author} {\bibfnamefont {A.}~\bibnamefont {Burkov}},\ }\href@noop {}
  {\bibfield  {journal} {\bibinfo  {journal} {Physical Review Research}\
  }\textbf {\bibinfo {volume} {3}},\ \bibinfo {pages} {043067} (\bibinfo {year}
  {2021})}\BibitemShut {NoStop}%
\bibitem [{\citenamefont {Teo}\ and\ \citenamefont
  {Hughes}(2013)}]{teo2013existence}%
  \BibitemOpen
  \bibfield  {author} {\bibinfo {author} {\bibfnamefont {J.~C.}\ \bibnamefont
  {Teo}}\ and\ \bibinfo {author} {\bibfnamefont {T.~L.}\ \bibnamefont
  {Hughes}},\ }\href@noop {} {\bibfield  {journal} {\bibinfo  {journal}
  {Physical review letters}\ }\textbf {\bibinfo {volume} {111}},\ \bibinfo
  {pages} {047006} (\bibinfo {year} {2013})}\BibitemShut {NoStop}%
\bibitem [{\citenamefont {Gopalakrishnan}\ \emph {et~al.}(2013)\citenamefont
  {Gopalakrishnan}, \citenamefont {Teo},\ and\ \citenamefont
  {Hughes}}]{gopalakrishnan2013disclination}%
  \BibitemOpen
  \bibfield  {author} {\bibinfo {author} {\bibfnamefont {S.}~\bibnamefont
  {Gopalakrishnan}}, \bibinfo {author} {\bibfnamefont {J.~C.}\ \bibnamefont
  {Teo}}, \ and\ \bibinfo {author} {\bibfnamefont {T.~L.}\ \bibnamefont
  {Hughes}},\ }\href@noop {} {\bibfield  {journal} {\bibinfo  {journal}
  {Physical review letters}\ }\textbf {\bibinfo {volume} {111}},\ \bibinfo
  {pages} {025304} (\bibinfo {year} {2013})}\BibitemShut {NoStop}%
\bibitem [{\citenamefont {Benalcazar}\ \emph {et~al.}(2014)\citenamefont
  {Benalcazar}, \citenamefont {Teo},\ and\ \citenamefont
  {Hughes}}]{benalcazar2014classification}%
  \BibitemOpen
  \bibfield  {author} {\bibinfo {author} {\bibfnamefont {W.~A.}\ \bibnamefont
  {Benalcazar}}, \bibinfo {author} {\bibfnamefont {J.~C.}\ \bibnamefont {Teo}},
  \ and\ \bibinfo {author} {\bibfnamefont {T.~L.}\ \bibnamefont {Hughes}},\
  }\href@noop {} {\bibfield  {journal} {\bibinfo  {journal} {Physical Review
  B}\ }\textbf {\bibinfo {volume} {89}},\ \bibinfo {pages} {224503} (\bibinfo
  {year} {2014})}\BibitemShut {NoStop}%
\bibitem [{\citenamefont {Li}\ \emph {et~al.}(2020)\citenamefont {Li},
  \citenamefont {Zhu}, \citenamefont {Benalcazar},\ and\ \citenamefont
  {Hughes}}]{li2020fractional}%
  \BibitemOpen
  \bibfield  {author} {\bibinfo {author} {\bibfnamefont {T.}~\bibnamefont
  {Li}}, \bibinfo {author} {\bibfnamefont {P.}~\bibnamefont {Zhu}}, \bibinfo
  {author} {\bibfnamefont {W.~A.}\ \bibnamefont {Benalcazar}}, \ and\ \bibinfo
  {author} {\bibfnamefont {T.~L.}\ \bibnamefont {Hughes}},\ }\href@noop {}
  {\bibfield  {journal} {\bibinfo  {journal} {Physical Review B}\ }\textbf
  {\bibinfo {volume} {101}},\ \bibinfo {pages} {115115} (\bibinfo {year}
  {2020})}\BibitemShut {NoStop}%
\bibitem [{\citenamefont {Manjunath}\ and\ \citenamefont
  {Barkeshli}(2021)}]{manjunath2021crystalline}%
  \BibitemOpen
  \bibfield  {author} {\bibinfo {author} {\bibfnamefont {N.}~\bibnamefont
  {Manjunath}}\ and\ \bibinfo {author} {\bibfnamefont {M.}~\bibnamefont
  {Barkeshli}},\ }\href@noop {} {\bibfield  {journal} {\bibinfo  {journal}
  {Physical Review Research}\ }\textbf {\bibinfo {volume} {3}},\ \bibinfo
  {pages} {013040} (\bibinfo {year} {2021})}\BibitemShut {NoStop}%
\bibitem [{\citenamefont {Schindler}\ \emph {et~al.}(2022)\citenamefont
  {Schindler}, \citenamefont {Tsirkin}, \citenamefont {Neupert}, \citenamefont
  {Bernevig},\ and\ \citenamefont {Wieder}}]{schindler2022topological}%
  \BibitemOpen
  \bibfield  {author} {\bibinfo {author} {\bibfnamefont {F.}~\bibnamefont
  {Schindler}}, \bibinfo {author} {\bibfnamefont {S.~S.}\ \bibnamefont
  {Tsirkin}}, \bibinfo {author} {\bibfnamefont {T.}~\bibnamefont {Neupert}},
  \bibinfo {author} {\bibfnamefont {B.~A.}\ \bibnamefont {Bernevig}}, \ and\
  \bibinfo {author} {\bibfnamefont {B.~J.}\ \bibnamefont {Wieder}},\
  }\href@noop {} {\bibfield  {journal} {\bibinfo  {journal} {arXiv preprint
  arXiv:2207.10112}\ } (\bibinfo {year} {2022})}\BibitemShut {NoStop}%
\bibitem [{\citenamefont {Hirsbrunner}\ \emph
  {et~al.}(2023{\natexlab{a}})\citenamefont {Hirsbrunner}, \citenamefont
  {Gray},\ and\ \citenamefont {Hughes}}]{hirsbrunner2023crystalline}%
  \BibitemOpen
  \bibfield  {author} {\bibinfo {author} {\bibfnamefont {M.~R.}\ \bibnamefont
  {Hirsbrunner}}, \bibinfo {author} {\bibfnamefont {A.~D.}\ \bibnamefont
  {Gray}}, \ and\ \bibinfo {author} {\bibfnamefont {T.~L.}\ \bibnamefont
  {Hughes}},\ }\href@noop {} {\bibfield  {journal} {\bibinfo  {journal} {arXiv
  preprint arXiv:2308.05796}\ } (\bibinfo {year}
  {2023}{\natexlab{a}})}\BibitemShut {NoStop}%
\bibitem [{\citenamefont {Hirsbrunner}\ \emph
  {et~al.}(2023{\natexlab{b}})\citenamefont {Hirsbrunner}, \citenamefont
  {Dubinkin}, \citenamefont {Burnell},\ and\ \citenamefont
  {Hughes}}]{hirsbrunner2023anomalous}%
  \BibitemOpen
  \bibfield  {author} {\bibinfo {author} {\bibfnamefont {M.~R.}\ \bibnamefont
  {Hirsbrunner}}, \bibinfo {author} {\bibfnamefont {O.}~\bibnamefont
  {Dubinkin}}, \bibinfo {author} {\bibfnamefont {F.~J.}\ \bibnamefont
  {Burnell}}, \ and\ \bibinfo {author} {\bibfnamefont {T.~L.}\ \bibnamefont
  {Hughes}},\ }\href@noop {} {\bibfield  {journal} {\bibinfo  {journal} {arXiv
  preprint arXiv:2309.10840}\ } (\bibinfo {year}
  {2023}{\natexlab{b}})}\BibitemShut {NoStop}%
\bibitem [{\citenamefont {Wen}\ and\ \citenamefont {Zee}(1992)}]{wen1992shift}%
  \BibitemOpen
  \bibfield  {author} {\bibinfo {author} {\bibfnamefont {X.-g.}\ \bibnamefont
  {Wen}}\ and\ \bibinfo {author} {\bibfnamefont {A.}~\bibnamefont {Zee}},\
  }\href@noop {} {\bibfield  {journal} {\bibinfo  {journal} {Physical review
  letters}\ }\textbf {\bibinfo {volume} {69}},\ \bibinfo {pages} {953}
  (\bibinfo {year} {1992})}\BibitemShut {NoStop}%
\bibitem [{\citenamefont {R{\"u}egg}\ and\ \citenamefont
  {Lin}(2013)}]{ruegg2013bound}%
  \BibitemOpen
  \bibfield  {author} {\bibinfo {author} {\bibfnamefont {A.}~\bibnamefont
  {R{\"u}egg}}\ and\ \bibinfo {author} {\bibfnamefont {C.}~\bibnamefont
  {Lin}},\ }\href@noop {} {\bibfield  {journal} {\bibinfo  {journal} {Physical
  Review Letters}\ }\textbf {\bibinfo {volume} {110}},\ \bibinfo {pages}
  {046401} (\bibinfo {year} {2013})}\BibitemShut {NoStop}%
\bibitem [{\citenamefont {R{\"u}egg}\ \emph {et~al.}(2013)\citenamefont
  {R{\"u}egg}, \citenamefont {Coh},\ and\ \citenamefont
  {Moore}}]{ruegg2013corner}%
  \BibitemOpen
  \bibfield  {author} {\bibinfo {author} {\bibfnamefont {A.}~\bibnamefont
  {R{\"u}egg}}, \bibinfo {author} {\bibfnamefont {S.}~\bibnamefont {Coh}}, \
  and\ \bibinfo {author} {\bibfnamefont {J.~E.}\ \bibnamefont {Moore}},\
  }\href@noop {} {\bibfield  {journal} {\bibinfo  {journal} {Physical Review
  B}\ }\textbf {\bibinfo {volume} {88}},\ \bibinfo {pages} {155127} (\bibinfo
  {year} {2013})}\BibitemShut {NoStop}%
\bibitem [{\citenamefont {Liu}\ \emph {et~al.}(2019)\citenamefont {Liu},
  \citenamefont {Vishwanath},\ and\ \citenamefont {Khalaf}}]{liu2019shift}%
  \BibitemOpen
  \bibfield  {author} {\bibinfo {author} {\bibfnamefont {S.}~\bibnamefont
  {Liu}}, \bibinfo {author} {\bibfnamefont {A.}~\bibnamefont {Vishwanath}}, \
  and\ \bibinfo {author} {\bibfnamefont {E.}~\bibnamefont {Khalaf}},\
  }\href@noop {} {\bibfield  {journal} {\bibinfo  {journal} {Physical Review
  X}\ }\textbf {\bibinfo {volume} {9}},\ \bibinfo {pages} {031003} (\bibinfo
  {year} {2019})}\BibitemShut {NoStop}%
\bibitem [{\citenamefont {Han}\ \emph {et~al.}(2019)\citenamefont {Han},
  \citenamefont {Wang},\ and\ \citenamefont {Ye}}]{han2019generalized}%
  \BibitemOpen
  \bibfield  {author} {\bibinfo {author} {\bibfnamefont {B.}~\bibnamefont
  {Han}}, \bibinfo {author} {\bibfnamefont {H.}~\bibnamefont {Wang}}, \ and\
  \bibinfo {author} {\bibfnamefont {P.}~\bibnamefont {Ye}},\ }\href@noop {}
  {\bibfield  {journal} {\bibinfo  {journal} {Physical Review B}\ }\textbf
  {\bibinfo {volume} {99}},\ \bibinfo {pages} {205120} (\bibinfo {year}
  {2019})}\BibitemShut {NoStop}%
\bibitem [{\citenamefont {Manjunath}\ and\ \citenamefont
  {Barkeshli}(2020)}]{manjunath2020classification}%
  \BibitemOpen
  \bibfield  {author} {\bibinfo {author} {\bibfnamefont {N.}~\bibnamefont
  {Manjunath}}\ and\ \bibinfo {author} {\bibfnamefont {M.}~\bibnamefont
  {Barkeshli}},\ }\href@noop {} {\bibfield  {journal} {\bibinfo  {journal}
  {arXiv preprint arXiv:2012.11603}\ } (\bibinfo {year} {2020})}\BibitemShut
  {NoStop}%
\bibitem [{\citenamefont {May-Mann}\ and\ \citenamefont
  {Hughes}(2022)}]{may2022crystalline}%
  \BibitemOpen
  \bibfield  {author} {\bibinfo {author} {\bibfnamefont {J.}~\bibnamefont
  {May-Mann}}\ and\ \bibinfo {author} {\bibfnamefont {T.~L.}\ \bibnamefont
  {Hughes}},\ }\href@noop {} {\bibfield  {journal} {\bibinfo  {journal}
  {Physical Review B}\ }\textbf {\bibinfo {volume} {106}},\ \bibinfo {pages}
  {L241113} (\bibinfo {year} {2022})}\BibitemShut {NoStop}%
\bibitem [{\citenamefont {Peterson}\ \emph {et~al.}(2021)\citenamefont
  {Peterson}, \citenamefont {Li}, \citenamefont {Jiang}, \citenamefont
  {Hughes},\ and\ \citenamefont {Bahl}}]{peterson2021trapped}%
  \BibitemOpen
  \bibfield  {author} {\bibinfo {author} {\bibfnamefont {C.~W.}\ \bibnamefont
  {Peterson}}, \bibinfo {author} {\bibfnamefont {T.}~\bibnamefont {Li}},
  \bibinfo {author} {\bibfnamefont {W.}~\bibnamefont {Jiang}}, \bibinfo
  {author} {\bibfnamefont {T.~L.}\ \bibnamefont {Hughes}}, \ and\ \bibinfo
  {author} {\bibfnamefont {G.}~\bibnamefont {Bahl}},\ }\href@noop {} {\bibfield
   {journal} {\bibinfo  {journal} {Nature}\ }\textbf {\bibinfo {volume}
  {589}},\ \bibinfo {pages} {376} (\bibinfo {year} {2021})}\BibitemShut
  {NoStop}%
\bibitem [{\citenamefont {Zhang}\ \emph {et~al.}(2022)\citenamefont {Zhang},
  \citenamefont {Manjunath}, \citenamefont {Nambiar},\ and\ \citenamefont
  {Barkeshli}}]{zhang2022fractional}%
  \BibitemOpen
  \bibfield  {author} {\bibinfo {author} {\bibfnamefont {Y.}~\bibnamefont
  {Zhang}}, \bibinfo {author} {\bibfnamefont {N.}~\bibnamefont {Manjunath}},
  \bibinfo {author} {\bibfnamefont {G.}~\bibnamefont {Nambiar}}, \ and\
  \bibinfo {author} {\bibfnamefont {M.}~\bibnamefont {Barkeshli}},\ }\href@noop
  {} {\bibfield  {journal} {\bibinfo  {journal} {arXiv preprint
  arXiv:2204.05320}\ } (\bibinfo {year} {2022})}\BibitemShut {NoStop}%
\bibitem [{\citenamefont {May-Mann}\ \emph {et~al.}(2022)\citenamefont
  {May-Mann}, \citenamefont {Hirsbrunner}, \citenamefont {Cao},\ and\
  \citenamefont {Hughes}}]{may2022topological}%
  \BibitemOpen
  \bibfield  {author} {\bibinfo {author} {\bibfnamefont {J.}~\bibnamefont
  {May-Mann}}, \bibinfo {author} {\bibfnamefont {M.~R.}\ \bibnamefont
  {Hirsbrunner}}, \bibinfo {author} {\bibfnamefont {X.}~\bibnamefont {Cao}}, \
  and\ \bibinfo {author} {\bibfnamefont {T.~L.}\ \bibnamefont {Hughes}},\
  }\href@noop {} {\bibfield  {journal} {\bibinfo  {journal} {arXiv preprint
  arXiv:2209.00026}\ } (\bibinfo {year} {2022})}\BibitemShut {NoStop}%
\bibitem [{\citenamefont {Benalcazar}\ \emph {et~al.}(2019)\citenamefont
  {Benalcazar}, \citenamefont {Li},\ and\ \citenamefont
  {Hughes}}]{benalcazar2019quantization}%
  \BibitemOpen
  \bibfield  {author} {\bibinfo {author} {\bibfnamefont {W.~A.}\ \bibnamefont
  {Benalcazar}}, \bibinfo {author} {\bibfnamefont {T.}~\bibnamefont {Li}}, \
  and\ \bibinfo {author} {\bibfnamefont {T.~L.}\ \bibnamefont {Hughes}},\
  }\href@noop {} {\bibfield  {journal} {\bibinfo  {journal} {Physical Review
  B}\ }\textbf {\bibinfo {volume} {99}},\ \bibinfo {pages} {245151} (\bibinfo
  {year} {2019})}\BibitemShut {NoStop}%
\bibitem [{\citenamefont {Khalaf}\ \emph {et~al.}(2021)\citenamefont {Khalaf},
  \citenamefont {Benalcazar}, \citenamefont {Hughes},\ and\ \citenamefont
  {Queiroz}}]{khalaf2021boundary}%
  \BibitemOpen
  \bibfield  {author} {\bibinfo {author} {\bibfnamefont {E.}~\bibnamefont
  {Khalaf}}, \bibinfo {author} {\bibfnamefont {W.~A.}\ \bibnamefont
  {Benalcazar}}, \bibinfo {author} {\bibfnamefont {T.~L.}\ \bibnamefont
  {Hughes}}, \ and\ \bibinfo {author} {\bibfnamefont {R.}~\bibnamefont
  {Queiroz}},\ }\href@noop {} {\bibfield  {journal} {\bibinfo  {journal}
  {Physical Review Research}\ }\textbf {\bibinfo {volume} {3}},\ \bibinfo
  {pages} {013239} (\bibinfo {year} {2021})}\BibitemShut {NoStop}%
\bibitem [{\citenamefont {Yang}\ and\ \citenamefont
  {Nagaosa}(2014)}]{Nagaosa2014}%
  \BibitemOpen
  \bibfield  {author} {\bibinfo {author} {\bibfnamefont {B.-J.}\ \bibnamefont
  {Yang}}\ and\ \bibinfo {author} {\bibfnamefont {N.}~\bibnamefont {Nagaosa}},\
  }\href {\doibase 10.1038/ncomms5898} {\bibfield  {journal} {\bibinfo
  {journal} {Nature Communications}\ }\textbf {\bibinfo {volume} {5}},\
  \bibinfo {pages} {4898} (\bibinfo {year} {2014})}\BibitemShut {NoStop}%
\bibitem [{\citenamefont {Ramamurthy}\ and\ \citenamefont
  {Hughes}(2015)}]{ramamurthy2015patterns}%
  \BibitemOpen
  \bibfield  {author} {\bibinfo {author} {\bibfnamefont {S.~T.}\ \bibnamefont
  {Ramamurthy}}\ and\ \bibinfo {author} {\bibfnamefont {T.~L.}\ \bibnamefont
  {Hughes}},\ }\href@noop {} {\bibfield  {journal} {\bibinfo  {journal}
  {Physical Review B}\ }\textbf {\bibinfo {volume} {92}},\ \bibinfo {pages}
  {085105} (\bibinfo {year} {2015})}\BibitemShut {NoStop}%
\bibitem [{\citenamefont {Gioia}\ and\ \citenamefont
  {Wang}(2022)}]{PhysRevX.12.031007}%
  \BibitemOpen
  \bibfield  {author} {\bibinfo {author} {\bibfnamefont {L.}~\bibnamefont
  {Gioia}}\ and\ \bibinfo {author} {\bibfnamefont {C.}~\bibnamefont {Wang}},\
  }\href {\doibase 10.1103/PhysRevX.12.031007} {\bibfield  {journal} {\bibinfo
  {journal} {Phys. Rev. X}\ }\textbf {\bibinfo {volume} {12}},\ \bibinfo
  {pages} {031007} (\bibinfo {year} {2022})}\BibitemShut {NoStop}%
\bibitem [{\citenamefont {Dubinkin}\ \emph {et~al.}(2021)\citenamefont
  {Dubinkin}, \citenamefont {Burnell},\ and\ \citenamefont
  {Hughes}}]{dubinkin2021higher}%
  \BibitemOpen
  \bibfield  {author} {\bibinfo {author} {\bibfnamefont {O.}~\bibnamefont
  {Dubinkin}}, \bibinfo {author} {\bibfnamefont {F.}~\bibnamefont {Burnell}}, \
  and\ \bibinfo {author} {\bibfnamefont {T.~L.}\ \bibnamefont {Hughes}},\
  }\href@noop {} {\bibfield  {journal} {\bibinfo  {journal} {arXiv preprint
  arXiv:2102.08959}\ } (\bibinfo {year} {2021})}\BibitemShut {NoStop}%
\bibitem [{\citenamefont {Kane}\ and\ \citenamefont
  {Mele}(2005)}]{kane2005quantum}%
  \BibitemOpen
  \bibfield  {author} {\bibinfo {author} {\bibfnamefont {C.~L.}\ \bibnamefont
  {Kane}}\ and\ \bibinfo {author} {\bibfnamefont {E.~J.}\ \bibnamefont
  {Mele}},\ }\href@noop {} {\bibfield  {journal} {\bibinfo  {journal} {Physical
  review letters}\ }\textbf {\bibinfo {volume} {95}},\ \bibinfo {pages}
  {226801} (\bibinfo {year} {2005})}\BibitemShut {NoStop}%
\bibitem [{\citenamefont {Bernevig}\ \emph {et~al.}(2006)\citenamefont
  {Bernevig}, \citenamefont {Hughes},\ and\ \citenamefont
  {Zhang}}]{bernevig2006quantum}%
  \BibitemOpen
  \bibfield  {author} {\bibinfo {author} {\bibfnamefont {B.~A.}\ \bibnamefont
  {Bernevig}}, \bibinfo {author} {\bibfnamefont {T.~L.}\ \bibnamefont
  {Hughes}}, \ and\ \bibinfo {author} {\bibfnamefont {S.-C.}\ \bibnamefont
  {Zhang}},\ }\href@noop {} {\bibfield  {journal} {\bibinfo  {journal}
  {science}\ }\textbf {\bibinfo {volume} {314}},\ \bibinfo {pages} {1757}
  (\bibinfo {year} {2006})}\BibitemShut {NoStop}%
\bibitem [{\citenamefont {Fu}\ and\ \citenamefont
  {Kane}(2007)}]{fu2007topologicalIn}%
  \BibitemOpen
  \bibfield  {author} {\bibinfo {author} {\bibfnamefont {L.}~\bibnamefont
  {Fu}}\ and\ \bibinfo {author} {\bibfnamefont {C.~L.}\ \bibnamefont {Kane}},\
  }\href@noop {} {\bibfield  {journal} {\bibinfo  {journal} {Physical Review
  B}\ }\textbf {\bibinfo {volume} {76}},\ \bibinfo {pages} {045302} (\bibinfo
  {year} {2007})}\BibitemShut {NoStop}%
\bibitem [{\citenamefont {Lieb}\ \emph {et~al.}(1961)\citenamefont {Lieb},
  \citenamefont {Schultz},\ and\ \citenamefont {Mattis}}]{lieb1961two}%
  \BibitemOpen
  \bibfield  {author} {\bibinfo {author} {\bibfnamefont {E.}~\bibnamefont
  {Lieb}}, \bibinfo {author} {\bibfnamefont {T.}~\bibnamefont {Schultz}}, \
  and\ \bibinfo {author} {\bibfnamefont {D.}~\bibnamefont {Mattis}},\
  }\href@noop {} {\bibfield  {journal} {\bibinfo  {journal} {Annals of
  Physics}\ }\textbf {\bibinfo {volume} {16}},\ \bibinfo {pages} {407}
  (\bibinfo {year} {1961})}\BibitemShut {NoStop}%
\bibitem [{\citenamefont {Coh}\ and\ \citenamefont
  {Vanderbilt}(2009)}]{coh2009}%
  \BibitemOpen
  \bibfield  {author} {\bibinfo {author} {\bibfnamefont {S.}~\bibnamefont
  {Coh}}\ and\ \bibinfo {author} {\bibfnamefont {D.}~\bibnamefont
  {Vanderbilt}},\ }\href@noop {} {\bibfield  {journal} {\bibinfo  {journal}
  {Phys. Rev. Lett.}\ }\textbf {\bibinfo {volume} {102}},\ \bibinfo {pages}
  {107603} (\bibinfo {year} {2009})}\BibitemShut {NoStop}%
\bibitem [{\citenamefont {Vaidya}\ \emph {et~al.}(2023)\citenamefont {Vaidya},
  \citenamefont {Rechtsman},\ and\ \citenamefont {Benalcazar}}]{vaidya2023}%
  \BibitemOpen
  \bibfield  {author} {\bibinfo {author} {\bibfnamefont {S.}~\bibnamefont
  {Vaidya}}, \bibinfo {author} {\bibfnamefont {M.~C.}\ \bibnamefont
  {Rechtsman}}, \ and\ \bibinfo {author} {\bibfnamefont {W.~A.}\ \bibnamefont
  {Benalcazar}},\ }\href@noop {} {\bibfield  {journal} {\bibinfo  {journal}
  {arXiv preprint arXiv:2304.13118}\ } (\bibinfo {year} {2023})}\BibitemShut
  {NoStop}%
\bibitem [{\citenamefont {Su}\ \emph {et~al.}(1979)\citenamefont {Su},
  \citenamefont {Schrieffer},\ and\ \citenamefont {Heeger}}]{su1979solitons}%
  \BibitemOpen
  \bibfield  {author} {\bibinfo {author} {\bibfnamefont {W.}~\bibnamefont
  {Su}}, \bibinfo {author} {\bibfnamefont {J.}~\bibnamefont {Schrieffer}}, \
  and\ \bibinfo {author} {\bibfnamefont {A.~J.}\ \bibnamefont {Heeger}},\
  }\href@noop {} {\bibfield  {journal} {\bibinfo  {journal} {Physical review
  letters}\ }\textbf {\bibinfo {volume} {42}},\ \bibinfo {pages} {1698}
  (\bibinfo {year} {1979})}\BibitemShut {NoStop}%
\bibitem [{\citenamefont {Su}\ \emph {et~al.}(1983)\citenamefont {Su},
  \citenamefont {Schrieffer},\ and\ \citenamefont {Heeger}}]{su1983erratum}%
  \BibitemOpen
  \bibfield  {author} {\bibinfo {author} {\bibfnamefont {W.}~\bibnamefont
  {Su}}, \bibinfo {author} {\bibfnamefont {J.}~\bibnamefont {Schrieffer}}, \
  and\ \bibinfo {author} {\bibfnamefont {A.}~\bibnamefont {Heeger}},\
  }\href@noop {} {\bibfield  {journal} {\bibinfo  {journal} {Physical Review
  B}\ }\textbf {\bibinfo {volume} {28}},\ \bibinfo {pages} {1138} (\bibinfo
  {year} {1983})}\BibitemShut {NoStop}%
\bibitem [{\citenamefont {Fang}\ and\ \citenamefont
  {Cano}(2021)}]{fang2021filling}%
  \BibitemOpen
  \bibfield  {author} {\bibinfo {author} {\bibfnamefont {Y.}~\bibnamefont
  {Fang}}\ and\ \bibinfo {author} {\bibfnamefont {J.}~\bibnamefont {Cano}},\
  }\href@noop {} {\bibfield  {journal} {\bibinfo  {journal} {Physical Review
  B}\ }\textbf {\bibinfo {volume} {103}},\ \bibinfo {pages} {165109} (\bibinfo
  {year} {2021})}\BibitemShut {NoStop}%
\bibitem [{\citenamefont {Rao}\ and\ \citenamefont
  {Bradlyn}(2023)}]{rao2023effective}%
  \BibitemOpen
  \bibfield  {author} {\bibinfo {author} {\bibfnamefont {P.}~\bibnamefont
  {Rao}}\ and\ \bibinfo {author} {\bibfnamefont {B.}~\bibnamefont {Bradlyn}},\
  }\href@noop {} {\bibfield  {journal} {\bibinfo  {journal} {Physical Review
  B}\ }\textbf {\bibinfo {volume} {107}},\ \bibinfo {pages} {195153} (\bibinfo
  {year} {2023})}\BibitemShut {NoStop}%
\bibitem [{\citenamefont {Zhang}\ \emph {et~al.}(2020)\citenamefont {Zhang},
  \citenamefont {Gu}, \citenamefont {Sun}, \citenamefont {Luo}, \citenamefont
  {Liu}, \citenamefont {Chen}, \citenamefont {Shao}, \citenamefont {Zhang},
  \citenamefont {Li}, \citenamefont {Sun} \emph {et~al.}}]{zhang2020eightfold}%
  \BibitemOpen
  \bibfield  {author} {\bibinfo {author} {\bibfnamefont {X.}~\bibnamefont
  {Zhang}}, \bibinfo {author} {\bibfnamefont {Q.}~\bibnamefont {Gu}}, \bibinfo
  {author} {\bibfnamefont {H.}~\bibnamefont {Sun}}, \bibinfo {author}
  {\bibfnamefont {T.}~\bibnamefont {Luo}}, \bibinfo {author} {\bibfnamefont
  {Y.}~\bibnamefont {Liu}}, \bibinfo {author} {\bibfnamefont {Y.}~\bibnamefont
  {Chen}}, \bibinfo {author} {\bibfnamefont {Z.}~\bibnamefont {Shao}}, \bibinfo
  {author} {\bibfnamefont {Z.}~\bibnamefont {Zhang}}, \bibinfo {author}
  {\bibfnamefont {S.}~\bibnamefont {Li}}, \bibinfo {author} {\bibfnamefont
  {Y.}~\bibnamefont {Sun}},  \emph {et~al.},\ }\href@noop {} {\bibfield
  {journal} {\bibinfo  {journal} {Physical Review B}\ }\textbf {\bibinfo
  {volume} {102}},\ \bibinfo {pages} {035125} (\bibinfo {year}
  {2020})}\BibitemShut {NoStop}%
\bibitem [{\citenamefont {Bradlyn}\ \emph {et~al.}(2016)\citenamefont
  {Bradlyn}, \citenamefont {Cano}, \citenamefont {Wang}, \citenamefont
  {Vergniory}, \citenamefont {Felser}, \citenamefont {Cava},\ and\
  \citenamefont {Bernevig}}]{bradlyn2016beyond}%
  \BibitemOpen
  \bibfield  {author} {\bibinfo {author} {\bibfnamefont {B.}~\bibnamefont
  {Bradlyn}}, \bibinfo {author} {\bibfnamefont {J.}~\bibnamefont {Cano}},
  \bibinfo {author} {\bibfnamefont {Z.}~\bibnamefont {Wang}}, \bibinfo {author}
  {\bibfnamefont {M.}~\bibnamefont {Vergniory}}, \bibinfo {author}
  {\bibfnamefont {C.}~\bibnamefont {Felser}}, \bibinfo {author} {\bibfnamefont
  {R.~J.}\ \bibnamefont {Cava}}, \ and\ \bibinfo {author} {\bibfnamefont
  {B.~A.}\ \bibnamefont {Bernevig}},\ }\href@noop {} {\bibfield  {journal}
  {\bibinfo  {journal} {Science}\ }\textbf {\bibinfo {volume} {353}},\ \bibinfo
  {pages} {aaf5037} (\bibinfo {year} {2016})}\BibitemShut {NoStop}%
\bibitem [{\citenamefont {Lv}\ \emph {et~al.}(2021)\citenamefont {Lv},
  \citenamefont {Qian},\ and\ \citenamefont {Ding}}]{lv2021experimental}%
  \BibitemOpen
  \bibfield  {author} {\bibinfo {author} {\bibfnamefont {B.}~\bibnamefont
  {Lv}}, \bibinfo {author} {\bibfnamefont {T.}~\bibnamefont {Qian}}, \ and\
  \bibinfo {author} {\bibfnamefont {H.}~\bibnamefont {Ding}},\ }\href@noop {}
  {\bibfield  {journal} {\bibinfo  {journal} {Reviews of Modern Physics}\
  }\textbf {\bibinfo {volume} {93}},\ \bibinfo {pages} {025002} (\bibinfo
  {year} {2021})}\BibitemShut {NoStop}%
\bibitem [{\citenamefont {Lin}\ and\ \citenamefont
  {Hughes}(2018)}]{lin2018topological}%
  \BibitemOpen
  \bibfield  {author} {\bibinfo {author} {\bibfnamefont {M.}~\bibnamefont
  {Lin}}\ and\ \bibinfo {author} {\bibfnamefont {T.~L.}\ \bibnamefont
  {Hughes}},\ }\href@noop {} {\bibfield  {journal} {\bibinfo  {journal}
  {Physical Review B}\ }\textbf {\bibinfo {volume} {98}},\ \bibinfo {pages}
  {241103} (\bibinfo {year} {2018})}\BibitemShut {NoStop}%
\bibitem [{\citenamefont {Ghorashi}\ \emph {et~al.}(2020)\citenamefont
  {Ghorashi}, \citenamefont {Li},\ and\ \citenamefont
  {Hughes}}]{ghorashi2020higher}%
  \BibitemOpen
  \bibfield  {author} {\bibinfo {author} {\bibfnamefont {S.~A.~A.}\
  \bibnamefont {Ghorashi}}, \bibinfo {author} {\bibfnamefont {T.}~\bibnamefont
  {Li}}, \ and\ \bibinfo {author} {\bibfnamefont {T.~L.}\ \bibnamefont
  {Hughes}},\ }\href@noop {} {\bibfield  {journal} {\bibinfo  {journal} {Phys.
  Rev. Lett.}\ }\textbf {\bibinfo {volume} {125}},\ \bibinfo {pages} {266804}
  (\bibinfo {year} {2020})}\BibitemShut {NoStop}%
\bibitem [{\citenamefont {Wang}\ \emph {et~al.}(2020)\citenamefont {Wang},
  \citenamefont {Lin}, \citenamefont {Jiang}, \citenamefont {Guo},\ and\
  \citenamefont {Jiang}}]{wang2020higher}%
  \BibitemOpen
  \bibfield  {author} {\bibinfo {author} {\bibfnamefont {H.-X.}\ \bibnamefont
  {Wang}}, \bibinfo {author} {\bibfnamefont {Z.-K.}\ \bibnamefont {Lin}},
  \bibinfo {author} {\bibfnamefont {B.}~\bibnamefont {Jiang}}, \bibinfo
  {author} {\bibfnamefont {G.-Y.}\ \bibnamefont {Guo}}, \ and\ \bibinfo
  {author} {\bibfnamefont {J.-H.}\ \bibnamefont {Jiang}},\ }\href@noop {}
  {\bibfield  {journal} {\bibinfo  {journal} {Phys. Rev. Lett.}\ }\textbf
  {\bibinfo {volume} {125}},\ \bibinfo {pages} {146401} (\bibinfo {year}
  {2020})}\BibitemShut {NoStop}%
\bibitem [{\citenamefont {Wieder}\ \emph
  {et~al.}(2020{\natexlab{b}})\citenamefont {Wieder}, \citenamefont {Wang},
  \citenamefont {Cano}, \citenamefont {Dai}, \citenamefont {Schoop},
  \citenamefont {Bradlyn},\ and\ \citenamefont {Bernevig}}]{wieder2020strong}%
  \BibitemOpen
  \bibfield  {author} {\bibinfo {author} {\bibfnamefont {B.~J.}\ \bibnamefont
  {Wieder}}, \bibinfo {author} {\bibfnamefont {Z.}~\bibnamefont {Wang}},
  \bibinfo {author} {\bibfnamefont {J.}~\bibnamefont {Cano}}, \bibinfo {author}
  {\bibfnamefont {X.}~\bibnamefont {Dai}}, \bibinfo {author} {\bibfnamefont
  {L.~M.}\ \bibnamefont {Schoop}}, \bibinfo {author} {\bibfnamefont
  {B.}~\bibnamefont {Bradlyn}}, \ and\ \bibinfo {author} {\bibfnamefont
  {B.~A.}\ \bibnamefont {Bernevig}},\ }\href@noop {} {\bibfield  {journal}
  {\bibinfo  {journal} {Nat. Comms.}\ }\textbf {\bibinfo {volume} {11}},\
  \bibinfo {pages} {627} (\bibinfo {year} {2020}{\natexlab{b}})}\BibitemShut
  {NoStop}%
\end{thebibliography}


%


\appendix 
\section{Disclination charge parity and the disclination filling anomaly in layered systems}\label{app:DiscChargeParity}
In this Appendix, we will discuss why the disclination charge parity is not a meaningful quantity for inversion-symmetric insulators, as it depends on the choice of inversion center for systems with open boundaries. However, as we shall see, the difference between the disclination charge parity for open and periodic boundary conditions (referred to in the main text as the disclination filling anomaly) does not depend on the inversion center, and is therefore a more physically relevant quantity. We shall establish this by considering layered insulators. Although we are only considering specific systems, the dependence of the disclination charge parity on the choice of inversion center is general. We recall from the main text that what we mean by a layered insulator is one that is adiabatically connected to a limit of decoupled layers where electrons are not hybrdized between different unit cells in the $z$-direction.

\begin{figure}
    \centering
   \includegraphics[width=0.45\textwidth]{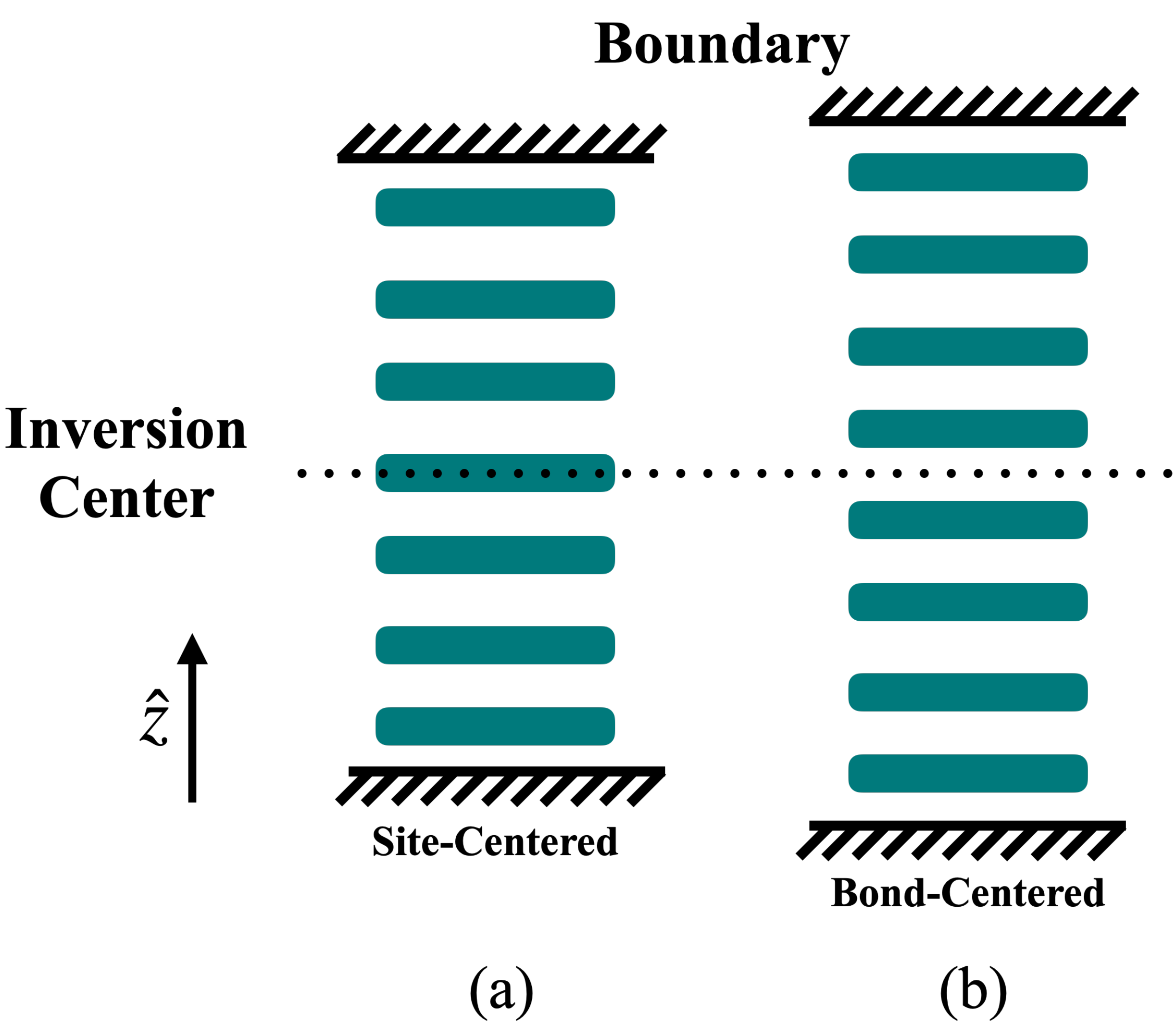}
    \caption[]{ An inversion-symmetric stack of 2D layers (green) for site-centered inversion symmetry (a) and bond-centered inversion symmetry (b). For site-centered inversion symmetry, one layer maps onto itself under inversion symmetry, while other layers map onto a partner, such that the total number of layers, $N_z$ is odd. For bond-centered inversion symmetry, all layers map onto a partner, such that the total number of layers, $N_z$ is even. 
    }\label{fig:DecoupledLayers}
\end{figure}

To this end, let us consider an inversion-symmetric 3D insulator composed of $N_z$ layers of 2D insulators, stacked along the $z$-direction. We further take the system to have open boundary conditions in the $z$-direction. For such a system, there are two choices of inversion symmetry that differ with respect to their inversion center. First, there is inversion symmetry where the inversion center is in a given layer, such that this layer maps to itself under inversion symmetry. This was referred to as site-centered inversion symmetry in the main text. Second, there is inversion symmetry where the inversion center is in-between two adjacent layers. This was referred to as bond-centered inversion symmetry in the main text. For site-centered inversion symmetry and open boundaries, $N_z$ must be odd, since under inversion symmetry, one layer maps to itself, while all other layers must map onto an inversion symmetry related partner. Similarly, for bond-centered inversion symmetry, $N_z$ must be even, since all layers map onto an inversion symmetry related partner. The two inversion-symmetric stacking configurations are shown in Fig.~\ref{fig:DecoupledLayers}(a),(b). 

\begin{figure}
    \centering
    \includegraphics[width=0.5\textwidth]{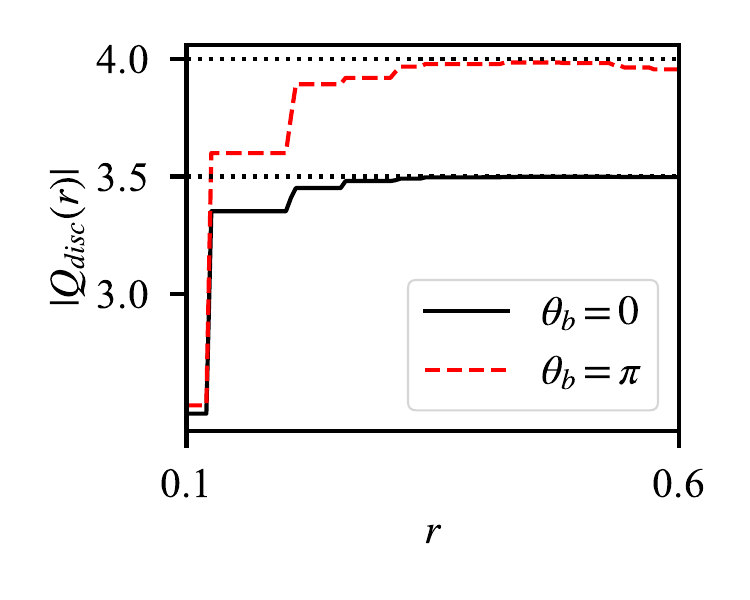}
    \caption{The charge bound to a $\Theta_F=\pi/2$ disclination for a Dirac-CDW insulator as a function of the integration radius distance $r$ with $Q=\pi/2$, $b_{xy}=1$, $\Delta_{\text{surf}}=0.25$, $\Delta_s=0$, $\Delta_b=0.5$, $N_x=15$, and $N_z=14$. The radius $r$ is scaled such that the farthest point from the disclination core is at $r=1$. Note that the $r$-axis only extends from $0.1$ to $0.6$.}
    \label{fig:q_disc_vs_r}
\end{figure}

We now consider the disclination response of the layered insulator. For a 3D layered insulator, the disclination response of the full system is simply the sum of responses of the 2D layers. Each 2D layer is described by a 2D dWZ term, with discrete shift $\mathcal{S}_{\text{2D}}$. For spin-1/2 fermions with TRS, $\mathcal{S}_{\text{2D}}$ is quantized as an even integer. Because of the 2D dWZ response, a $\Theta_F$ disclination line binds charge $Q_{\text{disc-2D}} = \mathcal{S}_{\text{2D}} \frac{\Theta_F}{2\pi}\mod(2)$ on each layer, where the $\mod(2)$ factor reflects that it is possible to locally add a Kramers degenerate pair of electrons to the disclination core. The total disclination charge of the full 3D system is then $Q_{\text{disc-3D}} = N\mathcal{S}_{\text{2D}} \frac{\Theta_F}{2\pi}\mod(2)$. For a finite size system of spin-1/2 fermions with time-reversal symmetry (TRS), the disclination charge parity is equal to the total charge on disclination line $\mod(2\frac{\Theta_F}{\pi})$,
\begin{equation}
    Q_{\text{disc-3D}}\,\,\text{mod}\left(2\frac{\Theta_F}{\pi}\right) = \mathcal{S}_{\text{2D}} N_z \frac{\Theta_F}{2\pi} \,\,\text{mod}\left(2\frac{\Theta_F}{\pi}\right),
\end{equation}
where we have used the fact that $\Theta_F$ is a multiple of $2\pi/n$ for $C_n$ symmetric insulators. The disclination charge parity is therefore zero when $N_z\mathcal{S}_{\text{2D}}/2$ is even, and non-zero when $N_z\mathcal{S}_{\text{2D}}/2$ is odd. From this, we can conclude the following: when $\mathcal{S}_{\text{2D}}/2$ is even, the disclination charge parity always vanishes, but when $\mathcal{S}_{\text{2D}}/2$ is odd, the disclination charge parity vanishes when $N_z$ is even and is non-zero when $N_z$ is odd. So when $\mathcal{S}_{\text{2D}}/2$ is odd, the disclination charge parity always vanishes for bond centered inversion symmetry, and is always non-zero for site-centered inversion symmetry. This occurs for an insulator composed of quantum spin Hall layers, each of which have $\mathcal{S}_{\text{2D}}=2$.

Having established that the disclination charge parity depends on the choice of inversion center, let us now consider the disclination filling anomaly. To do this, we will take the layered system and ``sew" the top and bottom layers to one another such that the system has periodic boundaries in the $z$-direction. For a layered system, the sewing procedure is trivial by definition and will not change the total disclination charge, or, by extension, the disclination charge parity. The disclination filling anomaly will therefore necessarily vanish for all layered systems, regardless of the choice of inversion center or the value of $\mathcal{S}_{\text{2D}}$.

\section{Details of the numerics}\label{app:Numerics}
Here we discuss the details of how we calculate the charge bound to a disclination and the extrapolations in system size that we performed. To calculate the charge bound to a disclination, we sum the charge on all lattice sites in each layer that are within some radius from the disclination core. In Fig.~\ref{fig:q_disc_vs_r} we plot the charge bound to a disclination as a function of the integration radius for a Dirac-CDW insulator with $Q=\pi/2$, $N_x=15$, $N_z=14$, $b_{xy}=1$, $\Delta_{\text{surf}}=0.25$, $\Delta_s=0$, $\Delta_b=0.5$, and both $\theta_b=0$ and $\theta_b=\pi$. The radius $r$ is scaled such that the farthest point from the disclination core in the lattice is at $r=1$. The disclination charges approach the theoretical predictions, marked by dashed horizontal lines, for roughly $r>0.3$, indicating the exponential localization of the disclination charge. The disclination charge approaches zero for both small and large $r$, and the largest values of the disclination charge generally approach the theoretical predictions, except for very small system sizes. However, the integration radius at which the disclination charge most closely obtains the theoretical prediction varies significantly as a function of system size and other parameters. As such, we always calculate the disclination charge using a range of integration radius and report the maximal absolute value.

To diminish the impact of finite-size effects, we calculated the charges for each case enumerated in Table~\ref{table:NumericResults} over a range of system sizes and extrapolated to the infinite system-size limit. We found that effects arising from the finite extent of the lattice in the $z$-direction were minimal, so for each calculation we fixed $N_z$ and varied $N_x$, the side length of the disclinated lattice. In Table~\ref{table:system_size_scaling} we plot the disclination charges as a function of $N_x$ along with fits to a decaying exponential, $|Q_{\text{disc}}(N_x)| = Q_\infty + b e^{-cN_x}$. We plot the infinite system size limit, $Q_\infty$, with red dashed lines and report the values and uncertainties in the insets. The values reported in Table~\ref{table:NumericResults} are the values of $Q_\infty$ rounded to the nearest half integer, which produces less than a $1\%$ error in all cases.

\newcolumntype{M}[1]{>{\centering\arraybackslash}m{#1}}
\def\figw{3.6cm}

\begin{table*}
\centering
\begin{tabular}{ M{1.4cm} | M{\figw} M{\figw} | M{\figw} M{\figw}} 
    \multicolumn{1}{c|}{\multirow{2}{*}{$\mathcal{K}a_z$}} &
    \multicolumn{2}{c|}{OBC} &
    \multicolumn{2}{c}{PBC} \\
    &
    $\theta_{s/b}=0$ &
    $\theta_{s/b}=\pi$ &
    $\theta_{s/b}=0$ &
    $\theta_{s/b}=\pi$ \\
    \hline
    \rule{0pt}{1.5\normalbaselineskip}
    $\pi/2$ & \parbox[m]{\figw}{\includegraphics[width=\figw]{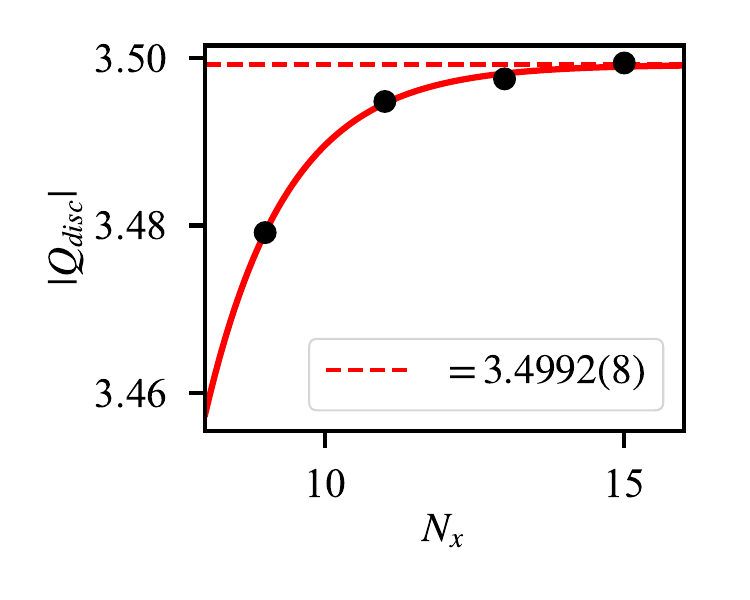}} & \parbox[m]{\figw}{\includegraphics[width=\figw]{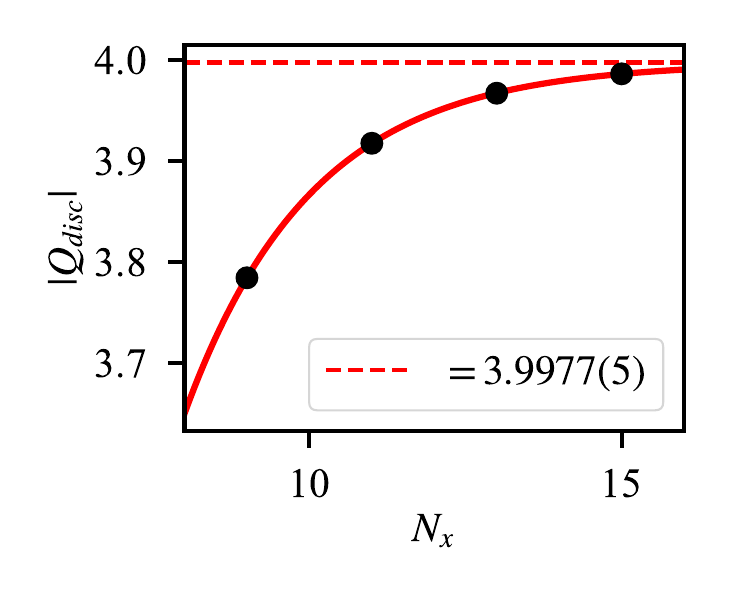}} & \parbox[m]{\figw}{\includegraphics[width=\figw]{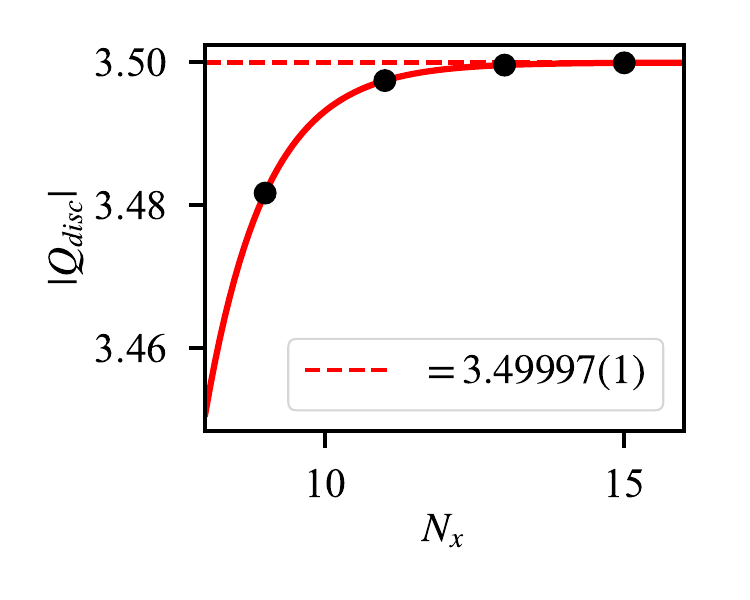}} & \parbox[m]{\figw}{\includegraphics[width=\figw]{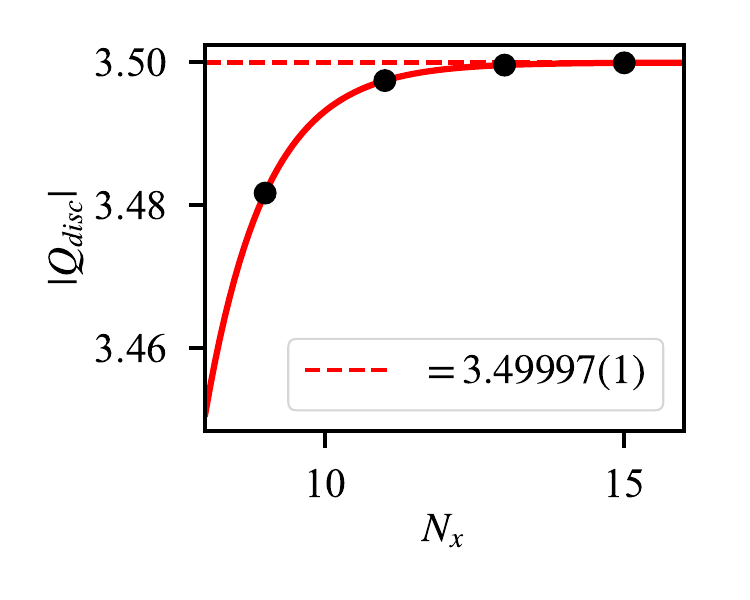}} \\
    \rule{0pt}{1.5\normalbaselineskip}
    $\pi/3$ & \parbox[m]{\figw}{\includegraphics[width=\figw]{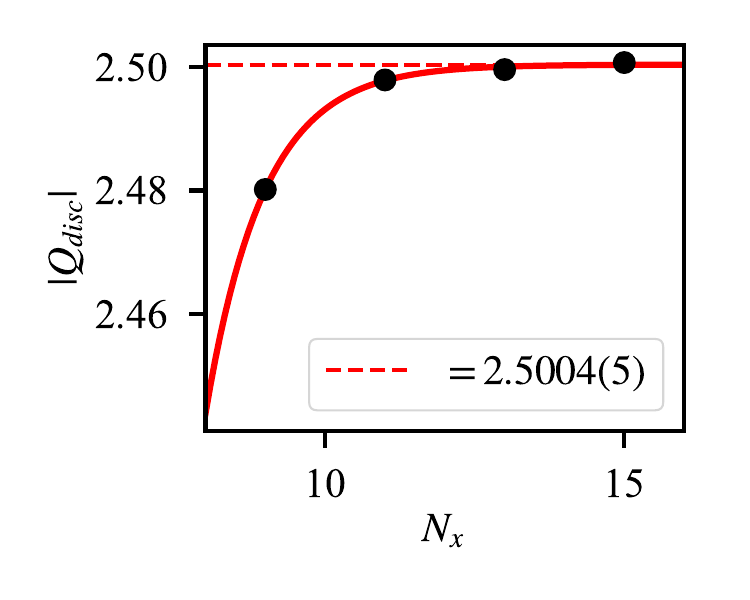}} & \parbox[m]{\figw}{\includegraphics[width=\figw]{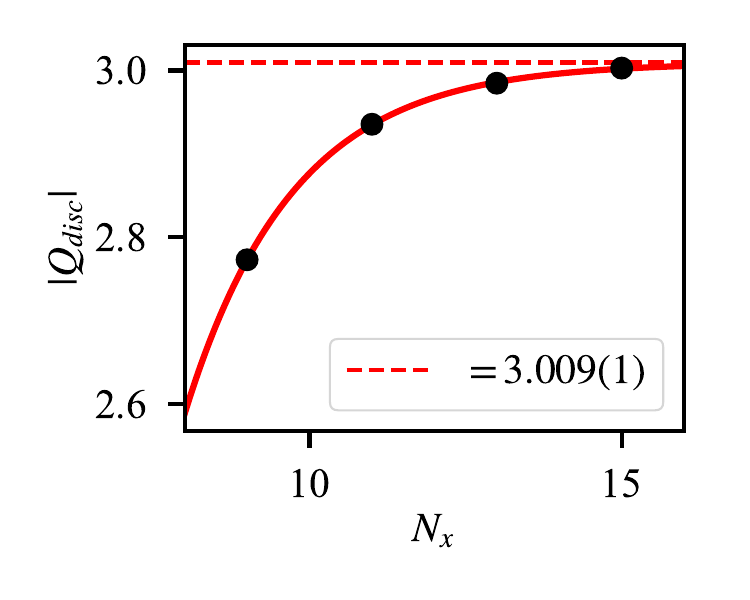}} & \parbox[m]{\figw}{\includegraphics[width=\figw]{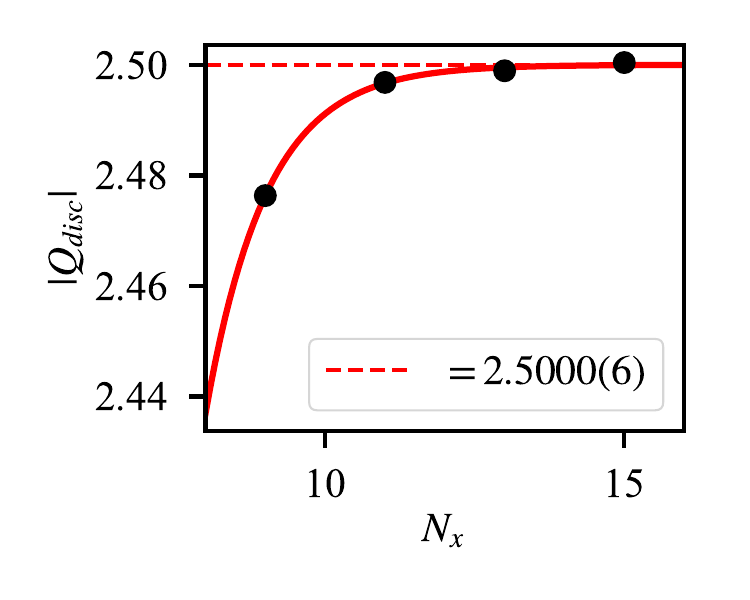}} & \parbox[m]{\figw}{\includegraphics[width=\figw]{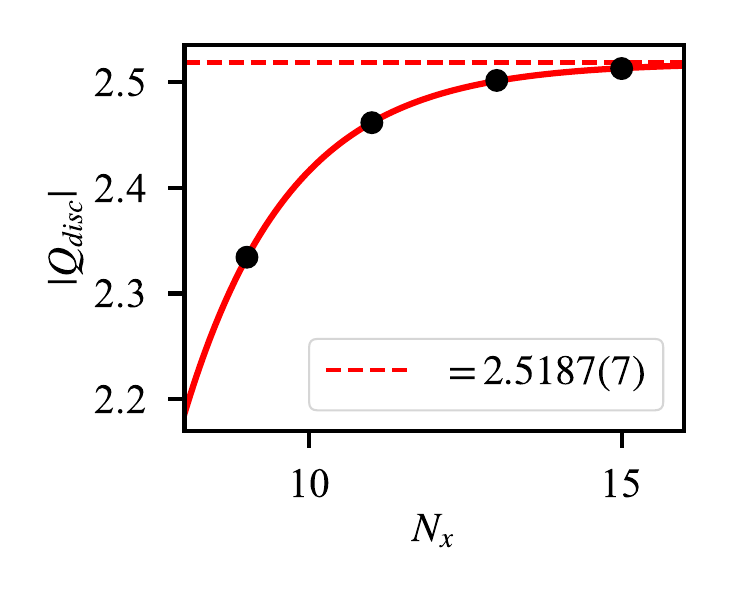}} \\
    \rule{0pt}{1.5\normalbaselineskip}
    $\pi/2\sqrt{2}$  & \parbox[m]{\figw}{\includegraphics[width=\figw]{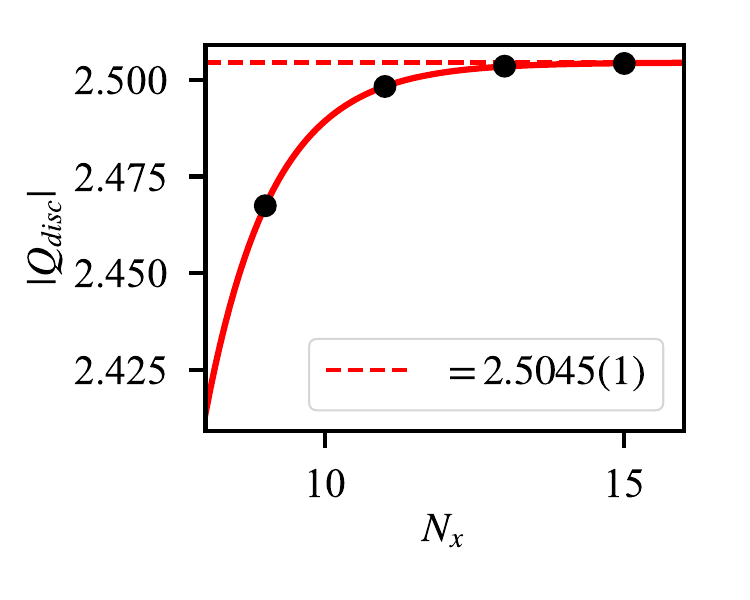}} & \parbox[m]{\figw}{\includegraphics[width=\figw]{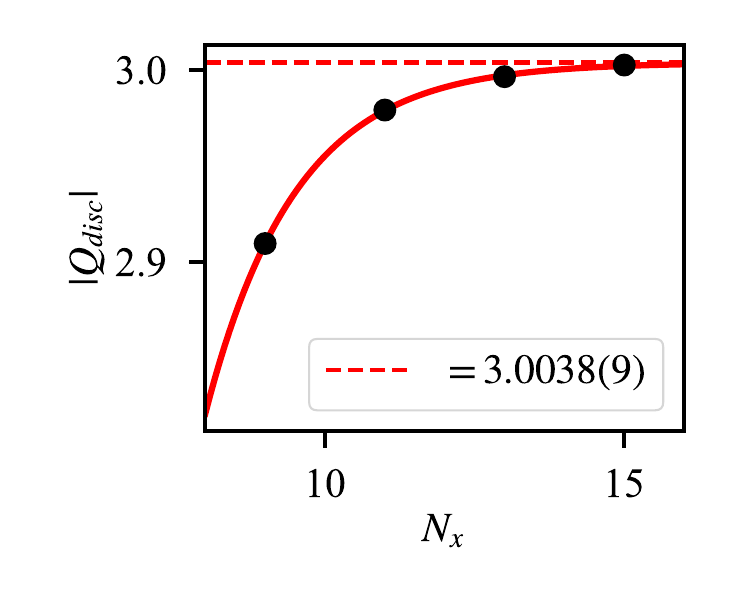}} & \parbox[m]{\figw}{} & \parbox[m]{\figw}{}
\end{tabular}
\caption{System size extrapolations of the charge bound to a $\Theta_F=\pi/2$ disclination for $\mathcal{K}=\pi/2$, $\pi/3$, and $\pi/2\sqrt{2}$ with $\theta_{s/b}=0$ and $\pi$ for both open and periodic boundary conditions. For each calculation we used $b_{xy}=1$, $\Delta_{\text{surf}}=0.25$, and $\Delta_s=0$. The system size along the disclination was set to $N_z=14$ for $\mathcal{K}=\pi/2$ and $N_z=15$ for $\mathcal{K}=\pi/3$ and $\pi/2\sqrt{2}$. The solid red lines are fits to a decaying exponential, $|Q_{\text{disc}}(N_x) = Q_\infty + b e^{-cN_x}$. The value and uncertainty of the large-$N_x$ limit, $Q_\infty$, is reported in the legend and plotted with a red dashed line.}
\label{table:system_size_scaling}
\end{table*}

\end{document}